\documentclass[a4paper]{jpconf}\usepackage{amsmath,amsfonts,amssymb,bm}
\usepackage{dsfont}
\usepackage{graphicx}
\usepackage{color}
\usepackage{soul}
\usepackage[comma,square,sort,compress,numbers]{natbib}

\usepackage[mathscr,scaled=1.15]{urwchancal}
\DeclareFontFamily{OT1}{pzc}{}
\DeclareFontShape{OT1}{pzc}{m}{it}%
{<-> s * [1.15] pzcmi7t}{}
\DeclareMathAlphabet{\mathpzc}{OT1}{pzc}{m}{it}

\definecolor{purple}{rgb}{0.5,0,0.5}
\definecolor{blue}{rgb}{0.0,0,0.9}

\newcommand{\Aslash}{\mbox{$\not \!\! A$}}
\newcommand{\dslash}{\mbox{$\not \! \partial$}}

\newcommand{\psib}{\mbox{$\overline{\psi}$}}

\begin{document}
\title{Three Lectures on Hadron Physics}

\author{Craig D. Roberts}

\address{Physics Division, Argonne National Laboratory, Argonne, Illinois 60439, USA}

\ead{cdroberts@anl.gov}

\begin{abstract}
These lectures explain that comparisons between experiment and theory can expose the impact of running couplings and masses on hadron observables and thereby aid materially in charting the momentum dependence of the interaction that underlies strong-interaction dynamics.  The series begins with a primer on continuum QCD, which introduces some of the basic ideas necessary in order to understand the use of Schwinger functions as a nonperturbative tool in hadron physics.  It continues with a discussion of confinement and dynamical symmetry breaking (DCSB) in the Standard Model, and the impact of these phenomena on our understanding of condensates, the parton structure of hadrons, and the pion electromagnetic form factor.  The final lecture treats the problem of grand unification; namely, the contemporary use of Schwinger functions as a symmetry-preserving tool for the unified explanation and prediction of the properties of both mesons and baryons.  It reveals that DCSB drives the formation of diquark clusters in baryons and sketches a picture of baryons as bound-states with Borromean character.  Planned experiments are capable of validating the perspectives outlined in these lectures.
\end{abstract}


\setcounter{equation}{0}
\renewcommand{\theequation}{\arabic{section}.\arabic{equation}}

\section{Continuum QCD: A Primer}
\subsection{Quantum chromodynamics}
\label{sec:QCD}
Quantum chromodynamics (QCD) holds many fascinations, which are somehow hidden in the seemingly simple local Lagrangian density that defines the theory:
\begin{eqnarray}
\label{LQCD}
{\cal L}_{\rm QCD} &=& \bar q_i [i \gamma^\mu [D_\mu]_{ij} - M \delta_{ij} ] q_j - \frac{1}{4} G_{\mu\nu}^a G_a^{\mu\nu}\\
&=& \bar q_i [i \gamma^\mu \partial_\mu - m] q_i - g G_\mu^a\bar q_i \gamma^\mu T^a_{ij} q_j - \frac{1}{4} G_{\mu\nu}^a G_a^{\mu\nu}
\label{LQCD2}
\end{eqnarray}
where $\{T^a|a=1,\ldots,8\}$ are the generators of $SU_c(3)$ in the fundamental representation; $G$ represents the gluon gauge-fields; $q$, the quark matter-fields; $g$ is the coupling constant; and the gluon field-strength tensor is
\begin{equation}
\label{gluonFST}
G_{\mu\nu}^a = \partial_\mu G_\nu^a - \partial_\nu G_\mu^a + \underline{g f^{abc} G_\mu^b G_\nu^c}.
\end{equation}

A first observation is that \emph{QCD works}: there is no confirmed breakdown over an enormous energy domain: $0<E<8\,$TeV.  Furthermore, as these notes will explain, it is probable that QCD is a self-contained, nonperturbatively renormalisable and hence well-defined quantum field theory.\footnote{This would make QCD unique.  In contrast, it is certainly not true of quantum electrodynamics (QED), an Abelian gauge theory, which is perturbatively simple but, alone, undefined nonperturbatively: four-fermion operators become relevant in strong-coupling QED and must be included in order to obtain a well-defined continuum limit \cite{Rakow:1990jv, Reenders:1999bg, Akram:2012jqS}.}  If that is so, then QCD is a true theory, not an effective theory.  This possibility and the lack of any evidence for supersymmetry in experiments at the large hadron collider (LHC) have led to a resurgence of interest in the possibility that any reasonable extension of the Standard Model will be based on the paradigm established by QCD.  One possibility is Extended Technicolour, in which electroweak symmetry breaks via a fermion bilinear operator in a strongly-interacting non-Abelian theory \cite{Andersen:2011yj, Sannino:2013wla} and the Higgs sector of the Standard Model becomes an effective description of a more fundamental fermionic theory, similar to the Ginzburg-Landau theory of superconductivity.  These features and possibilities lend urgency to the problem of solving QCD, and hence also to modern programs in hadron physics.

The underlined term in Eq.\,\eqref{gluonFST} is the source of everything that is remarkable about QCD because it generates gauge-field self-interactions, which have extraordinary consequences.  This is readily elucidated through a comparison with QED.  A characteristic feature of QED is that gauge-boson self interactions do not occur at tree level, \emph{i.e}.\ at lowest order in perturbation theory. The leading contribution occurs at order $\alpha^4$; and since $\alpha \sim 1/137$, such interactions take place with extremely small probability.  In contrast, the non-Abelian character of QCD produces tree-level interactions between gauge bosons -- the three- and four-gluon vertices generated by $G_{\mu\nu}^a G_a^{\mu\nu}$ in Eq.\,\eqref{LQCD}.  One might guess that these terms should have a big impact; and doing better than guessing led to the 2004 Nobel Prize in Physics \cite{Politzer:1973fx, Politzer:1974fr, Gross:1973id}.

\begin{figure}[t]
\vspace*{-1ex}

\includegraphics[clip,width=0.5\textwidth,height=0.222\textheight]{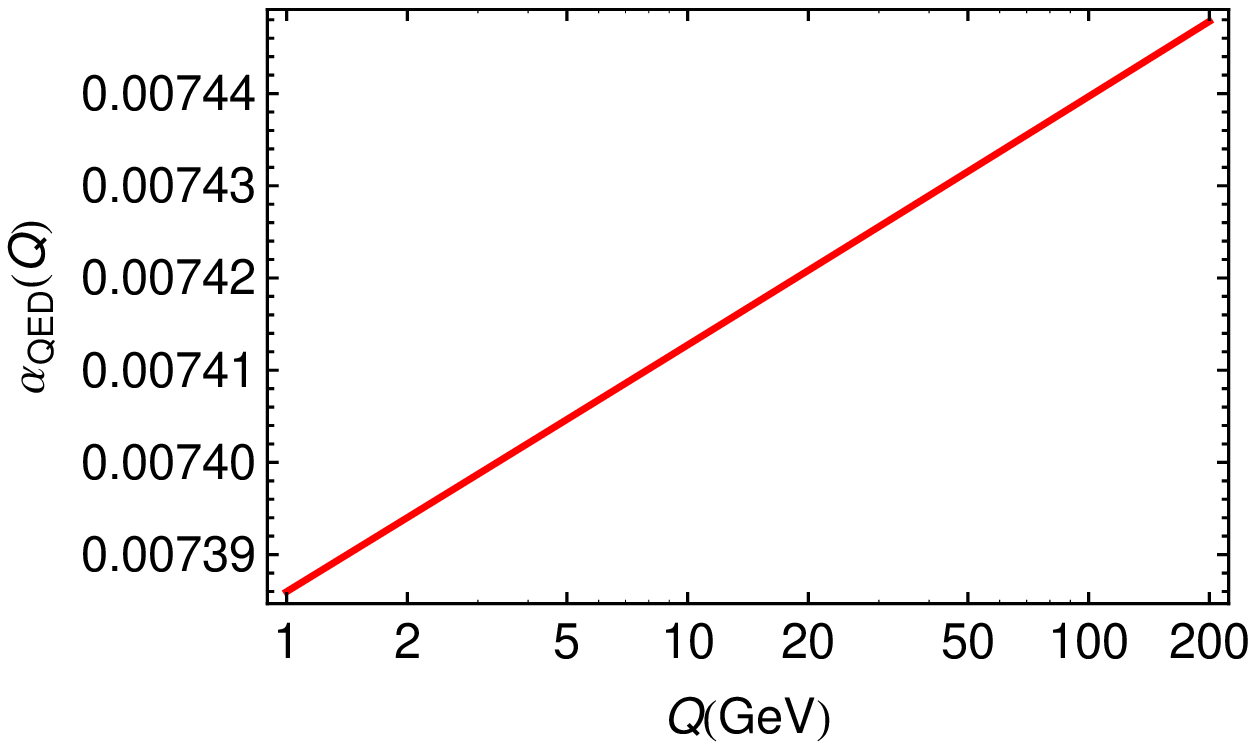}
\includegraphics[clip,width=0.455\textwidth]{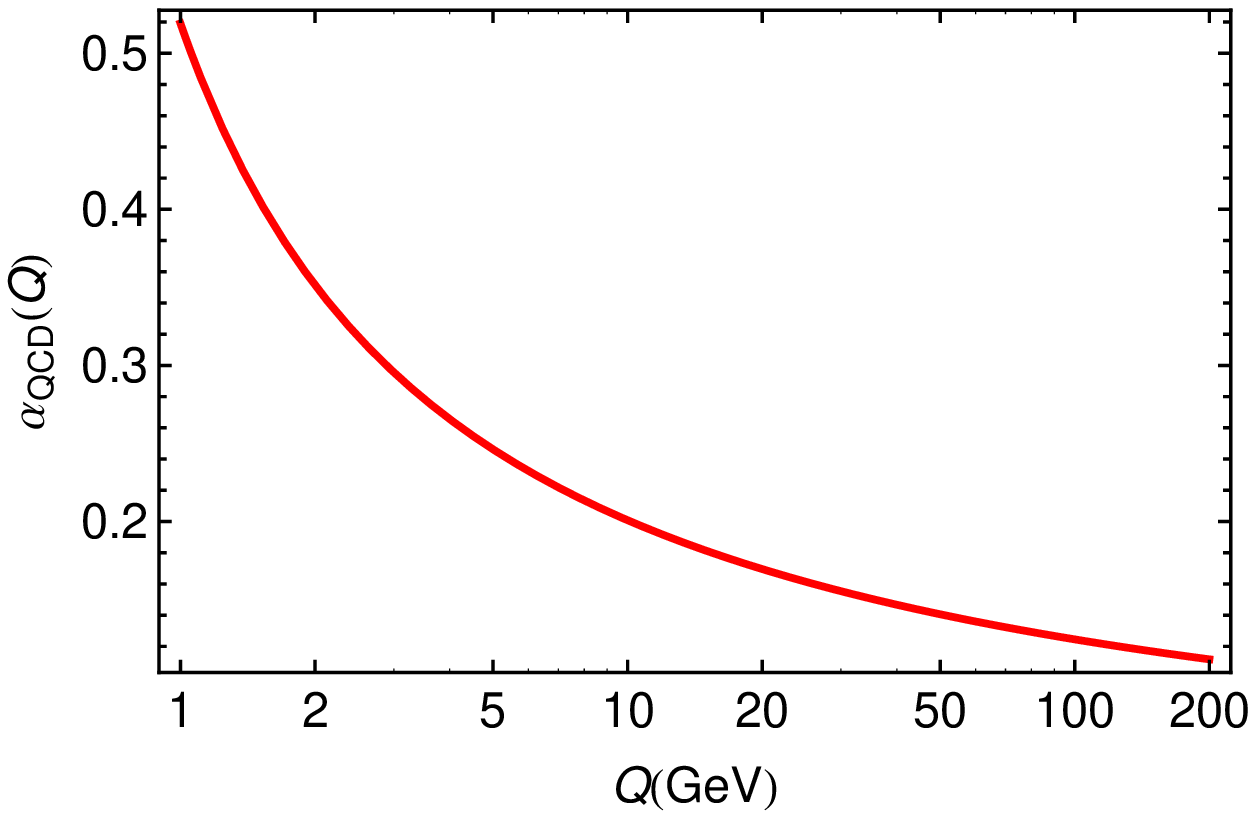}

\caption{\label{F3} \small \emph{Left panel} -- Running coupling in QED; and \emph{right panel} -- running coupling in QCD.  \emph{N.B}.\ On the same momentum domain, the QCD coupling changes 500\,000-times more than the QED coupling and runs in the opposite direction.}
\end{figure}

Relativistic quantum gauge-field theories are typified by the feature that nothing is constant.  The distribution of charge and mass, the number of particles, \emph{etc}.; indeed, all the things that quantum mechanics holds fixed come to depend upon the wavelength of the tool being used to measure them.  Couplings and masses are renormalised via processes involving virtual-particles.  Such effects make these quantities depend on the energy scale at which one observes them, and produce the running couplings illustrated in Fig.\,\ref{F3}.  At one-loop order, the QED coupling is
\begin{equation}
\alpha_{\rm QED}(Q) = \frac{\alpha}{\displaystyle 1-\frac{2 \alpha}{3\pi}\ln \frac{Q}{m_e}},
\end{equation}
where $\alpha$ and $m_e$ are renormalised such that they take their observed values when measured at the physical electron mass, $Q=m_e$.  The negative sign in the denominator signals that fermions screen electric charge.  In QCD, the analogous one-loop quantity is
\begin{equation}
\label{eqalphaQCD}
\alpha_{\rm QCD}(Q) = \frac{6 \pi}{\displaystyle (33 - 2 N_f) \ln \frac{Q}{\Lambda_{\rm QCD}}},
\end{equation}
where $N_f$ is the number of quark flavours that are active in the process under consideration and $\Lambda_{\rm QCD}\sim 300\,$MeV is a mass-scale.\footnote{$\Lambda_{\rm QCD}$ is the natural mass-scale of QCD.  It is dynamically generated, via quantisation.  The appearance of $\Lambda_{\rm QCD}$ spoils the conformal invariance of the classical massless theory \cite{Collins:1976yq, Nielsen:1977sy, tarrach}.  The value of $\Lambda_{\rm QCD}$ must be determined empirically within the Standard Model; and it sets the scale for all dynamically generated masses described herein.}
Plainly, quarks screen the colour charge.  However, owing to their self-interactions, gluons antiscreen this charge.  The enormous impact of that effect is evident in Fig.\,\ref{F3}.

The running coupling depicted in the right panel of Fig.\,\ref{F3} can be translated into an interaction between quarks and gluons that depends strongly on separation, \emph{i.e}.\ an interaction which grows rapidly with separation between the sources.  Taking the experimental value of $\Lambda_{\rm QCD}$, the coupling is found to be huge at a separation $r \simeq \frac{1}{4}r_p$, where $r_p$ is the proton's charge radius.  This is a peculiar circumstance, \emph{viz}.\ an interaction that becomes stronger as the participants try to separate.  It leads one to explore some fascinating possibilities: if the coupling grows so strongly with separation, then perhaps it is unbounded; and conceivably it would require an infinite amount of energy in order to extract a quark or gluon from the interior of a hadron?
This manner of thinking has led to the \\[0.7ex]
\hspace*{2em}\parbox[t]{0.9\textwidth}{\textit{Confinement Hypothesis}: \emph{Colour-charged particles cannot be isolated and therefore cannot be directly observed.  They clump together in colour-neutral bound-states}.}\\

\hspace*{-\parindent}This is an empirical fact; but there is no mathematical demonstration.  Possibly in the hope of focusing attention in that direction, at the turn of the Millennium the Clay Mathematics Institute offered a prize of \$1-million for a proof that $SU_c(3)$ gauge theory is mathematically well-defined \cite{Jaffe:Clay}, one necessary consequence of which will be an answer to the question of whether or not the confinement conjecture is correct.  I will return to this issue.

\subsection{Hadron Physics}
\label{secHP}
The term ``hadron,'' first used in 1962 \cite{1962hep..conf..845O}, refers to a class of subatomic particles that are composed of quarks and/or gluons and take part in the strong interaction.  Well known examples are the proton, neutron and pion.  The world of hadrons is subdivided into baryons, which are hadrons with half-integer spin that contain three valence-quarks; and mesons, hadrons with integer spin that contain a valence-quark and -antiquark.  Hadron physics is unique at the cutting edge of modern science because Nature has provided us with just one instance of a fundamental strongly-interacting theory, \emph{viz}.\ QCD: basic science has never before confronted such a challenge as solving this theory.

In facing this task, hadron physics has developed into an international research endeavour of remarkable scope.  Indeed, before the end of this decade the field will be operating a host of upgraded or new facilities and detectors.
An illustrative list may readily be compiled: Beijing's electron-positron collider;
J-PARC, the Japan Proton Accelerator Research Complex, 150km NE of Tokyo;
the ALICE and COMPASS detectors at CERN;
%
%
and, in the USA, both RHIC (Relativistic Heavy Ion Collider), at Brookhaven National Laboratory, and JLab (Thomas Jefferson National Accelerator Facility), in Newport News, VA.  RHIC focuses primarily on the strong-interaction phase transition, \emph{i.e}.\ physics just 10$\mu$s after the Big Bang;
and JLab explores the nature of cold hadronic matter.  Regarding the upgrade of JLab \cite{Dudek:2012vr}, commissioning of Hall-A and Hall-D are now taking place, a process that is itself expected to yield new physics results, and project completion is expected in 2017.
In addition to these existing facilities and those under construction, excitement is also generated by the discovery potential of proton-nucleus collisions at the LHC \cite{Salgado:2011wcS} and new machines, such as an electron ion collider (EIC) \cite{Accardi:2012qut}.

A theoretical understanding of the phenomena of hadron physics requires use of the full machinery of relativistic quantum field theory.  Based on the relativistic dynamics elucidated by Dirac \cite{Dirac:1949cp}, it is the \emph{only} known way to reconcile quantum mechanics with special relativity.

It is worth highlighting that the unification of relativity with quantum mechanics took some time; and questions still remain as to a practical implementation of a Hamiltonian formulation of the relativistic quantum mechanics of interacting systems.  The Poincar\'e group has ten generators: six associated with the Lorentz transformations (rotations and boosts); and four associated with translations.  Quantum mechanics describes the time evolution of a system with interactions and that evolution is generated by the Hamiltonian, or some generalisation thereof.  However, the Hamiltonian is one of the generators of the Poincar\'e group, and it is apparent from the Poincar\'e algebra that boosts do not generally commute with the Hamiltonian.
This is partly because the existence of antiparticles is concomitant with relativistic quantum mechanics, \emph{i.e}.\ the equations of relativistic quantum mechanics admit \emph{negative energy} solutions.  However, once one allows for negative energy, then particle number conservation is lost:
\begin{equation}
\label{eqParticleNumber}
E_{\rm system} = E_{\rm system} + (E_{p_1} + E_{\bar{p}_1}) + \ldots\ \mbox{\emph{ad~infinitum}},
\end{equation}
where $E_{\bar p}= - E_{p}$.  This poses a fundamental problem for relativistic quantum mechanics: few particle systems can be studied, but the study of (infinitely) many bodies is difficult and no general theory currently exists.
Following from these observations, the state vector calculated in one momentum frame will not be kinematically related to the state vector for the original system in another frame, a fact that makes a new calculation necessary in every momentum frame.  The discussion of scattering, which takes a state of momentum $p$ to a different state with momentum $p^\prime$ is therefore problematic \cite{Dirac:1949cp, Keister:1991sb, Coester:1992cg, Brodsky:1997de}.

Relativistic quantum field theory provides a way forward.  In this framework the fundamental entities are fields, which can simultaneously represent infinitely many particles.  The neutral scalar field, $\phi(x)$, provides an example.  One may write
\begin{equation}
\phi(x) = \int\frac{d^3 k}{(2\pi)^3 }\frac{1}{2 \omega_k} \left[ a(k) {\rm e}^{-i k\cdot x} + a^\dagger(k) {\rm e}^{i k\cdot x} \right],
\end{equation}
where: $\omega_k=\sqrt{|\vec{k}|^2+m^2}$ is the relativistic dispersion relation for a massive particle; the four-vector $(k^\mu)= (\omega_k,\vec{k})$; $a(k)$ is an annihilation (creation) operator for a particle (antiparticle) with four-momentum $k$ ($-k$); and $a^\dagger(k)$ is a creation (annihilation) operator for a particle (antiparticle) with four-momentum $k$ ($-k$).  With this plane-wave expansion of the field one may proceed to develop a framework in which the nonconservation of particle number is not a problem.  That is crucial because key observable phenomena in hadron physics are essentially connected with the existence of \emph{virtual} particles.

Relativistic quantum field theory has its own problems, however.  For example, the question of whether a given quantum field theory is rigorously well defined is an \emph{unsolved} mathematical problem.  All relativistic quantum field theories admit analysis via perturbation theory, and perturbative renormalisation is a well-defined procedure that has long been used in QED and QCD.  However, the rigorous definition of a theory means proving that the theory makes sense \emph{nonperturbatively}.  This is equivalent to proving that all the theory's renormalisation constants are nonperturbatively well-behaved.

Understanding the properties of hadrons requires solving QCD.  However, despite the possibility highlighted above and discussed further below, QCD is not now known to be a rigorously well-defined theory.  It might not have a solution.  Thus, more properly, the task of hadron physics is to determine whether QCD is truly the theory of the strong interaction.

Physics is an experimental science and experiment advances quickly.  Hence, developing an understanding of observable phenomena cannot wait on mathematical rigour.  Assumptions must be made and their consequences explored.  Hadron physics practitioners therefore assume that QCD is (somehow) well-defined and follow where that idea may lead.  This means exploring and mapping the hadron physics landscape with well-understood probes, such as the electron at JLab; and employing established mathematical tools, refining and inventing others, in order to use the Lagrangian density of QCD to predict what should be observable real-world phenomena.

A primary aim of the world's current hadron physics programmes in experiment and theory is to determine whether there are any contradictions with what can actually be \emph{proved} in QCD.  There are none today; but that doesn't mean there are neither puzzles nor controversies.  The interplay between experiment and theory is the engine of discovery and progress: the discovery potential of both is high.  Much has been learnt in the last ten years; and these lectures will describe some of those discoveries and provide a perspective on their meaning.  Many of the most important questions in basic science are the purview of Hadron Physics.

\subsection{Green Functions and Propagators}
The Dirac equation provides the starting point for a Lagrangian formulation of the quantum field theory for fermions interacting via gauge boson exchange. For a free fermion with mass $m$,
\begin{equation}
\label{freeDirac} [ i \partial\!\!\!\slash_{x}\, - m ] \, \psi(x) = 0\,,
\end{equation}
where $\psi(x)$ is the fermion's ``spinor'' -- a four component column vector, while in the presence of an external electromagnetic field the fermion's wave function obeys
\begin{equation}
\label{ADirac}
[ i \partial \!\!\!\slash_{x}\, - e A\!\!\!\slash (x)\,- m ] \, \psi(x) = 0
\,,
\end{equation}
which is obtained, as usual, via ``minimal substitution:'' $\mbox{\boldmath $p$}^\mu \to \mbox{\boldmath $p$}^\mu - e \mbox{\boldmath $A$}^\mu$ in Eq.\,(\ref{freeDirac}).  These equations have a manifestly covariant appearance and a proof of covariance is given in the early chapters of Refs.\,\cite{BD64, IZ80}.

The Dirac equation is a partial differential equation.  A general method for solving such equations is to use a Green function, which is the inverse of the differential operator that appears in the equation.  The analogy with matrix equations is obvious and can be exploited heuristically.

Equation~(\ref{ADirac}) yields the wave function for a fermion in an external electromagnetic field.  In this connection, consider the operator obtained as a solution of the following equation
\begin{equation}
\label{SAprop} [ i \partial\!\!\!\slash_{x^\prime}\, - e
A\!\!\!\slash (x^\prime) \,- m ]\,S(x^\prime,x) = \mbox{\boldmath
$1$}\,\delta^4(x^\prime -x)\,.
\end{equation}
It is immediately apparent that if, at a given spacetime point $x$, $\psi(x)$ is a solution of Eq.\,(\ref{ADirac}), then
\begin{equation}
\label{Spsi} \psi(x^\prime) := \int d^4 x\,S(x^\prime,x)\,\psi(x)
\end{equation}
is a solution of
\begin{equation}
[ i \partial\!\!\!\slash_{x^\prime}\, - e A\!\!\!\slash (x^\prime)\,- m ] \,
\psi(x^\prime) = 0 \,,
\end{equation}
\emph{i.e}.\ $S(x^\prime,x)$ has propagated the solution at $x$ to the point $x^\prime$.

This effect is equivalent to the application of Huygens' principle in wave mechanics: if the wave function at $x$, $\psi(x)$, is known, then the wave function at $x^\prime$ is obtained by considering $\psi(x)$ as a source of spherical waves that propagate outward from $x$.  The amplitude of the wave at $x^\prime$ is proportional to the amplitude of the original wave, $\psi(x)$, and the constant of proportionality is the propagator (Green function), $S(x^\prime,x)$.  The total amplitude of the wave at $x^\prime$ is the sum over all points on the wavefront, \emph{viz}.\ Eq.\,(\ref{Spsi}).

Such an approach is practical because all physically reasonable external fields can only be nonzero on a compact subdomain of spacetime.  Therefore the solution of the complete equation is transformed into solving for the Green function, which can then be used to propagate the free-particle solution, already found, to arbitrary spacetime points. However, obtaining the exact
form of $S(x^\prime,x)$ is impossible for all but the simplest cases (see, \emph{e.g}.\ Ref.\,\cite{DR85,DR86}).

In the absence of an external field Eq.\,(\ref{SAprop}) becomes
\begin{equation}
\label{S0prop} [ i \partial\!\!\!\slash_{x^\prime}\, \,- m ]\,S(x^\prime,x) =
\mbox{\boldmath $1$}\,\delta^4(x^\prime -x)\,.
\end{equation}
Assume a solution of the form:
\begin{equation}
\label{S0x} S_0(x^\prime,x) = S_0(x^\prime -x) = \int\frac{d^4 p}{(2\pi)^4} \,
{\rm e}^{-i (p,x^\prime-x)} \, S_0(p)\,,
\end{equation}
so that substituting yields
\begin{equation}
\label{S0undef} (p \!\!\!\slash\, - m)\, S_0(p) = \mbox{\boldmath $1$}\,;\;{\it
i.e.}\;S_0(p)= \frac{p \!\!\!\slash\, + m}{p^2-m^2}\,.
\end{equation}

To obtain the result in configuration space one must adopt a prescription for handling the on-shell singularities in $S(p)$, \emph{viz}.\ the pole at $p^2=m^2$, and that convention is linked with the boundary conditions applied to Eq.\,(\ref{S0prop}). An obvious and physically sensible definition of the Green function is that it should propagate positive-energy-fermions and -antifermions forward in time but not backwards in time, and vice versa for negative energy states.

Consider now that the wave function for a positive energy free-fermion is
\begin{equation}
\psi^{(+)}(x)= u(p)\, {\rm e}^{-i (p,x)}\,.
\end{equation}
The wave function for a positive-energy antifermion is the charge-conjugate of the negative-energy fermion solution:
\begin{equation}
\psi_c^{(+)}(x) = C \, \gamma^0\,\left( v(p) \, {\rm e}^{i (p,x)} \right)^\ast
= C \, \bar v(p)^{\rm T}\,{\rm e}^{-i (p,x)}\,,
\end{equation}
where $C= i\gamma^2\gamma^0$ and $(\cdot)^{\rm T}$ denotes matrix transpose.\footnote{This defines the operation of charge conjugation.  \emph{N.B}.\  The form of $C$ introduced here is appropriate to a particular Dirac matrix representation. Nevertheless, the unitary equivalence of Clifford algebra representations ensures that generality is not lost \protect\cite{BD64}.}
It is thus evident that the physically sensible $S_0(x^\prime -x)$ must only contain positive-frequency components for $x_0^\prime -x_0 >0$, \emph{i.e}.\ it must be proportional to the positive-energy projection operator $\Lambda_+(p) = (p \!\!\!\slash\,+m)/(2m)$.  One can ensure this via a small modification of the denominator in Eq.\,(\ref{S0undef}):
\begin{equation}
\label{S0def} S_0(p)= \frac{p \!\!\!\slash\, + m}{p^2-m^2} \to
\frac{p \!\!\!\slash\, + m}{p^2-m^2 + i \eta}\,,
\end{equation}
with $\eta \to 0^+$ at the end of all calculations.  Inserting Eq.\,\eqref{S0def} into Eq.\,(\ref{S0x}) is equivalent to evaluating the $p^0$ integral by employing a contour in the complex-$p^0$ plane that is below the real-$p^0$ axis for $p^0<0$, and above it for $p^0>0$.  This prescription defines the \textbf{Feynman} propagator.

Let's now return to Eq.~(\ref{SAprop}):
\begin{equation}
\label{SAprop2} [ i \partial\!\!\!\slash_{x^\prime}\, - e A\!\!\!\slash(x^\prime) \,- m ]\,S(x^\prime,x) = \mbox{\boldmath
$1$}\,\delta^4(x^\prime -x)\,,
\end{equation}
which defines the Green function for a fermion in an external electromagnetic field.  As mentioned, a closed form solution of this equation is impossible in all but the simplest field configurations.  Is there, nevertheless, a way to construct an approximate solution that can be systematically improved?

To achieve that one rewrites the equation thus:
\begin{equation}
\label{SAprop3} [ i \partial\!\!\!\slash_{x^\prime} \,- m ]\,S(x^\prime,x) =
\mbox{\boldmath $1$}\,\delta^4(x^\prime -x) + e A\!\!\!\slash(x^\prime)
\,S(x^\prime,x)\,,
\end{equation}
which, as easily seen by substitution, is solved by
\begin{eqnarray}
S(x^\prime,x) & = & S_0(x^\prime-x)
+ e \int d^4y \,S_0(x^\prime-y)A\!\!\!\slash(y) \,S(y,x)\\
\nonumber & = & S_0(x^\prime-x)
+ e \int d^4y \,S_0(x^\prime-y)A\!\!\!\slash(y) \,S_0(y-x) \\
& & \nonumber + e^2 \, \int d^4y_1 \int d^4 y_2
\,S_0(x^\prime-y_1)A\!\!\!\slash(y_1) \,S_0(y_1-y_2) A\!\!\!\slash(y_2)
\,S_0(y_2-x) \\
&& + [\ldots]\,. \label{DSEone}
\end{eqnarray}
This perturbative expansion of the full propagator in terms of the free propagator provides an archetype for perturbation theory in quantum field theory.  One obvious application is the scattering of an electron/positron by a Coulomb field, which is an example explored in Sec.\,2.5.3 of Ref.\,\cite{IZ80}.  Equation~(\ref{DSEone}) is a first example of a \emph{Dyson-Schwinger equation} (DSE) \cite{Roberts:1994dr}.

In quantum field theory this Green function has the following interpretation: \\[-1.3em]
\begin{enumerate}
\item It creates a positive energy fermion (antifermion) at spacetime point $x$;
\item Propagates the fermion to spacetime point $x^\prime$, \emph{i.e}.\ forward in
time;
\item Annihilates this fermion at $x^\prime$.
\end{enumerate}
The process can equally well be viewed as \\[-1.3em]
\begin{enumerate}
\item The creation of a negative energy antifermion (fermion) at spacetime
point $x^\prime$;
\item Propagation of the antifermion to the spacetime point $x$, \emph{i.e}.\
backward in time;
\item Annihilation of this antifermion at $x$.
\end{enumerate}
Other propagators have similar interpretations.

\subsection{Dyson Schwinger Equations: \mbox{\rm e.g.} Photon Vacuum Polarisation}
\label{secVacPol}
Local gauge theories are the keystone of contemporary hadron and high-energy physics.  Such theories are difficult to quantise because the gauge dependence is an extra non-dynamical degree of freedom that must be handled.  The modern approach is to quantise the theories using the method of functional integrals, attributed to Feynman and Kac.  References~\cite{IZ80,tarrach} provide useful overviews of the technique, which replaces canonical second-quantisation.

Beginning with the field equations of quantum field theory, it has long been known that one can derive a system of coupled integral equations interrelating all of a theory's Green functions \cite{Dyson:1949ha, Schwinger:1951ex}.  This collection of a countable infinity of equations is called the complex of DSEs.  It is an intrinsically nonperturbative complex, which is vitally important in proving the renormalisability of quantum field theories, and, at its simplest level, the complex provides a generating tool for perturbation theory.  I will illustrate a nonperturbative derivation of one equation in this complex within the context of quantum electrodynamics QED.  The derivation of others follows the same pattern.

\subsubsection{Generating Functional and Maxwell's Equations.}
Let's begin with the action for QED with $N_f$ flavours of electromagnetically active fermions:
\begin{equation}
S[A_\mu,\psi,\bar\psi] = \int\, d^4 x\;
\left[\,\sum_{f=1}^{N_f}\, \bar\psi^f  \left( i\dslash - m_0^f + e_0^f
\Aslash \right)\psi^f
 - \frac{1}{4} F_{\mu\nu}F^{\mu\nu} -\frac{1}{2\lambda_0}\,
 \partial^\mu  A_\mu(x)   \,\partial^\nu A_\nu(x) \right]. \label{Sqed}
\end{equation}
In this manifestly Poincar\'e covariant action: $\bar\psi^f(x)$, $\psi^f(x)$ are the elements of the Grassmann algebra that describe the fermion degrees of freedom, so that, \emph{e.g}.\ $\forall f,g$, $\{\psi^f,\psi^g\}=0$; $m_0^f$ are the fermions' bare masses and $e_0^f$ their bare charges; and $A_\mu(x)$ describes the gauge boson [photon] field, with
\begin{equation}
F_{\mu\nu} = \partial_\mu A_\nu - \partial_\nu A_\mu\,,
\end{equation}
and $\lambda_0$ is the bare Lorentz gauge fixing parameter.  (\emph{N.B}.\ To describe an electron, the physical charge $e_f < 0$.)

The generating QED functional is defined via the action in Eq.\,(\ref{Sqed}):
\begin{align}
\nonumber
 W[J_\mu,\xi,\bar\xi] = &  \int\![{\cal D}A_\mu] \, [{\cal D}\psi] [{\cal D}\bar\psi] \,
\exp \bigg\{i\int\! d^4 x\, \bigg[-\tfrac{1}{4}
F^{\mu\nu}(x) \,  F_{\mu\nu}(x) -\frac{1}{2\lambda_0}\,\partial^\mu A_\mu(x)
\,\partial^\nu  A_\nu(x)\\
&
+\, \sum_{f=1}^{N_f}\, \bar\psi^f \left( i\dslash - m_0^f + e_0^f
\Aslash \right)\psi^f
+ \, J^\mu(x) A_\mu(x) + \bar\xi^f(x) \psi^f(x) + \bar\psi^f(x)
\xi^f(x) \rule{0em}{3.0ex}\bigg]\bigg\}\,, \label{WGFqed}
\end{align}
where $J_\mu$ is an external source for the electromagnetic field, and $\xi^f$, $\bar\xi^f$ are external sources for the fermion fields that are also elements in the Grassmann algebra.  An $n^{\rm th}$-order functional derivative of $W[J_\mu,\xi,\bar\xi]$ with respect to any combination of its arguments yields an expectation value of a product of fields, \emph{i.e}.\ a moment of the measure.  It follows that so long as the integral is well defined, then all the theory's physical content is expressed in Eq.\,\eqref{WGFqed}; but is the integral well defined?

The product $[{\cal D}A_\mu] \, [{\cal D}\psi] [{\cal D}\bar\psi]$ in Eq.\,\eqref{WGFqed} indicates that one is supposed to integrate over all possible values of all the fields at all the spacetime point in the Universe.  \emph{A priori}, that would seem an impossible task.  Consequently, any meaningful definition of relativistic quantum field theory is formulated in Euclidean space.  The oscillating weight factor then becomes an exponential damping factor, which acts to suppress contributions from all configurations that are distant from the action's stationary point(s).  For the remainder of this section I will nevertheless continue to work in Minkowski space, ignoring the problem because most readers will be more familiar with Minkowski space conventions.  Nothing is lost in doing this: all the manipulations performed in this sketch of the derivation of the DSE for the photon vacuum polarisation will be purely formal, with no attempt at mathematical rigour.

The derivation of a DSE follows simply from the observation that the integral of a total derivative vanishes, \emph{viz}.\
\begin{equation}
\int_a^b dz\, \frac{d}{dz} f(z) = f(b)-f(a) =0 \,,
\end{equation}
given appropriate boundary conditions.  Consider, therefore,
{\allowdisplaybreaks
\begin{eqnarray}
\nonumber 0 & =&  \int\![{\cal D}A_\mu] \, [{\cal D}\psi] [{\cal
D}\bar\psi]
\;\frac{\delta}{\delta A_\mu(x)}\;
    {\rm e}^{i\left( S[A_\mu,\psi,\bar\psi] +
    \int\,d^4 x\;\left[ \psib^f\xi^f + \bar\xi^f\psi^f + A_\mu J^\mu \right]
                \right)} \\
%
%
\nonumber & = &\int\![{\cal D}A_\mu] \, [{\cal D}\psi] [{\cal
    D}\bar\psi]\,\left\{ \frac{\delta S}{\delta A_\mu(x)} + J_\mu(x) \right\} \ldots 
    \\
& & \times
        \exp\left\{i\left( S[A_\mu,\psi,\bar\psi] +
    \int\,d^4 x\;\left[ \psib^f\xi^f + \bar\xi^f\psi^f + A_\mu J^\mu \right]
                \right)\right\} \nonumber\\
& = & \left\{ \frac{\delta S}{\delta A_\mu(x)}
        \left[\frac{\delta}{i\delta J },
        \frac{\delta}{i\delta\bar\xi}, -\frac{\delta}{i\delta\xi}\right]
        + J_\mu(x) \right\} W[J_\mu,\xi,\bar\xi] ~,
\label{ELeqn}
\end{eqnarray}}
\hspace*{-0.4\parindent}where the last line has meaning as a functional differential operator acting on the generating functional.  It may be interpreted as an implementation in QFT of the ``principle of stationary action'', where the generating functional replaces the classical action, \emph{i.e}.\ as an Euler-Lagrange equation in QFT.

To proceed from here it is advantageous to introduce the generating functional of connected Green functions, \emph{viz}.\ $Z[J_\mu,\bar\xi,\xi]$ defined via
\begin{equation}
\label{WZdef} W[J_\mu,\xi,\bar\xi] =: \exp\left\{ i
Z[J_\mu,\xi,\bar\xi]\right\}\,.
\end{equation}
Then, making use of the fact that differentiating Eq.~(\ref{Sqed}) gives
\begin{equation}
\frac{\delta S}{\delta A_\mu(x)} =\left[ \partial_\rho \partial^\rho g_{\mu\nu}
- \left( 1- \frac{1}{\lambda_0}\right) \partial_\mu \partial_\nu\right]
A^\nu(x) + \sum_f e_0^f \, \psib^f (x)\gamma_\mu \psi^f(x) \,,
\end{equation}
Eq.\,(\ref{ELeqn}) becomes
\begin{align}
\nonumber
-J_\mu(x)  & = \left[ \partial_\rho \partial^\rho g_{\mu\nu}
- \left( 1-  \frac{1}{\lambda_0}\right) \partial_\mu \partial_\nu\right]
\frac{\delta  Z}{\delta J_\nu(x)} \\
& +  \sum_f e_0^f\,\left(  - \frac{\delta Z}{\delta\xi^f(x)}
        \gamma_\mu \frac{\delta Z}{\delta \bar\xi^f(x)}
+ \frac{\delta}{\delta\xi^f(x)}
        \left[\gamma_\mu \frac{\delta \, i Z}{\delta \bar\xi^f(x)}
        \right]\right) , \label{FldEqn}
\end{align}
where I have divided through by $W[J_\mu,\xi,\bar\xi]$.  Equation~(\ref{FldEqn}) represents a compact form of the nonperturbative equivalent of Maxwell's equations.

\subsubsection{One Particle Irreducible Green Functions.}
The next step is to introduce the generating functional for one-particle-irreducible (1PI) Green functions: $\Gamma[A_\mu,\psi,\bar\psi]$, which is obtained from $Z[J_\mu,\xi,\bar\xi]$ via a Legendre transformation
\begin{equation}
Z[J_\mu,\xi,\bar\xi] = \Gamma[A_\mu,\psi,\bar\psi] + \int\,d^{4} x\;\left[
\psib^f\xi^f + \bar\xi^f\psi^f + A_\mu J^\mu \right] . \label{Legendre_transf}
\end{equation}
A one-particle-irreducible $n$-point function or ``proper vertex'' contains no contributions that become disconnected when a single connected $m$-point Green function is removed, \emph{e.g}.\ via functional differentiation.  This is equivalent to the statement that no diagram representing or contributing to a given proper vertex separates into two disconnected diagrams if only one connected propagator is cut.  (A detailed explanation is provided in Ref.\,\cite{IZ80}, pp.\,289-294.)

The following identities are plain from the definition of the generating functional, Eq.\,\eqref{WGFqed}:
\begin{equation}
\label{deltaZ} \frac{\delta Z}{\delta J^\mu(x)} = A_\mu(x)\,,\;
\frac{\delta Z}{\delta \bar\xi(x)} = \psi(x)\,,\;
\frac{\delta Z}{\delta \xi(x)} = -\bar\psi(x)\,,
\end{equation}
where here the external sources are \textbf{nonzero}.  Hence $\Gamma$ in Eq.\,(\ref{Legendre_transf}) must satisfy
\begin{equation}
\label{deltaGamma} \frac{\delta \Gamma}{\delta A^\mu(x)} = - J_\mu(x)\,,\;
\frac{\delta \Gamma}{\delta \bar\psi^f(x)} = - \xi^f(x)\,,\;
\frac{\delta \Gamma}{\delta \psi^f(x)} = \bar\xi^f(x)\,.
\end{equation}
Notably, since the sources are not zero:
\begin{equation}
A_\rho(x)= A_\rho(x;[J_\mu,\xi,\bar\xi]) \;\Rightarrow\; \frac{\delta
A_\rho(x)}{\delta J^\mu(y)} \neq 0\,,
\end{equation}
with analogous statements for the Grassmannian functional derivatives.

It is easy to see that setting $\bar\psi = 0 = \psi$ after differentiating $\Gamma$ gives zero \textit{unless} there are equal numbers of $\bar\psi$ and $\psi$ derivatives: any integral that is odd under $\psi\to -\psi$, $\bar\psi \to -\bar\psi$ must vanish.
Similarly, with $J_\mu=0$ the generating functional for 1PI Green functions is even under $A_\mu\to -A_\mu$, from which follows Furry's Theorem, \emph{i.e}.\  the vacuum expectation value of any odd number of electromagnetic currents is zero.

Consider the operator and matrix product (with spinor labels $r$, $s$, $t$)
\begin{equation}
\label{SGamma}
\left. - \int\! d^4z \, \frac{\delta^2 Z}{\delta \xi_r^f(x)
      \bar\xi_t^h(z)}
       \, \frac{\delta^2\Gamma}{\delta \psi_t^h(z) \psib_s^g(y)}
        \right|_{\begin{array}{c} \xi=\bar\xi=0 \\
                                \psi=\overline{\psi}=0
                \end{array}}\,,
\end{equation}
which, using Eqs.\,(\ref{deltaZ}), (\ref{deltaGamma}), simplifies as follows:
\begin{equation}
= \left. \int\! d^4z \,
\frac{\delta \psi^h_t(z)}{\delta \xi_r^f(x)}\, \frac{\delta\xi^g_s(y)}{\delta
\psi_t^h(z)}
        \right|_{\begin{array}{c} \xi=\bar\xi=0 \\
                                \psi=\overline{\psi}=0
                \end{array}}\, \\
 =  \left. \frac{\delta\xi^g_s(y)}{\delta \xi_r^f(x)}
        \right|_{\begin{array}{c} \psi=\overline{\psi}=0
                \end{array}}\,
= \delta_{rs}\, \delta^{fg}\, \delta^4(x-y)\,.
\end{equation}

Returning to Eq.\,(\ref{FldEqn}) and setting $\bar\xi=0=\xi$ one obtains
\begin{equation}
\left. \frac{\delta\Gamma}{\delta A^\mu(x)}\right|_{\psi=\overline{\psi}=0} =
\left[ \partial_\rho \partial^\rho g_{\mu\nu} - \left( 1-
\frac{1}{\lambda_0}\right)
\partial_\mu \partial_\nu\right] A^\nu(x)
- \, i \,\sum_f e_0^f {\rm tr}\left[
\gamma_\mu S^f(x,x;[A_\mu]) \right] \;, \label{FEppta}
\end{equation}
after making the identification
\begin{equation}
S^f(x,y;[A_\mu]) =  - \, \frac{\delta^2 Z}{\delta \xi^f(y) \bar\xi^f(x)}
= \frac{\delta^2 Z}{\delta \bar\xi^f(x) \xi^f(y)}\; (\mbox{no~summation~on~}f)
\label{SfA}\,,
\end{equation}
which is the connected Green function that describes the propagation of a fermion with flavour $f$ in an external electromagnetic field $A_\mu$. (Recall the free fermion Green function in Eq.~(\ref{S0x}).)  Notably, as a direct consequence of Eq.\,(\ref{SGamma}) the inverse of this Green function is given by
\begin{equation}
\label{SfinverseA} S^f(x,y;[A])^{-1} =  \left. \frac{\delta^2
\Gamma}{\delta\psi^f(x) \delta\bar\psi^f(y)} \right|_{\psi=\overline{\psi}=0}
\;.
\end{equation}
It is a general property that such functional derivatives of the generating
functional for 1PI Green functions are related to the associated propagator's
inverse.  Clearly the vacuum fermion propagator or connected fermion
$2$-point function is
\begin{equation}
\label{SAeq0} S^f(x,y):=S^f(x,y;[A_\mu=0])\,.
\end{equation}
Such vacuum Green functions are keystones in quantum field theory.

At this point one differentiates Eq.\,(\ref{FEppta}) with respect to $A_\nu(y)$ and sets $J_\mu(x)=0$:
\begin{align}
\nonumber \left. \frac{\delta^2 \Gamma}{\delta A^\mu(x) \delta
A^\nu(y)}  \right|_{\begin{array}{c} A_\mu=0 \\ \psi=\overline{\psi}=0
\end{array}}
& = \left[ \partial_\rho \partial^\rho g_{\mu\nu} - \left( 1-
\frac{1}{\lambda_0}\right) \partial_\mu \partial_\nu\right] \, \delta^4(x-y) \\
& -  i \sum_f e_0^f {\rm tr}\left[
\gamma_\mu \frac{\delta }{\delta A_\nu(y)} \, \left( \left. \frac{\delta^2
\Gamma}{\delta\psi^f(x) \delta\bar\psi^f(x)}
\right|_{\psi=\overline{\psi}=0}\right)^{-1}\right]\,.
\label{deltaFEppta}
\end{align}

The left-hand-side (lhs) is readily understood.  Just as Eqs.\,(\ref{SfinverseA}),
(\ref{SAeq0}) define the inverse of the fermion propagator, here one has
\begin{equation}
\label{Dinverse} (D^{-1})^{\mu\nu}(x,y) := \left.\frac{\delta^2 \Gamma}{\delta
A^\mu(x) \delta A^\nu(y)} \right|_{\begin{array}{c} A_\mu=0 \\
\psi=\overline{\psi}=0
\end{array}},
\end{equation}
\emph{viz}.\ the inverse of the photon propagator.

On the other hand, the rhs must be simplified and interpreted.  First observe that
\begin{align}
\nonumber
& \frac{\delta }{\delta A_\nu(y)} \, \left( \left. \frac{\delta^2 \Gamma}{\delta\psi^f(x) \delta\bar\psi^f(x)}
\right|_{\psi=\overline{\psi}=0}\right)^{-1} \\
& =   - \, \int\!d^4 u d^4 w\, \left(
\left.\frac{\delta^2 \Gamma}{\delta\psi^f(x) \delta\bar\psi^f(w)}
\right|_{\psi=\overline{\psi}=0} \right)^{-1}
\frac{\delta }{\delta A_\nu(y)}
\frac{\delta^2 \Gamma}{\delta\psi^f(u) \delta\bar\psi^f(w)}\,
\left( \left.\frac{\delta^2 \Gamma}{\delta\psi^f(w) \delta\bar\psi^f(x)}
\right|_{\psi=\overline{\psi}=0} \right)^{-1} \! ,  \label{dAAinv}
\end{align}
which is an analogue of the result for finite dimensional matrices:
\begin{eqnarray}
\nonumber \frac{d}{dx} \left[ A(x) A^{-1}(x) = \mbox{\boldmath $I$} \right] & =
0  = &  \frac{d A(x)}{dx} \, A^{-1}(x) + A(x)\,\frac{d A^{-1}(x)}{dx}
 \\
& \Rightarrow & \frac{d A^{-1}(x)}{dx} = - A^{-1}(x) \,\frac{d
A(x)}{dx}\,A^{-1}(x)\,.
\end{eqnarray}

Equation\,(\ref{dAAinv}) involves the 1PI $3$-point function (no summation on $f$)
\begin{equation}
\label{Gammadef} e_0^f \Gamma_\mu^f(x,y;z) := \frac{\delta }{\delta A_\nu(z)}
\frac{\delta^2 \Gamma}{\delta\psi^f(x) \delta\bar\psi^f(y)}\,.
\end{equation}
This is the proper fermion-gauge-boson vertex.  At leading order in perturbation theory
\begin{equation}
\Gamma_\nu^f(x,y;z) = \gamma_\nu\,\delta^4(x-z) \, \delta^4(y-z)\,,
\end{equation}
a result which can be obtained via explicit calculation of the functional derivatives in Eq.\,(\ref{Gammadef}).

\subsubsection{Photon Self Energy.}  
\begin{figure}[t]
\begin{minipage}[t]{\textwidth}
\begin{minipage}{0.49\textwidth}
\centerline{\includegraphics[clip,width=0.9\textwidth]{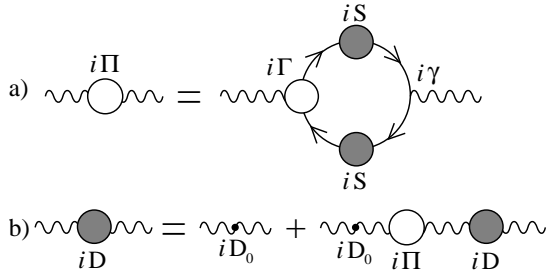}}
\end{minipage}
\begin{minipage}{0.49\textwidth}
\caption{\label{photonDSE} \small
\emph{Upper panel} -- photon vacuum polarisation in Eq.\,\eqref{PPTeq}.
\emph{Lower panel} -- DSE for the photon propagator, whose kernel is the photon vacuum polarisation.}
\end{minipage}
\end{minipage}
\end{figure}

Defining the gauge-boson \textit{vacuum polarisation}:
\begin{equation}
\Pi_{\mu\nu}(x,y) =  i \sum_f (e_0^f)^2  \int\,d^{4}z_1\,d^{4}z_2\,
 {\rm tr}\left[ \gamma_\mu S^f(x,z_1)\Gamma^f_\nu(z_1,z_2;y)
        S^f(z_2,x)\right],
\label{PPTeq}
\end{equation}
which is illustrated in Fig.\,\ref{photonDSE}, it is immediately apparent that Eq.~(\ref{deltaFEppta}) may be expressed as
\begin{equation}
(D^{-1})^{\mu\nu}(x,y) = \left[ \partial_\rho \partial^\rho g_{\mu\nu} - \left(
1- \frac{1}{\lambda_0}\right)
\partial_\mu \partial_\nu\right] \delta^4(x-y) + \Pi_{\mu\nu}(x,y)\,.
\label{fullDinverse}
\end{equation}
In general, the gauge-boson vacuum polarisation, or ``photon self-energy,'' describes the modification of the gauge-boson's propagation characteristics owing to the presence of virtual particle-antiparticle pairs in quantum field
theory.  In particular, the photon vacuum polarisation is an important element in the description of many physical processes, for instance $\rho^0\to e^+ e^- $.

The propagator for a free gauge boson is obtained when $\Pi_{\mu\nu}\equiv 0$ and may be written, in momentum space:
\begin{equation}
\label{freegluon} D_0^{\mu\nu}(x-y)
= \left( -
g^{\mu\nu} + (1-\lambda)  \frac{q^\mu q^\nu}{q^2+i\eta^+} \right)
\frac{1}{q^2+i \eta^+} \,.
\end{equation}
In the presence of interactions, \emph{i.e}.\ for $\Pi_{\mu\nu}\neq 0$ in Eq.\,(\ref{fullDinverse}), this becomes
\begin{equation}
D^{\mu\nu}(q)=\frac{-g^{\mu\nu}+(q^\mu q^\nu/[q^2+i\eta])}{q^2+i\eta}
\frac{1}{1+\Pi(q^2)} - \lambda_0\frac{q^\mu q^\nu}{(q^2+i\eta)^2}\;,
\label{photon_propagator}
\end{equation}
where I have used the Ward-Green-Takahashi identity \cite{Ward:1950xp, Green:1953te, Takahashi:1957xn, Takahashi:1985yz, Kondo:1996xn, He:2000we, He:2002jg, Pennington:2005mw, He:2006ce, He:2006my, He:2007zza, Qin:2013mtaS, Qin:2014vya}:
\begin{equation}
q_\mu\, \Pi_{\mu\nu}(q) = 0  = \Pi_{\mu\nu}(q) \, q_\nu\,,
\end{equation}
which means that one can write
\begin{equation}
\label{PiWTI} \Pi^{\mu\nu}(q) = \left(-g^{\mu\nu}q^2+q^\mu q^\nu\right) \,
\Pi(q^2)\,.
\end{equation}
$\Pi(q^2)$ is the polarisation scalar.  In QED, it is independent of the gauge parameter, $\lambda_0$.  The choice $\lambda_0=1$ is called ``Feynman gauge''; and it is useful in perturbative calculations because it simplifies the $\Pi(q^2)=0$ gauge boson propagator enormously.  In nonperturbative applications, however, $\lambda_0=0$, ``Landau gauge,'' is most useful because, \emph{inter alia}, it ensures that the gauge boson propagator is itself transverse.  Equation\,\eqref{photon_propagator} corresponds to the DSE expressed in the lower panel of Fig.\,\ref{photonDSE}.

The longitudinal Ward-Green-Takahashi identities (WGTIs) \cite{Ward:1950xp, Green:1953te, Takahashi:1957xn} are relations satisfied by $n$-point Green functions, relations which are an essential consequence of a theory's local gauge invariance, \emph{i.e}.\ local current conservation.  They can be proved directly from the generating functional and have physical implications.  For example, Eq.\,(\ref{PiWTI}) ensures that the photon remains massless in the presence of charged fermions.\footnote{A discussion of longitudinal WGTIs can be found in Ref.\,\cite{IZ80}, pp.\,407-411, and Ref.\,\cite{BD65}, pp.\,299-303; and their generalisation to non-Abelian theories as ``Slavnov-Taylor'' identities is described in Ref.\,\cite{tarrach}, Chap.\,2.  There are also transverse WGTIs \cite{Takahashi:1985yz, Kondo:1996xn, He:2000we, He:2002jg, Pennington:2005mw, He:2006ce, He:2006my, He:2007zza, Qin:2013mtaS, Qin:2014vya}, which have additional, important consequences.}

In the absence of external sources for fermions and gauge bosons, Eq.\,(\ref{PPTeq}) can easily be represented in momentum space, for then the $2$-and $3$-point functions appearing therein must be translationally invariant and hence they can be expressed in terms of Fourier amplitudes, \emph{i.e}.\ one has
\begin{equation}
\label{PiMom} i\Pi_{\mu\nu}(q)= - \,\sum_f (e_0^f)^2\int \frac{d^4
\ell}{(2\pi)^{\rm d}} {\rm
tr}[(i\gamma_\mu)(iS^f(\ell))(i\Gamma^f(\ell,\ell+q))(iS(\ell+q))] \,.
\end{equation}
It is the reduction to a single integral that makes momentum space representations most widely used in continuum calculations.

The vacuum polarisation in QED is directly related to the running coupling constant, which is a connection that makes its importance obvious.  The connection is not so direct in QCD; but, nevertheless, the polarisation scalar is a key component in the evaluation of the strong running coupling.

In the above analysis we saw that second derivatives of the generating functional, $\Gamma[A_\mu,\psi,\bar\psi]$, give the inverse-fermion and -photon propagators and that the third derivative gave the proper photon-fermion vertex.  In general, all derivatives of this generating functional, higher than two, produce a corresponding proper vertex, where the number and type of derivatives give the number and type of proper Green functions that it can serve to connect.

\setcounter{equation}{0}
\section{The Ins and Outs of Mesons}
\label{InOutMesons}
\subsection{Elegance Is Not Everything}
It is worth reiterating at this point that physics is an empirical science.  I stress this because, faced with challenges in applying fundamental theories to the observed Universe, some researchers have called for a change in how theoretical physics is done \cite{Nature:2014}.  They have begun to argue that if a theory is sufficiently \emph{elegant}, it need not be tested experimentally.  Prominent amongst the ``elegance will suffice'' advocates are some string theorists and cosmologists.  This is plainly because their elegant theories have not found experimental confirmation at the LHC.  This is not science; and it brings me to recollect the words of perhaps the world's most famous PI \cite{ConanDoyle:1891}:

\centerline{\parbox{0.95\textwidth}{\flushleft\emph{I have no data yet.  It is a capital mistake to theorize before one has data.  Insensibly one begins to twist facts to suit theories, instead of theories to suit facts}.}}

\medskip

Thus, once more, physics is an empirical science.  If a claim cannot be tested empirically, then it does no lie within the realm of physics; and no claim can be considered proven unless it's verified experimentally.  Viewed from this perspective, the theoretical physicist's task is to develop theories and elicit their observable content, \emph{viz}.\ elucidate a diverse array of the theory's most basic measurable predictions.  In due time experimental outcomes will then decide which facets of Nature are explained by the theory and where lie the boundaries of its applicability.  Perhaps there is no \emph{theory of everything}; or, perhaps, the theory of everything will derive its \emph{elegance} by following the paradigm established by QCD?

\subsection{Jefferson Lab}
\label{secJLab}
A context for the remainder of these lectures is provided by the experimental programmes underway and planned at JLab in Newport News, Virginia.
Funds for the development of research plans and designs for the Continuous Electron Beam Accelerator Facility (CEBAF) at JLab were initially provided in 1984, and construction began in 1987.  Seven years later, in 1994, the facility achieved its design capability, delivering 4\,GeV electron beams on targets in the three associated experimental halls.  The goal was to ``write the book'' about the strongest known force in Nature -- the force that holds nuclei together -- and determine how that force can be explained in terms of the gluons and quarks of QCD.

An aim of the original JLab programme was verification of the following prediction: at energy-scales greater than some minimum value, $Q_0$, whose value was not determined by the proof, hadron exclusive form should behave as follows: \cite{Brodsky:1973kr, Brodsky:1974vy, Farrar:1979aw, Efremov:1979qk, Lepage:1979zb, Lepage:1980fj}
\begin{equation}
\label{QCDtest}
{\mathpzc F}(\sigma^2) \stackrel{\sigma\gg Q_0}{\propto}
\left[\frac{Q_0^2}{\sigma^2}\right]^{{N}}
\ln\left[\frac{Q_0^2}{\sigma^2}\right]^{\gamma_{\mathpzc F}}
\end{equation}
where the integer $N=n_{\rm quarks} - 1$, with $n_{\rm quarks}$ counting the lowest possible number of quarks and/or antiquarks in the target.  The power-law term gives rise to parton-model scaling and can be explained on dimensional grounds.  The distinctive signature of QCD lies in the logarithmic factor, the exponent on which, ${\gamma_{\mathpzc F}}$, can be computed and whose appearance signals scaling violation, which is well documented in deep inelastic scattering \cite{ESW96}.  It was optimistically imagined that $Q_0$ might be as small as 1\,GeV and so CEBAF was designed to reach 4\,GeV.

In the ten years following achievement of design capacity, numerous fascinating experimental results were obtained at JLab, including an empirical demonstration that the distribution of charge and magnetization within the proton are completely different \cite{Jones:1999rz, Gayou:2001qtS, Gayou:2001qd, Punjabi:2005wqS, Puckett:2010ac, Puckett:2011xgS}, \emph{viz}.\
\begin{equation}
\label{GEGMRatio}
\mbox{on}\; Q^2 \in [1,9]\, m_p^2,\; \mu_p G_E^p(Q^2)/G_M^p(Q^2)< 1\,
\end{equation}
where $m_p$ is the proton mass.  This fact, which overturned a longstanding particle physics paradigm, along with a range of related observations, is explained by the presence of strong, nonpointlike, electromagnetically-active scalar and axial-vector diquark correlations within the nucleon (see Refs.\,\cite{Eichmann:2008ef, Cloet:2008re, Chang:2011tx, Cloet:2011qu, Wilson:2011aa, Cloet:2013gva, Cloet:2014rja, Segovia:2014aza, Segovia:2015ufa} and Sec.\,\ref{secBaryons} herein).  However, notwithstanding the breadth of JLab's programme and the associated excitement, no experiment has produced an unambiguous signal for the realisation of Eq.\,\eqref{QCDtest}, \emph{i.e}.\ as yet we have no clear sign of parton model scaling and certainly no sign of scaling violations.  Partly owing to this but also with a vast array of new experimental tests of the Standard Model's strong interaction sector in mind \cite{Dudek:2012vr}, in 2004 it was decided that CEBAF should be upgraded so that it could deliver 12\,GeV electrons on targets.  That upgrade is now complete and commissioning beams are being sent to the experimental halls.

The JLab\,12 facility and, indeed, an array of modern accelerators worldwide, will confront a diverse range of scientific challenges.
In the foreseeable future, we will know the results of a search for hybrid and exotic hadrons, the discovery of which would force a dramatic reassessment of the distinction between matter and force fields in Nature.\footnote{Exotic mesons are states which possess quantum numbers that are not possible in two-body quark-antiquark systems that can be described by quantum mechanics whereas hybrid mesons have quark model quantum numbers but unusual decay patterns.  Both types of systems are possible in QCD if one admits the concept of ``constituent'' or valence gluon content.}
Opportunities provided by new data on hadron elastic and transition form factors will be exploited, yielding insights into the infrared running of QCD's coupling and dressed-masses, revealing those correlations which are key to hadron structure, exposing the facts or fallacies in contemporary descriptions, and seeking verification of Eq.\,\eqref{QCDtest} in numerous processes.
Precise experimental studies of the valence-quark region will proceed, with the results being used to confront predictions from theoretical computations of distribution functions and distribution amplitudes -- computation is critical here because without it no amount of data will reveal anything about the theory underlying the phenomena of strong interaction physics.
In addition, the international community will seek and exploit opportunities to use precision-QCD as a probe for physics beyond the Standard Model.
Each of the pieces in this body of exploration, however, can be viewed as steps directed toward understanding a single overarching puzzle within the Standard Model, \emph{viz}.\ what is confinement and how is it related to dynamical chiral symmetry breaking (DCSB) -- the origin of the vast bulk of visible mass in the Universe?

\subsection{Dyson-Schwinger Equations for Hadron Physics}
In order to match the rate at which progress is being made experimentally and, better, to guide and enhance those programmes, flexible, responsive theoretical approaches are needed: methods that are capable both of rapidly providing an intuitive understanding of complex problems and illuminating a path toward answers and new discoveries.  In this milieu, notwithstanding its steady progress toward results with input parameters that approximate the real world, the numerical simulation of lattice-regularised QCD (lQCD) will not suffice. Approaches formulated in the continuum and inspired by, based upon, or connected directly with QCD are necessary.  Prominent amongst such tools are QCD Sum Rules \cite{Leinweber:1995fn, Nielsen:2009uh} and DSEs \cite{Maris:2003vk, Chang:2011vu, Boucaud:2011ugS, Bashir:2012fs, Cloet:2013jya}.

These lectures focus on the use of DSEs in cold, sparse hadron physics.  (Applications to hot, dense QCD may be traced from Ref.\,\cite{Roberts:2000aa}.)  The DSEs are well suited to the study of QCD because, as remarked above, their simplest application is as a generating tool for perturbation theory.  Since QCD is asymptotically free, that materially reduces model dependence in intelligent applications because the interaction kernel in each DSE is known for all momenta within the perturbative domain: $k^2\gtrsim 2\,$GeV$^2$.  Any model that needs building need then only make statements about the long-range behaviour of the kernels.  This is good because DSE solutions are Schwinger functions, \emph{i.e}.\ propagators and vertices; and since all cross-sections are built from Schwinger functions, the approach connects observables with the long-range behaviour of the theory's running coupling and masses.  Consequently, feedback between theoretical predictions and experimental tests can then refine the statements and lead to an understanding of these fundamental quantities.  Those predictions are wide-ranging because the DSEs provide a nonperturbative, continuum approach to hadron physics and can therefore address questions pertaining to, \emph{e.g}.: the gluon- and quark-structure of hadrons; and the roles of emergent phenomena -- confinement and DCSB -- and the connections between them.  Indeed, the past decade has seen the emergence of a novel understanding of gluon and quark confinement and its consequences; and we are arriving at a clear picture of how hadron masses emerge dynamically in a universe with light quarks, \emph{viz}.\ DCSB.  Furthermore, computations of ground-state hadron wave functions with a direct connection to QCD are now available.  They reveal that quark-quark correlations are crucial in baryon structure \cite{Segovia:2015ufa}; and experimental evidence in support of this prediction is accumulating \cite{Cates:2011pz}.

\subsection{Confinement}
\label{secConfinement}
This notion has already been mentioned.  However, in order to consider the concept further it is actually crucial to \emph{define} the subject.  That problem is canvassed in Sec.\,2.2 of Ref.\,\cite{Cloet:2013jya}, which explains that the potential between infinitely-heavy quarks measured in numerical simulations of quenched lQCD -- the so-called static potential \cite{Wilson:1974sk} -- is \emph{irrelevant} to the question of confinement in our Universe, in which light quarks are ubiquitous and the pion is unnaturally light.  This is because light-particle creation and annihilation effects are essentially nonperturbative in QCD and so it is impossible in principle to compute a quantum mechanical potential between two light quarks \cite{Bali:2005fuS, Prkacin:2005dc, Chang:2009ae}.  This means there is no flux tube in a Universe with light quarks and consequently that the flux tube is not the correct paradigm for confinement.

DCSB is critical to this property of the Standard Model because it ensures the existence of nearly-massless pseudo-Goldstone modes (pions), each constituted from a valence-quark and -antiquark whose individual Lagrangian current-quark masses are $<1$\% of the proton mass \cite{Maris:1997hd}.  In the presence of these modes, no flux tube between a static colour source and sink can have a measurable existence.  To verify this statement, consider such a tube being stretched between a source and sink.  The potential energy accumulated within the tube may increase only until it reaches that required to produce a particle-antiparticle pair of the theory's pseudo-Goldstone modes.  Simulations of lQCD show \cite{Bali:2005fuS, Prkacin:2005dc} that the flux tube then disappears instantaneously along its entire length, leaving two isolated colour-singlet systems.  The length-scale associated with this effect in QCD is $r_{\not\sigma} \simeq (1/3)\,$fm and hence if any such string forms, it would dissolve well within a hadron's interior.

An alternative perspective associates confinement with dramatic, dynamically-driven changes in the analytic structure of QCD's propagators and vertices.  In this realisation, confinement is a dynamical process.  In fact, as will subsequently be explained, contemporary theory predicts that both gluons and quarks acquire running mass distributions in QCD, which are large at infrared momenta (see, \emph{e}.\emph{g}.\ Refs.\,\cite{Bhagwat:2003vw, Bhagwat:2006tu, Bowman:2005vx, Boucaud:2011ugS, Ayala:2012pb, Binosi:2014aea}).  The generation of these masses leads to the emergence of a length-scale $\varsigma \approx 0.5\,$fm, whose existence and magnitude is evident in all existing studies of dressed-gluon and -quark propagators and which characterizes a dramatic change in their analytic structure.  In models based on such features \cite{Stingl:1994nk}, once a gluon or quark is produced, it begins to propagate in spacetime; but after each ``step'' of length $\varsigma$, on average, an interaction occurs so that the parton loses its identity, sharing it with others.  Finally a cloud of partons is produced, which coalesces into colour-singlet final states.  Such pictures of parton propagation, hadronisation and confinement can be tested in experiments at modern and planned facilities.

\begin{figure}[tbp]
\begin{minipage}[t]{\textwidth}
\begin{minipage}{0.5\textwidth}
\includegraphics[width=0.95\textwidth]{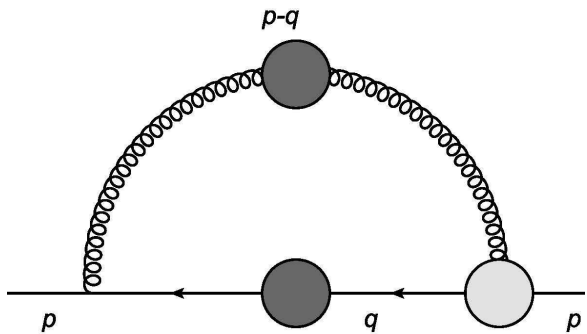}
\end{minipage}
\begin{minipage}{0.5\textwidth}{\small
\caption{\small
\label{FigSigma} Dressed-quark self energy, which is the dynamical content of QCD's most basic fermion gap equation.  The kernel is composed from the dressed-gluon propagator (spring with dark circle) and the dressed-quark-gluon vertex (light-circle).  The equation is nonlinear owing to the appearance of the dressed-quark propagator (line with dark circle).  This image encodes every valid Feynman diagram relevant to the quark dressing process and, additionally, \emph{e.g}.\  any topologically nontrivial contributions that might be important.  (Momentum flows from right-to-left.)}}
\end{minipage}
\end{minipage}
\end{figure}

\subsection{Dynamical chiral symmetry breaking}
\label{secDCSB}
Whilst the nature and realisation of confinement in empirical QCD is still being explored, DCSB; namely, the generation of \emph{mass} \emph{from nothing}, is a theoretically-established nonperturbative feature of QCD.  It is important to insist on the term ``dynamical,'' as distinct from spontaneous, because nothing is added to QCD in order to effect this remarkable outcome and there is no simple change of variables in the QCD action that will make it apparent.  Instead, through the act of quantising the classical chromodynamics of massless gluons and quarks, a large mass-scale is generated.  DCSB is the most important mass generating mechanism for visible matter in the Universe, being responsible for approximately $98$\% of the proton's mass.\footnote{This means that the Higgs boson plays almost no role in generating the mass of anyone reading this text.  Here ``almost'' means that there must be something like a Higgs mechanism in the Standard Model because at least one of the light quarks should have a nonzero mass so that the pion, too, acquires a nonzero mass, with magnitude $\lesssim \Lambda_{\rm QCD}$.  In the presence of an infinite-range strongly-interacting boson, the Universe could not exist in its present form.}

\begin{figure}[t]
\begin{minipage}[t]{\textwidth}
\begin{minipage}{0.5\textwidth}
\centerline{\includegraphics[width=0.9\textwidth]{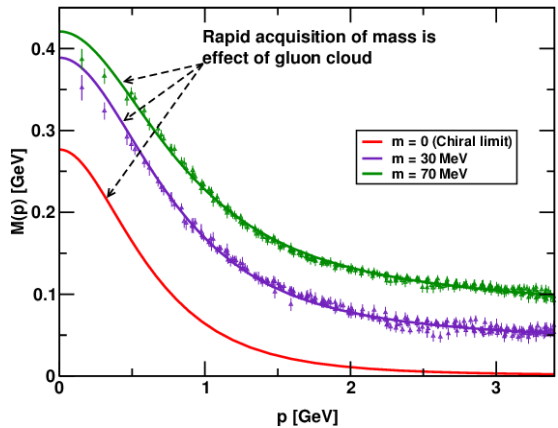}}
\end{minipage}
\begin{minipage}{0.5\textwidth}{\small
\caption{\label{gluoncloud} \small
Dressed-quark mass function, $M(p)$ in Eq.\,(\ref{SgeneralN}): \emph{solid curves} -- DSE results, explained in Refs.\,\protect\cite{Bhagwat:2003vw,Bhagwat:2006tu}, ``data'' -- numerical simulations of lattice-regularised QCD \protect\cite{Bowman:2005vx}.  (\emph{N.B}.\ $m=70\,$MeV is the uppermost curve and current-quark mass decreases from top to bottom.)  The current-quark of perturbative QCD evolves into a constituent-quark as its momentum becomes smaller.  The constituent-quark mass arises from a cloud of low-momentum gluons attaching themselves to the current-quark.  This is DCSB: an essentially nonperturbative effect that generates a quark \emph{mass} \emph{from nothing}; namely, it occurs even in the chiral limit.
}\vspace*{5ex}}
\end{minipage}
\end{minipage}
\end{figure}

A fundamental expression of DCSB is the behaviour of the quark mass-function, $M(p)$, which is a basic element in the dressed-quark propagator\footnote{I have now switched to a Euclidean metric, in which the Dirac matrices are defined via $\{\gamma_\mu,\gamma_\nu\} = 2\delta_{\mu\nu}$, $\gamma_\mu^\dagger = \gamma_\mu$, $\gamma_5= \gamma_4\gamma_1\gamma_2\gamma_3$, tr$[\gamma_5\gamma_\mu\gamma_\nu\gamma_\rho\gamma_\sigma]=-4 \epsilon_{\mu\nu\rho\sigma}$, $\sigma_{\mu\nu}=(i/2)[\gamma_\mu,\gamma_\nu]$; and $a \cdot b = \sum_{i=1}^4 a_i b_i$, with $P_\mu$ timelike $\Rightarrow$ $P^2<0$.}
\begin{equation}
\label{SgeneralN}
S(p) = 
1/[i\gamma\cdot p A(p^2) + B(p^2)] = Z(p^2)/[i\gamma\cdot p + M(p^2)]\,,
\end{equation}
which may be obtained as a solution to QCD's most basic fermion gap equation (see Fig.\,\ref{FigSigma}).  The nontrivial behaviour of the mass function, depicted in Fig.\,\ref{gluoncloud}, arises primarily because a dense cloud of gluons comes to clothe a low-momentum quark; and explains how an almost-massless parton-like quark at high energies transforms, at low energies, into a constituent-like quark, which possesses an effective ``spectrum mass'' $M_D \sim 350\,$MeV.  Consequently, the proton's mass is two orders-of-magnitude larger than the sum of the current-masses of its three valence-quarks.

One might ask just how the self-energy depicted in Fig.\,\ref{FigSigma} is capable of generating \emph{mass from nothing}, \emph{viz}.\ the $m=0$ curve in Fig.\,\ref{gluoncloud}, which cannot arise in the classical theory.  The answer lies in the fact that Fig.\,\ref{FigSigma} is a deceptively simply picture.  At the very least, it corresponds to a countable infinity of diagrams,\footnote{In principle, the dressed-quark-gluon vertex involves terms that express DCSB and consequently contains information about any topologically nontrivial configurations that are important in this connection \cite{Bhagwat:2007ha}.} all of which can potentially contribute.  To provide a context, quantum electrodynamics, an Abelian gauge theory, has 12\,672 diagrams at order $\alpha^5$ in the computation of the electron's anomalous magnetic moment \cite{Aoyama:2012wj}.  Owing to its foundation in the non-Abelian group $SU(3)$, the analogous perturbative computation of a quark's anomalous chromomagnetic moment has many more diagrams at this order in the strong coupling.  The number of diagrams represented by the self energy in Fig.\,\ref{FigSigma} grows equally rapidly, \emph{i}.\emph{e}.\ combinatorially with the number of propagators and vertices used at a given order.  Indeed, proceeding systematically, a computer will very quickly generate the first diagram in which the number of loops is so great that it is simply impossible to calculate in perturbation theory: impossible in the sense that we don't yet have the mathematical capacity to solve the problem.

Each of the diagrams which contributes to $M(p^2)$ in a weak-coupling expansion of Fig.\,\ref{FigSigma} is multiplied by the current-quark mass, $\hat m$.  Plainly, any finite sum of diagrams must therefore vanish as $\hat m\to 0$.  However, with \emph{infinitely many} diagrams the situation might be very different: one has ``$ 0 \times \infty$,'' a product whose limiting value is contingent upon the cumulative magnitude of each term in the sum.  Consider therefore the behaviour of $M(p^2)$ at large $p^2$.  QCD is asymptotically free \cite{Politzer:2005kc, Gross:2005kv, Wilczek:2005az}.  Hence, on this domain, each of the regularised loop diagrams must individually evaluate to a small number whose value depends on just how large is the coupling.  It will not be surprising, therefore, to learn that for a monotonically-decreasing running-coupling, $\alpha_S(k^2)$, there is a critical value of $\alpha_S(0)$ above which the magnitude of the sum of infinitely many diagrams is sufficient to balance the linear decrease of $\hat m\to 0$, so that the answer is nonzero and finite in this limit, \emph{viz}.,
\begin{equation}
\exists\, \alpha_S^c(0) \; |\; \forall \alpha_S(0) >\alpha_S^c(0),
M_0(p^2):= \lim_{\hat m\to 0} M(p^2;\hat m) \neq 0\,.
\end{equation}
The internal consistency of QCD appears to guarantee that the limit is always finite.  (As mentioned above, the case of Abelian theories is more complicated \cite{Roberts:1994dr} because they are not asymptotically free.)

\subsection{Challenge of Truncation}
\label{secTrunc}
The gap equation illustrates the features and flaws of each DSE.  It is a nonlinear integral equation for the dressed-quark propagator and hence can yield much-needed nonperturbative information.  However, the kernel involves the two-point function $D_{\mu\nu}$ and the three-point function $\Gamma^f_\nu$.  The gap equation is therefore coupled to the DSEs satisfied by these functions, which in turn involve higher $n$-point functions.  Hence the DSEs are a tower of coupled integral equations, with a tractable problem obtained only once a truncation scheme is specified.  It is unsurprising that the best known truncation scheme is the weak coupling expansion, which reproduces every diagram in perturbation theory.  This scheme is systematic and valuable in the analysis of large momentum transfer phenomena because QCD is asymptotically free but it precludes any possibility of obtaining nonperturbative information.

Given the importance of DCSB in QCD, it is significant that the dressed-quark propagator features in the axial-vector WGTI, which expresses chiral symmetry and its breaking pattern:
\begin{equation}
P_\mu \Gamma_{5\mu}^{fg}(k;P) + \, i\,[m_f(\zeta)+m_g(\zeta)] \,\Gamma_5^{fg}(k;P)
= S_f^{-1}(k_+) i \gamma_5 +  i \gamma_5 S_g^{-1}(k_-) \,,
\label{avwtimN}
\end{equation}
where $f$, $g$ label quark flavours, $P=p_1+p_2$ is the total-momentum entering the vertex and $k$ is the relative-momentum between the amputated quark legs.  To be explicit, $k=(1-\eta) p_1 + \eta p_2$, with $\eta \in [0,1]$, and hence $k_+ = p_1 = k + \eta P$, $k_- = p_2 = k - (1-\eta) P$.  In a Poincar\'e covariant approach, such as presented by a proper use of DSEs, no observable can depend on $\eta$, \emph{i.e}.\ the definition of the relative momentum.  (\emph{N.B}.\ Ref.\,\cite{Bhagwat:2007ha} discusses the important differences encountered in treating flavourless pseudoscalar mesons.)

In Eq.\,(\ref{avwtimN}), $\Gamma_{5\mu}^{fg}$ and $\Gamma_5^{fg}$ are, respectively, the amputated axial-vector and pseudoscalar vertices.  They are both obtained from an inhomogeneous Bethe-Salpeter equation (BSE), which is exemplified here using a textbook expression \cite{Salpeter:1951sz}:
\begin{equation}
[\Gamma_{5\mu}(k;P)]_{tu} = Z_2 [\gamma_5 \gamma_\mu]_{tu}+ \int_q^\Lambda [ S(q_+) \Gamma_{5\mu}(q;P) S(q_-) ]_{sr} K_{tu}^{rs}(q,k;P),
\label{bsetextbook}
\end{equation}
in which $K$ is the fully-amputated quark-antiquark scattering kernel, and the colour-, Dirac- and flavour-matrix structure of the elements in the equation is denoted by the indices $r,s,t,u$.  \emph{N.B}.\ By definition, $K$ does not contain quark-antiquark to single gauge-boson annihilation diagrams, nor diagrams that become disconnected by cutting one quark and one antiquark line.  In the parlance of Sec.\,\ref{secVacPol}, $K$ is two-particle irreducible (2PI).

The WGTI, Eq.\,(\ref{avwtimN}), entails that an intimate relation exists between the kernel in the gap equation and that in the BSE.  (This is another example of the coupling between DSEs.)  Therefore an understanding of chiral symmetry and its dynamical breaking can only be obtained with a truncation scheme that preserves this relation, and hence guarantees Eq.\,(\ref{avwtimN}) without a fine-tuning of model-dependent parameters.  Until 1995--1996 no one had a good idea about how to do this.  Equations were truncated, sometimes with good phenomenological results and sometimes with poor results.  Neither the successes nor the failures could be explained.

That changed with Refs.\,\cite{Munczek:1994zz, Bender:1996bb}, which described a procedure that generates a Bethe-Salpeter equation from the kernel of any gap equation whose diagrammatic content is known.  Its mere existence enabled the proof of exact nonperturbative results in QCD.  (See, e.g., Sec.\,V in Ref.\,\cite{Bashir:2012fs}.)  The scheme remains the most widely used today.  Its leading-order term provides the rainbow-ladder (RL) truncation, which is accurate for ground-state vector- and isospin-nonzero-pseudoscalar-mesons \cite{Maris:2003vk, Chang:2011vu, Bashir:2012fs, Cloet:2013jya}, and the properties of ground-state octet and decouplet baryons \cite{Eichmann:2011ej, Chen:2012qr, Segovia:2013rca, Segovia:2013ugaS} because corrections in these channels largely cancel, owing to parameter-free preservation of the WGTIs.  However, they do not cancel in other channels \cite{Roberts:1996jx, Roberts:1997vs, Bender:2002as, Bhagwat:2004hn}.  Hence studies based on the rainbow-ladder truncation, or low-order improvements thereof, have usually provided poor results for scalar- and axial-vector-mesons \cite{Cloet:2007pi, Burden:1996nh, Watson:2004kd, Maris:2006ea, Fischer:2009jm, Krassnigg:2009zh}, produced masses for exotic states that are too low in comparison with other estimates \cite{Cloet:2007pi, Qin:2011dd, Burden:1996nh, Qin:2011xq, Krassnigg:2009zh}, and exhibit gross sensitivity to model parameters for tensor-mesons \cite{Krassnigg:2010mh} and excited states \cite{Qin:2011dd, Qin:2011xq, Holl:2004fr, Holl:2004un}.  In these circumstances one must conclude that physics important to these states is omitted.

Fortunately, a recently developed truncation scheme overcomes these difficulties \cite{Chang:2009zb} and is beginning to have a material impact.  (An overview is presented in Sec.\,VI of Ref.\,\cite{Bashir:2012fs}.)  This scheme is also symmetry preserving.  Its additional strengths, however, are the capacities to work with an arbitrary dressed-quark gluon vertex and express DCSB nonperturbatively in the Bethe-Salpeter kernels.  The new scheme has enabled a range of novel nonperturbative features of QCD to be demonstrated.  For example, the existence of dressed-quark anomalous chromo- and electro-magnetic moments \cite{Chang:2010hb} and the key role they play in determining observable quantities \cite{Chang:2011tx}; elucidation of the causal connection between DCSB and the splitting between vector and axial-vector mesons \cite{Chang:2011ei} and the impact of this splitting on the baryon spectrum \cite{Chen:2012qr}; and most recently, as explained in Sec.\,\ref{secAbInitio} below, a crucial step toward the \emph{ab initio} prediction of hadron observables in continuum-QCD.

\subsection{Gluon Cannibalism}
\label{secCannibals}
I described DCSB and the dynamical generation of a running dressed-quark mass in Sec.\,\ref{secDCSB}.  Crucially, it is not just the propagation of quarks that is affected by strong interactions in QCD.  The propagation of gluons, too, is described by a gap equation \cite{Aguilar:2009nf}; and its solution shows that gluons are cannibals: they are a particle species whose members become massive by eating each other!  The associated gluon mass function, $m_g(k^2)$, is monotonically decreasing with increasing $k^2$ and recent work \cite{Binosi:2014aea} has established that
\begin{equation}
\label{gluonmassEq}
m_g(k^2=0) \approx 0.5\,{\rm GeV}.
\end{equation}
The value of the mass-scale in Eq.\,\eqref{gluonmassEq} is \emph{natural} in the sense that it is commensurate with but larger than the value of the dressed light-quark mass function at far infrared momenta: $M(0)\approx 0.3\,$GeV (see Fig.\,\ref{gluoncloud}). Moreover, the mass term appears in the transverse part of the gluon propagator, hence gauge-invariance is not tampered with; and the mass function falls as $1/k^2$ for $k^2\gg m_g(0)$ (up to logarithmic corrections), so the gluon mass is invisible in perturbative applications of QCD: it has dropped to less-than 5\% of it's infrared value by $k^2=4\,$GeV$^2$.

Gauge boson cannibalism presents a new physics frontier within the Standard Model. Asymptotic freedom means that the ultraviolet behaviour of QCD is controllable.  At the other extreme, dynamically generated masses for gluons and quarks entail that QCD creates its own infrared cutoffs.  Together, these effects eliminate both the infrared and ultraviolet problems that typically plague quantum field theories and thereby make reasonable the hope that QCD is nonperturbatively well defined.

The dynamical generation of gluon and quark masses provides a basis for understanding the notion of a maximum wavelength for gluons and quarks in QCD \cite{Brodsky:2008be}.  Indeed, given the magnitudes of the gluon and quark mass-scales, it is apparent that field modes with wavelengths $\lambda > \varsigma \approx 2/m_g(0) \approx 0.5\,$fm decouple from the dynamics.  They are screened in the sense described in Sec.\,\ref{secConfinement}.  This is just one consequence of the appearance of a dynamically generated gluon mass-scale.

There are many more.  For example, the exceptionally light pion degree-of-freedom becomes dominant in QCD at those length-scales above which dressed-gluons and -quarks decouple from the theory owing to the large magnitudes of their dynamically generated masses.  It is therefore conceivable that Gribov copies have no measurable impact on observables within the Standard Model because they affect only those gluonic modes whose wavelengths lie in the far infrared; and such modes are dynamically screened, by an exponential damping factor $\sim \exp(-\lambda/\varsigma)$, so that their role in hadron physics is superseded by the dynamics of light-hadrons.  This conjecture is consistent with the insensitivity to Gribov copies of the dressed-gluon and -quark two-point Schwinger functions observed in numerical simulations of QCD on fine lattices \cite{Bowman:2002fe,Zhang:2004gv}.  Another plausible conjecture is that dynamical generation of an infrared gluon mass-scale leads to saturation of the gluon parton distribution function at small Bjorken-$x$ within hadrons. The possible emergence of this phenomenon stirs great scientific interest and curiosity.  It is a key motivation in plans to construct an EIC that would be capable of producing a precise understanding of collective behaviour amongst gluons \cite{Accardi:2012qut}.

\subsection{Continuum-QCD and \mbox{\rm ab initio} predictions of hadron observables}
\label{secAbInitio}
Within contemporary hadron physics there are two common methods for determining the momentum-dependence of the interaction between quarks: the top-down approach, which works toward an \textit{ab initio} computation of the interaction via direct analysis of the gauge-sector gap equations; and the bottom-up scheme, which aims to infer the interaction by fitting data within a well-defined truncation of those equations in the matter sector that are relevant to bound-state properties.  These two approaches have recently been united \cite{Binosi:2014aea} by a demonstration that the renormalisation-group-invariant (RGI) running-interaction predicted by contemporary analyses of QCD's gauge sector coincides with that required in order to describe ground-state hadron observables using a nonperturbative truncation of QCD's Dyson-Schwinger equations in the matter sector, \emph{i.e}.\ the DCSB-improved (DSE-DB) kernel described briefly in Sec.\,\ref{secTrunc} and elucidated in Refs.\,\cite{Chang:2009zb, Chang:2010hb, Chang:2011ei, Bashir:2012fs}.

\begin{figure}[t]
\begin{minipage}[t]{\textwidth}
\begin{minipage}{0.48\textwidth}
\centerline{\includegraphics[clip,width=0.9\textwidth]{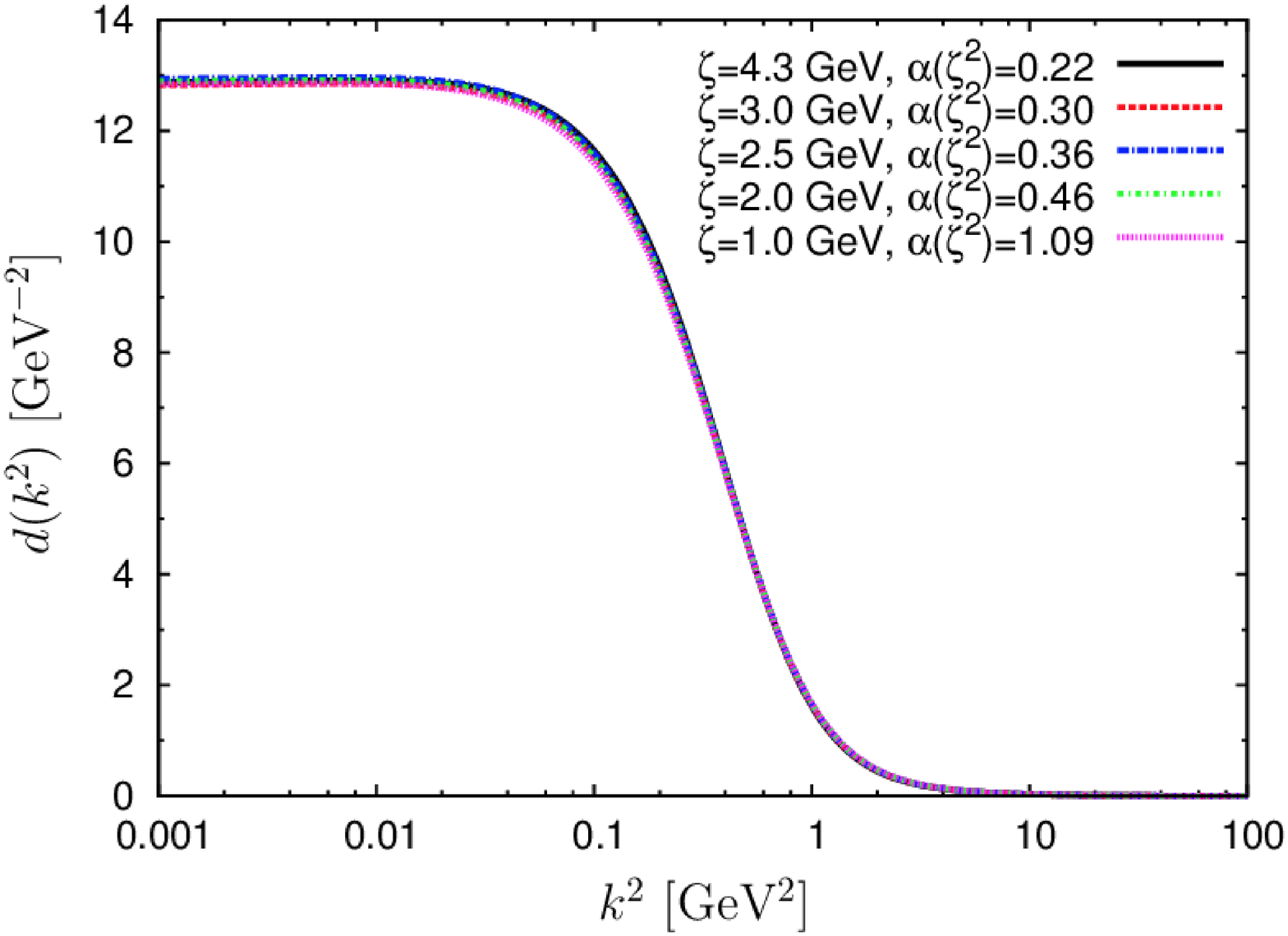}}
\end{minipage}
\begin{minipage}{0.02\textwidth}
\hspace*{-0.2em}\mbox{\LARGE \textbf{$\Rightarrow$}}
\end{minipage}
\begin{minipage}{0.48\textwidth}
\centerline{\includegraphics[clip,width=0.9\textwidth]{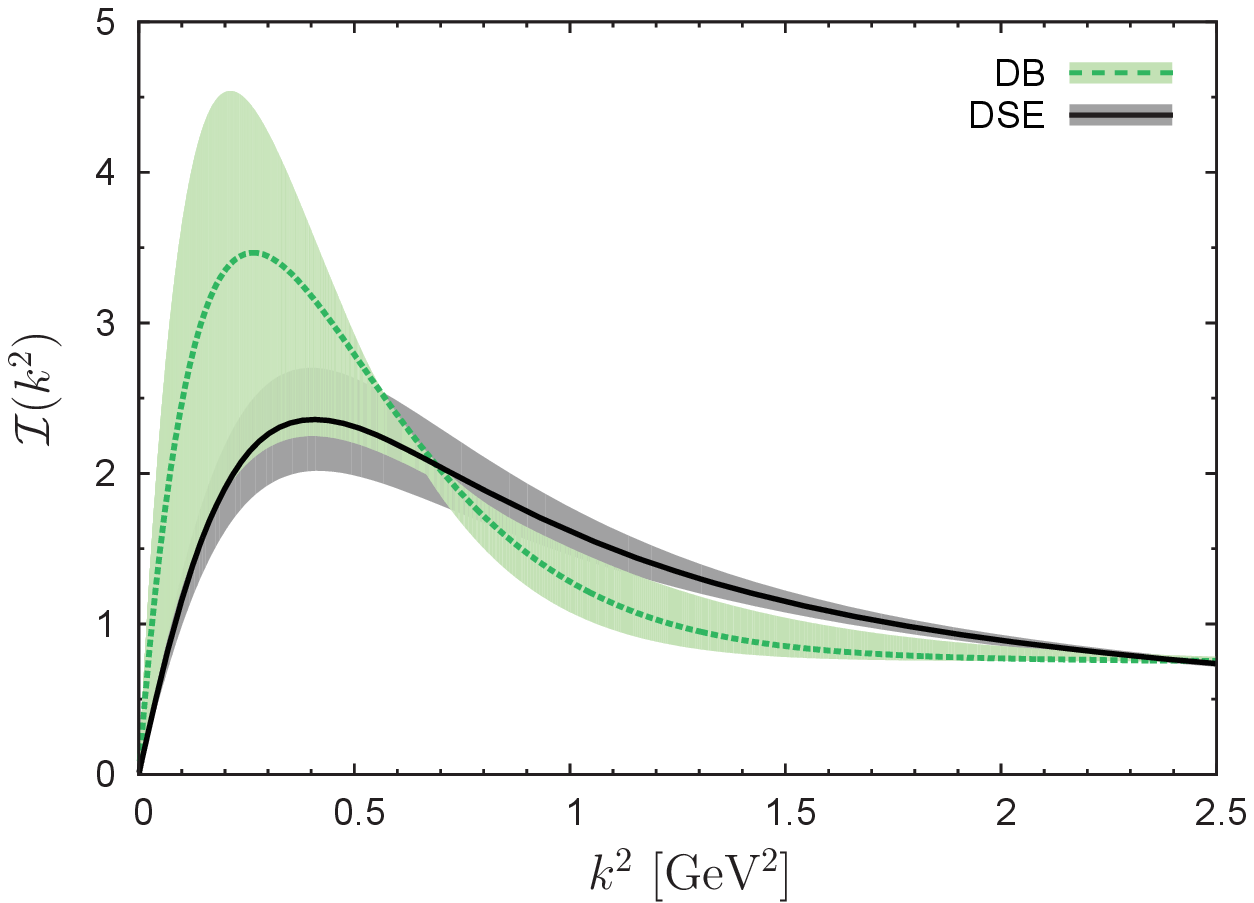}}
\end{minipage}
\end{minipage}
\caption{\label{figInteraction} \small
\emph{Left panel} -- RGI running interaction strength computed via a combination of DSE- and lattice-QCD analyses, as explained in  Ref.\,\cite{Aguilar:2009nf}.  The function obtained with five different values of the renormalisation point is depicted in order to highlight that the result is RGI.  The interaction is characterized by a value $\alpha_s(0) \approx 0.9\, \pi$ and the gluon mass-scale in Eq.\,\eqref{gluonmassEq}.
\emph{Right panel} -- Comparison between top-down results for the gauge-sector interaction (derived from the left-panel) with those obtained using the bottom-up approach based on hadron physics observables.  \underline{Solid curve} within \emph{grey band} -- top-down result for the RGI running interaction; and \underline{dashed curve} within \emph{pale-green band} -- advanced bottom-up result obtained using the most sophisticated truncation of the matter sector DSEs - the DSE-DB kernel.  The bands denote the domain of uncertainty in the determinations of the interaction.
}
\end{figure}

The unification is illustrated in Fig.\,\ref{figInteraction}: the right panel presents a comparison between the top-down RGI interaction (solid-black curve within grey band, derived from the left panel) and the DB-truncation bottom-up interaction (green band containing dashed curve).  Plainly, the interaction predicted by modern analyses of QCD's gauge sector is in near precise agreement with that required for a veracious description of measurable hadron properties using the most sophisticated matter-sector gap and Bethe-Salpeter kernels available today.  This is a remarkable result, given that there had previously been no serious attempt at communication between practitioners from the top-down and bottom-up hemispheres of continuum-QCD.  It bridges a gap that had lain between nonperturbative continuum-QCD and the \emph{ab initio} prediction of bound-state properties.

It should be noted that if the realistic interaction depicted in Fig.\,\ref{figInteraction} were employed as the seed for a RL-truncation study, it would fail completely because, \emph{inter alia}, DCSB would be absent.  We now know that a veracious description of DCSB and hence hadron properties in QCD requires a dressed-quark-gluon vertex.  Constraining its form is a topic of great contemporary interest; and in this connection it cannot be emphasised too strongly that little of value today will be produced by any attempt at a term-by-term diagrammatic construction of this vertex.

\subsection{Enigma of mass}
\label{secEnigma}
As noted in Sec.\,\ref{secConfinement}, the pion is Nature's lightest hadron.  In fact, it is peculiarly light, with a mass just one-fifth of that which quantum mechanics would lead one to expect.  This remarkable feature has its origin in DCSB.  In quantum field theory the pion's structure is described by a Bethe-Salpeter amplitude (here $k$ is the relative momentum between the  valence-quark and -antiquark constituents, and $P$ is their total momentum):
\begin{equation}
\Gamma_{\pi}(k;P) = \gamma_5 \left[
i E_{\pi}(k;P) + \gamma\cdot P F_{\pi}(k;P)  + \gamma\cdot k \, G_{\pi}(k;P) - \sigma_{\mu\nu} k_\mu P_\nu H_{\pi}(k;P)
\right],
\label{genGpi}
\end{equation}
which is simply related to an object that would be the pion's Schr\"odinger wave function if a nonrelativistic limit were appropriate.  In QCD if, and only if, chiral symmetry is dynamically broken, then in the chiral limit ($\hat m=0$) \cite{Maris:1997hd, Qin:2014vya}:
\begin{equation}
\label{gtrE}
f_\pi E_\pi(k;0) = B(k^2)\,,
\end{equation}
where $f_\pi$ is the pion's leptonic decay constant, a directly measurable quantity that connects the strong and weak interactions, and the rhs is a scalar function in the dressed-quark propagator, Eq.\,\eqref{SgeneralN}.  This identity is miraculous.  It means that the two-body problem is solved, almost completely, without lifting a finger, once the solution to the one body problem is known.  Eq.\,\eqref{gtrE} is a quark-level Goldberger-Treiman relation.  It is also the most basic expression of Goldstone's theorem in QCD, \emph{viz}.\\[-3ex]

\centerline{\parbox{0.95\textwidth}{\flushleft \emph{Goldstone's theorem is fundamentally an expression of equivalence between the one-body problem and the two-body problem in QCD's colour-singlet pseudoscalar channel.}}}

\medskip

\hspace*{-\parindent}Eq.\,\eqref{gtrE} emphasises that Goldstone's theorem has a pointwise expression in QCD; and, furthermore, that pion properties are an almost direct measure of the mass function depicted in Fig.\,\ref{gluoncloud}.  Thus, enigmatically, properties of the (nearly-)massless pion are the cleanest expression of the mechanism that is responsible for almost all the visible mass in the Universe.  Plainly, DCSB has a very deep and far-reaching impact on physics within the Standard Model.

\subsection{Nature's Vacuum}
\label{secVacuum}
\begin{figure}[t]
\begin{minipage}[t]{\textwidth}
\begin{minipage}{0.45\textwidth}
\centerline{\includegraphics[clip,width=0.9\textwidth]{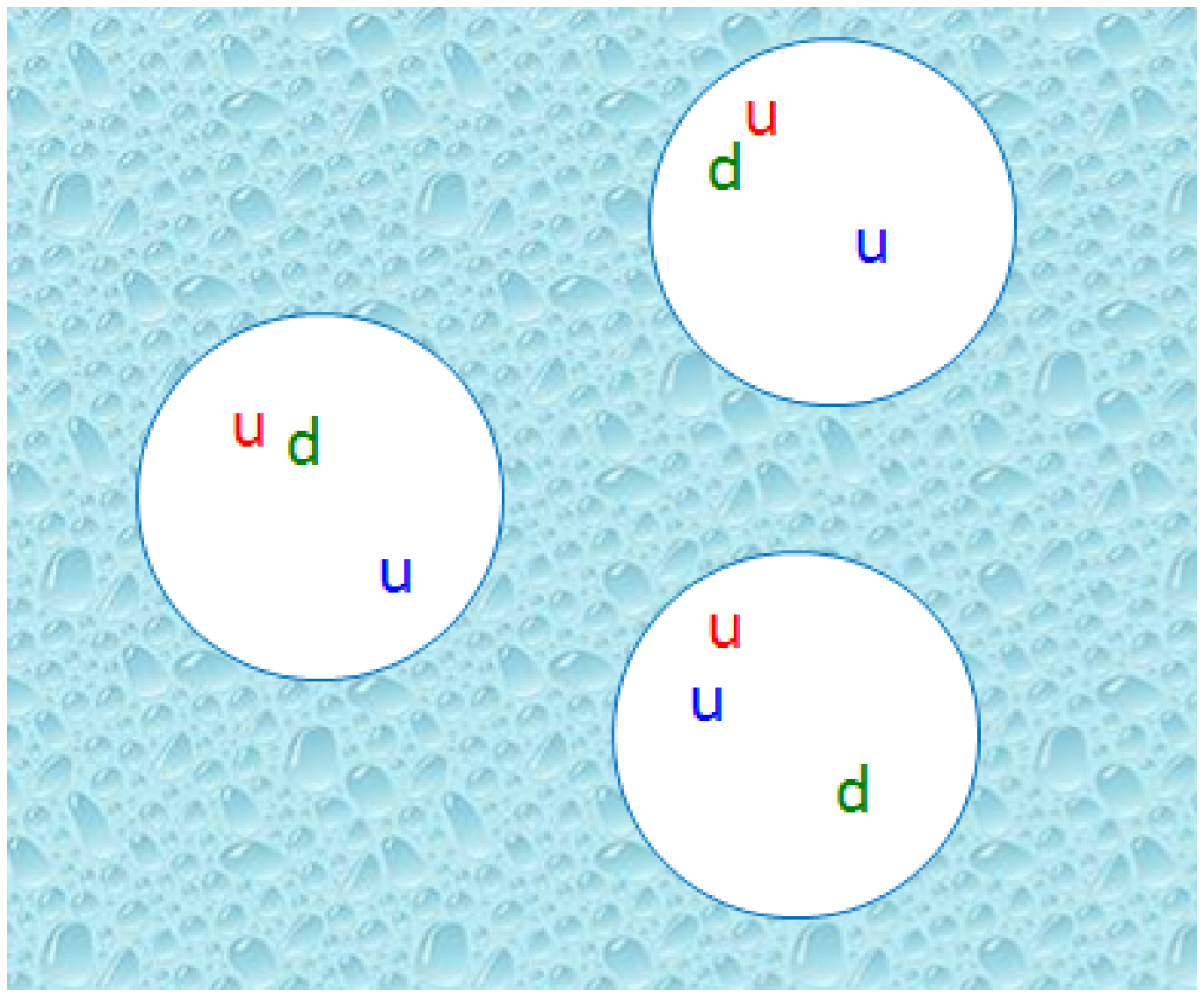}}
\end{minipage}
\begin{minipage}{0.10\textwidth}
\hspace*{2em}\mbox{\LARGE \textbf{$\Rightarrow$}}
\end{minipage}
\begin{minipage}{0.45\textwidth}
\centerline{\includegraphics[clip,width=0.9\textwidth]{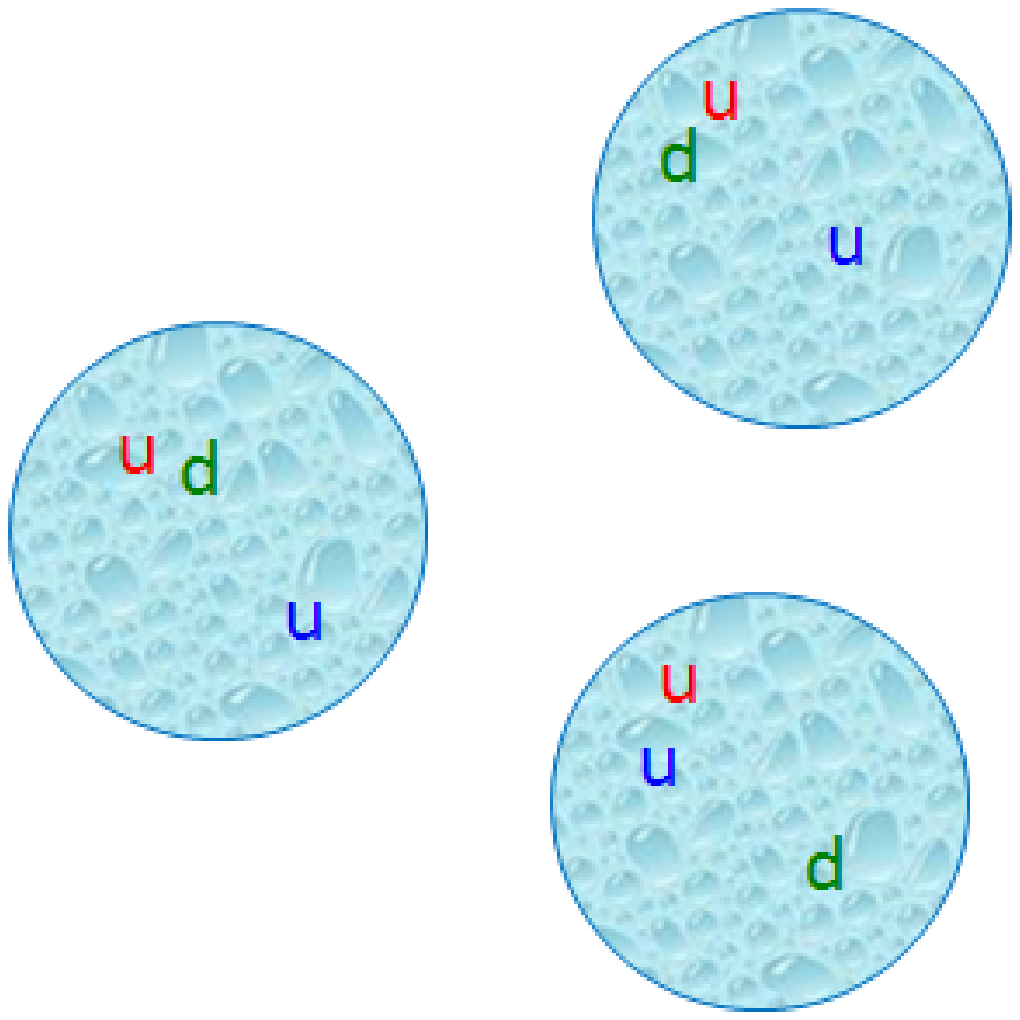}}
\end{minipage}
\end{minipage}
\caption{\label{vacuum} \small
\emph{Left panel} -- The QCD vacuum was historically imagined to be a ``frothing sea,'' with hadrons being merely bubbles of foam containing nothing but quarks and gluons interacting perturbatively throughout.  Only near the bubble's boundary did some sort of strong interaction occur, enforcing quark and gluon confinement.
\emph{Right panel} -- A modern view \cite{Maris:1997tm, Brodsky:2009zd, Brodsky:2010xf, Chang:2011mu, Brodsky:2012ku} flips the picture completely.  The space between hadrons is ``empty,'' except for the perturbative quantum fluctuations that characterise all interacting relativistic quantum field theories.  The interior of hadrons, however, is extremely complex, with nonperturbative dynamics dominating in $\sim 98$\% of the interior: the so-called condensates are spacetime-dependent and confined within the hadrons.
}
\end{figure}

It will now be evident that DCSB is a crucial emergent feature of the Standard Model.  It is very clearly expressed in the dressed-quark mass function of Fig.\,\ref{gluoncloud}.  However, this perspective is relatively recent.  DCSB was historically conflated with the existence of a spacetime-independent quark-antiquark condensate, $\langle \bar q q\rangle$, that permeates the Universe.  This notion was born with the introduction of QCD sum rules as a theoretical artifice to estimate nonperturbative strong-interaction matrix elements \cite{Shifman:1978bx, Leinweber:1995fn} and is typically tied to a belief, popularised in a  Wikipedia entry, that the QCD vacuum is characterised by an infinite number of mass-dimensioned, spacetime-independent condensates, as illustrated in the left panel of Fig.\,\ref{vacuum}.

It is strange to imagine that in order to understand the properties of any single one of the hadrons supported by QCD, it is necessary to compute and explain the values of an infinite number of mass-dimensioned parameters, none of which is a property of any of the hadrons supported by QCD dynamics and all of which exist throughout the Universe, even in the absence of hadrons.  This means, \emph{e.g}.\ that the values of all these condensates in domains of the Universe which are not causally connected with ours nevertheless provide the keys to understanding the proton's radius.  This belief is too bizarre for me.  Indeed, it is qualitatively equivalent to supposing that the Cooper pairs which explain superconductivity actually exist outside the superconducting metal instead of inside the lattice of positive ions that enables the electron-phonon interactions responsible for the phenomenon \cite{Brodsky:2009zd}.

There is an alternative, which can be stated succinctly as follows \cite{Maris:1997tm, Brodsky:2009zd, Brodsky:2010xf, Chang:2011mu, Brodsky:2012ku}: \\[0.7ex]
\hspace*{2em}\parbox[t]{0.9\textwidth}{\emph{If quark-hadron duality is a reality in QCD, then condensates, those quantities that were commonly viewed historically as constant empirical mass-scales that fill all spacetime, are instead wholly contained within hadrons, {\rm i.e}.\ they are a property of hadrons themselves and expressed, \mbox{\rm e.g.}, in their Bethe-Salpeter or light-front wave functions}.}
\smallskip

\hspace*{-\parindent}This standpoint is depicted in the right panel of Fig.\,\ref{vacuum} and canvassed fully in Sec.\,4 of Ref.\,\cite{Cloet:2013jya}.  It presents the reasonable view that the understanding of hadrons requires that one explain what lies within those hadrons in contrast to the historical alternative, which suggested that hadrons could only be understood by explaining the properties of the vast spacetime domains that contain no hadrons at all.

In order to provide a little more context, it is perhaps worth noting that the Gell-Mann--Oakes--Renner (GMOR) relation \cite{GellMann:1968rz} is often cited as ``proof'' of the existence of a vacuum quark condensate.  So let's consider the published form of that identity, \emph{viz}.\ Eq.\,(3.4) in Ref.\,\cite{GellMann:1968rz}:
\begin{equation}
\label{gmor}
m_\pi^2 = \lim_{P^\prime \to P \to 0} \langle \pi(P^\prime) | {\cal H}_{\chi{\rm sb}}|\pi(P)\rangle\,,
\end{equation}
where $m_\pi$ is the pion's mass and ${\cal H}_{\chi{\rm sb}}$ is that part of the hadronic Hamiltonian density which explicitly breaks chiral symmetry.  It is crucial to observe that the operator expectation value in Eq.\,(\ref{gmor}) is evaluated between pion states: it makes no reference to a vacuum quark condensate.
Moreover, the virtual low-energy limit expressed in Eq.\,\eqref{gmor} is purely formal.  It does not describe an achievable empirical situation. 

In terms of QCD quantities, Eq.\,(\ref{gmor}) entails
\begin{equation}
\label{gmor1}
\forall m_{ud} \sim 0\,,\;  m_{\pi^\pm}^2 =  m_{ud}^\zeta \, {\cal S}_\pi^\zeta(0)\,,
\quad
{\cal S}_\pi^\zeta(0) = - \langle \pi(P) | \mbox{\small $\frac{1}{2}$}(\bar u u + \bar d d) |\pi(P)\rangle\,,
\end{equation}
where $m_{ud}^\zeta = m_u^\zeta+m_d^\zeta$, with $m^\zeta$ the running quark mass of the subscripted quark flavour, and ${\cal S}^\zeta(0)$ is the pion's scalar form factor at zero momentum transfer, $Q^2=0$.  The rhs of Eq.\,(\ref{gmor1}) is proportional to the pion $\sigma$-term (see, e.g., Ref.\,\cite{Flambaum:2005kc}).  Consequently, using the connection between the $\sigma$-term and the Feynman-Hellmann theorem, Eq.\,(\ref{gmor}) is actually the statement
\begin{equation}
\label{pionmass2}
\forall m_{ud} \simeq 0\,,\; m_\pi^2 = m_{ud}^\zeta \frac{\partial }{\partial m^\zeta_{ud}} m_\pi^2.
\end{equation}

At this point one should recall the following mass formula for pseudoscalar mesons containing a valence $f$-quark and valence $g$-antiquark, which is exact in QCD \cite{Maris:1997hd,Maris:1997tm}:
\begin{equation}
\label{mrtrelation}
f_{H_{0^-}} m_{H_{0^-}}^2 = (m_{f}^\zeta + m_{g}^\zeta)\, \rho_{H_{0^-}}^\zeta,
\end{equation}
where, with the trace over colour and spinor indices,
\begin{eqnarray}
i f_{H_{0^-}} P_\mu  &=& \langle 0 | \bar q_{g} \gamma_5 \gamma_\mu q_{f} |H_{0^-}\rangle
= Z_2\; {\rm tr}_{\rm CD}
\int_{dk}^\Lambda i\gamma_5\gamma_\mu S_{f}(k_+) \Gamma_{H_{0^-}}(k;P) S_{g}(k_{-})\,, \label{fpigen} \\
i\rho_{H_{0^-}} &=& -\langle 0 | \bar q_{g} i\gamma_5 q_{f} |H_{0^-} \rangle
= Z_4\; {\rm tr}_{\rm CD}
\int_{dk}^\Lambda \gamma_5 S_{f}(k_+) \Gamma_{H_{0^-}}(k;P) S_{g}(k_{-})
\label{rhogen}
\end{eqnarray}
are the meson's pseudovector and pseudoscalar decay constants.

Now, using Eq.\,\eqref{mrtrelation}, one obtains
\begin{equation}
\label{gmor2}
{\cal S}_\pi^\zeta(0)
= \frac{\partial }{\partial m^\zeta_{ud}} m_\pi^2
=\frac{\partial }{\partial m^\zeta_{ud}} \left[ m_{ud}^\zeta\frac{\rho_\pi^\zeta}{f_\pi}\right].
\end{equation}
Equation~(\ref{gmor2}) is valid for any values of $m_{u,d}$, including the neighborhood of the chiral limit, wherein
\begin{equation}
\label{gmor3}
{\cal S}_\pi^\zeta(0)= \frac{\partial }{\partial m^\zeta_{ud}} \left[ m_{ud}^\zeta\frac{\rho_\pi^\zeta}{f_\pi} \right]_{m_{ud} = 0}
= \frac{\rho_\pi^{\zeta 0}}{f_\pi^0}\,.
\end{equation}
The superscript ``0'' indicates that the quantity is computed in the chiral limit.  With Eqs.\,(\ref{gmor1}), (\ref{gmor2}), (\ref{gmor3}), one has shown that in the neighborhood of the chiral limit
\begin{equation}
m_{\pi^\pm}^2 =  -m_{ud}^\zeta  \frac{\langle \bar q q \rangle^{\zeta 0}}{(f_\pi^0)^2} + {\rm O}(m_{ud}^2).
\label{truegmor}
\end{equation}
This is a QCD derivation of the commonly recognised form of the GMOR relation.  Neither the venerable tools of PCAC, \emph{i.e}.\ exploiting the consequences of partial conservation of the axial current, nor soft-pion theorems were employed in analysing the rhs of Eqs.\,\,\eqref{gmor}, (\ref{gmor1}).  In addition, the derivation shows that in the chiral limit, the matrix element describing the pion-to-vacuum transition through the pseudoscalar vertex is equal to the expectation value of the explicit chiral symmetry breaking term in the QCD Lagrangian \emph{computed between pion states} after normalisation by the pion's decay constant.

This recapitulation of the analysis in Ref.\,\cite{Chang:2011mu} emphasises anew that any connection between the pion mass and a ``vacuum'' quark condensate is purely a theoretical artifice.  The true connection is that which one would expect, \emph{viz}.\ the pion's mass is a property of the pion, determined by the interactions between its constituents.  One may further highlight the illogical nature of the vacuum connection by considering the expectation value
\begin{equation}
\label{HOH}
\langle H(P) | \bar q {\cal O} q | H(P)\rangle \,,
\end{equation}
where $H(P)$ represents some hadron with total momentum $P$.  If one chooses ${\cal O}=\gamma_\mu$, then all readers will accept that Eq.\,\eqref{HOH} yields the electric charge of the hadron $H$.  The choice ${\cal O}=\gamma_5\gamma_\mu$ in Eq.\,\eqref{HOH} will yield the axial charge of the hadron $H$; ${\cal O}=\sigma_{\mu\nu}$ will yield the tensor charge of the hadron $H$; and the choice ${\cal O}=\mathbf I$ will yield the scalar charge of the hadron $H$.  All these statements are true irrespective of the hadron involved, so that the scalar charge of the pion; i.e., ${\cal S}_\pi^\zeta(0)=\kappa_\pi^0/(f_\pi^0)^2$, the ``B-parameter'' in chiral perturbation theory, is no more a property of the vacuum than is the pion's electric charge.

The change in understanding highlighted by shifting from the right to the left panel in Fig.\,\ref{vacuum} does not undermine the utility of the QCD sum rules approach to the estimation of hadron observables.  Instead, it tames the condensates so that they return to being merely mass-dimensioned parameters in a useful computation scheme.  Its implications are, however, significant and wide-ranging, especially if one imagines that the theory of gravity is understood well enough so that it may reliably be coupled to quantum field theory.  Subscribers to this view argue \cite{Turner:2001yu, Bass:2011zz} that the energy-density of the Universe must receive a contribution from ``vacuum condensates'' and that the only possible covariant form for the energy of the (quantum) vacuum, \emph{viz}.
\begin{equation}
T_{\mu\nu}^{\rm VAC} = \rho_{\rm VAC}\, \delta_{\mu\nu}\,,
\end{equation}
is mathematically equivalent to the cosmological constant.  From this perspective, the quantum vacuum is \cite{Turner:2001yu} ``\ldots a perfect fluid and precisely spatially uniform \ldots'' so that ``Vacuum energy is almost the perfect candidate for dark energy.''  Now, if the ground state of QCD is really expressed in a nonzero spacetime-independent expectation value $\langle\bar q q\rangle$, then the energy difference between the symmetric and broken phases is roughly $M_{\rm QCD} \sim 0.3\,$GeV, as indicated by Fig.\,\ref{gluoncloud}.  One obtains therefrom:
\begin{equation}
\rho_\Lambda^{\rm QCD} = 10^{46} \rho_\Lambda^{\rm obs},
\end{equation}
\emph{i.e}.\ the contribution from the QCD vacuum to the energy density associated with the cosmological constant exceeds the observed value by forty-six orders-of-magnitude.  In fact, the discrepancy is far greater if the Higgs vacuum expectation value is treated similarly.

This mismatch has been called \cite{Zee:2008zza} ``\ldots one of the gravest puzzles of theoretical physics.''  However, it vanishes if one discards the notion that condensates have a physical existence, which is independent of the hadrons that express QCD's asymptotically realisable degrees of freedom \cite{Brodsky:2009zd}; namely, if one accepts that such condensates are merely mass-dimensioned parameters in one or another theoretical computation and truncation scheme.  This appears mandatory in a confining theory \cite{Chang:2011mu,Brodsky:2010xf,Brodsky:2012ku}, a perspective one may embed in a broader context by considering just what is observable in quantum field theory \cite{Weinberg:1978kz}: ``\ldots although individual quantum field theories have of course a good deal of content, quantum field theory itself has no content beyond analyticity, unitarity, cluster decomposition and symmetry.''  If QCD is a confining theory, then the principle of cluster decomposition is only realised for colour singlet states \cite{Krein:1990sf} and all observable consequences of the theory, including its ground state, can be expressed via a hadronic basis.  This is quark-hadron duality.  Furthermore, if technicolour-like theories \cite{Andersen:2011yj, Sannino:2013wla} are the correct scheme for explaining electroweak symmetry breaking, then the impact of the notion of in-hadron condensates is far greater still because it enables the Higgs vacuum expectation value to be understood and eliminated in the same manner.

\subsection{Parton Structure of Hadrons}
\label{sec-1}
Since the advent of the parton model and the first deep inelastic scattering experiments there has been a determined effort to deduce the parton distribution functions (PDFs) of the most stable hadrons \cite{Holt:2010vj}.  The behaviour of such distributions on the valence domain (Bjorken-$x> 0.5$) is of particular interest because this domain is definitive of hadrons, \emph{e.g}.\ quark content on the valence domain is how one distinguishes between a neutron and a proton: a neutron possesses one valence $u$-quark plus two valence $d$-quarks whereas the proton possesses two valence $u$-quarks plus one valence $d$-quark.  Indeed, all Poincar\'e-invariant properties of a hadron: baryon number, charge, total spin, \emph{etc}., are determined by the PDFs which dominate on the valence domain.  Moreover, via QCD evolution \cite{Dokshitzer:1977, Gribov:1972, Lipatov:1974qm, Altarelli:1977}, PDFs on the valence-quark domain determine backgrounds at the LHC.  There are also other questions, \emph{e.g}.\ regarding flavour content of a hadron's sea and whether that sea possesses an intrinsic component \cite{Brodsky:1980pb, Signal:1987gz}.  The answers to all these questions are essentially nonperturbative properties of QCD.

Recognising the significance of the valence domain, a new generation of experiments, focused on Bjorken-$x\gtrsim 0.5$, is planned at JLab, and under examination in connection with Drell-Yan studies at Fermilab and a possible EIC.  Consideration is also being given to experiments aimed at measuring parton distribution functions in mesons at J-PARC.  

A concentration on such measurements requires theory to move beyond merely parametrising distribution functions and amplitudes.  Computation within QCD-connected frameworks becomes critical because, without it, no amount of data will reveal anything about the theory underlying strong interaction phenomena.  This is made clear by the example of the pion's valence-quark PDF, $u_v^\pi(x)$, in connection with which a failure of QCD was suggested following a leading-order analysis of $\pi N$ Drell-Yan measurements \cite{Conway:1989fs}.  As explained in Ref.\,\cite{Holt:2010vj}, this confusion was fostered by the application of a diverse range of models.  On the other hand, a series of QCD-connected calculations  \cite{Hecht:2000xa, Aicher:2010cb, Nguyen:2011jy, Chang:2014lvaS} subsequently established that the leading-order analysis was misleading, so that $u_v^\pi(x)$ may now be seen as a success for the unification of nonperturbative and perturbative studies in QCD.

At this point it is worth noting that a meson's Bethe-Salpeter wave function, which is obtained from its Bethe-Salpeter amplitude (see Eq.\,\eqref{genGpi} for the pion example) by reattaching the dressed-quark legs, \emph{viz}.\
\begin{equation}
\chi^{fg}(k;P)= S_f(k_+) \Gamma^{fg}(k;P) S_g(k_-)\,,
\end{equation}
is the quantum field theory analogue of the Schr\"odinger wave function that would describe the system if it were simply quantum mechanical, and whenever a nonrelativistic limit makes sense, the Bethe-Salpeter and Schr\"odinger wave functions become the same in that limit \cite{Salpeter:1951sz}.  For those desiring a probability interpretation of wave functions, this would be reassuring except, of course, for the fact that a nonrelativistic limit is never appropriate when solving continuum bound-state equations for composite systems containing the light $u$-, $d$- and $s$-quarks.

To explain this comment, consider that the momentum-space wave function for a nonrelativistic quantum mechanical system, $\psi(p,t)$, is a probability amplitude, such that $|\psi(p,t)|^2$ is a non-negative density which expresses the probability that the system is described by momenta $p$ at a given equal-time instant $t$.  Although the replacement of certainty in classical mechanics by probability in quantum mechanics was disturbing for some, the step to relativistic quantum field theory is still more confounding.  As noted in connection with Eq.\,\eqref{eqParticleNumber}, much of the additional difficulty owes to the loss of particle number conservation when this step is made.  Two systems with equal energies need not have the same particle content because that is not conserved by Lorentz boosts and thus interpretation via probability densities is typically lost.  To exemplify: a charge radius cannot generally be defined via the overlap of two wave functions because the initial and final states do not possess the same four-momentum and hence are not described by the same wave function.

Such difficulties may be circumvented by formulating a theory on the light-front because the eigenfunctions of the light-front Hamiltonian are independent of the system's four-momentum \cite{Keister:1991sb, Coester:1992cg, Brodsky:1997de}.  The light-front wave function of an interacting quantum system therefore provides a connection between dynamical properties of the underlying relativistic quantum field theory and notions familiar from nonrelativistic quantum mechanics.  It can translate features arising purely through the infinitely-many-body nature of relativistic quantum field theory into images whose interpretation is seemingly more straightforward.  Put simply: quantum mechanics-like wave functions can be defined in quantum field theory; quantum-mechanics-like expectation values can then also be defined and evaluated; and parton distributions may be obtained therefrom via correlation functions evaluated at equal light-front time, namely within the initial surface $x^+ = x^0+x^3 = 0$, and can thus be expressed directly in terms of ground-state light-front wave functions.  Naturally, all that is only achieved if the light-front wave function can be calculated.

The simplest object to compute is a hadron's leading-twist two-particle parton distribution amplitude (PDA); and any framework that provides access to a hadron's Poincar\'e-covariant bound-state amplitude can also be employed to compute its PDAs.  For example, the pion's leading-twist two-particle PDA is given by the following projection of the pion's Bethe-Salpeter wave function onto the light-front \cite{Chang:2013pqS}
\begin{equation}
f_\pi\, \varphi_\pi(x;\zeta) = {\rm tr}_{\rm CD}
Z_2 \! \int_{dq}^\Lambda \!\!
\delta(n\cdot q_+ - x \,n\cdot P) \,\gamma_5\gamma\cdot n\,
S(q_+)\Gamma(q;P)S(q_{-})\,,
\label{pionPDA}
\end{equation}
where: $\int_{dq}^\Lambda$ is a Poincar\'e-invariant regularisation of the four-dimensional integral, with $\Lambda$ the ultraviolet regularization mass-scale; $Z_{2}(\zeta,\Lambda)$ is the quark wave-function renormalisation constant, with $\zeta$ the renormalisation scale; $n$ is a light-like four-vector, $n^2=0$; and $P$ is the pion's four-momentum, $P^2=-m_\pi^2$ and $n\cdot P = -m_\pi$.

\begin{figure}[t]
\begin{minipage}{\textwidth}
\begin{minipage}{0.5\textwidth}
\centerline{\includegraphics[width=0.95\textwidth,clip]{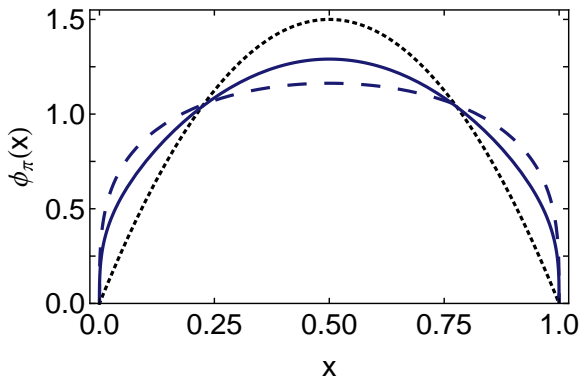}}
\end{minipage}
\begin{minipage}{0.5\textwidth}
\caption{\label{FigpionPDA} \small
Twist-two pion PDA computed us\-ing two very different DSE truncations at a scale $\zeta=2\,$GeV. \emph{Dashed curve} -- rainbow-ladder (RL), the leading-order in a systematic, sym\-metry-preserving scheme \protect\cite{Munczek:1994zz, Bender:1996bb}; and \emph{solid curve} -- the most sophisticated kernel that is currently available; namely, the DB kernel discussed in connection with Fig.\,\ref{figInteraction}, which incorporates nonperturbative effects generated by DCSB that are omitted in RL truncation and any stepwise improvement thereof \cite{Chang:2009zb, Chang:2010hb, Chang:2011ei}.  These results are consistent with contemporary lattice-QCD \cite{Cloet:2013tta, Segovia:2013ecaS, Shi:2014uwaS, Shi:2015esaS}.  The dotted curve is $\varphi^{\rm asy}_\pi(x)=6 x(1-x)$, the result obtained in conformal QCD \cite{Efremov:1979qk, Lepage:1979zb}.}
\end{minipage}
\end{minipage}
\end{figure}

The amplitude in Eq.\,\eqref{pionPDA} has been computed using two very different truncations of QCD's DSEs \cite{Chang:2013pqS}, with the result depicted in Fig.\,\ref{FigpionPDA}.  Both kernels agree: compared with the asymptotic form, which is valid when $\zeta$ is extremely large, there is a marked broadening of $\varphi_\pi(x;\zeta)$, which owes exclusively to DCSB.  This causal connection may be claimed because the PDA is computed at a low renormalisation scale in the chiral limit, whereat the quark mass function owes entirely to DCSB via Eq.\,\eqref{gtrE}.  Moreover, the dilation measures the rate at which a dressed-quark approaches the asymptotic bare-parton limit.  It can be verified empirically at JLab\,12, \emph{e.g}.\ in measurements of the pion's electromagnetic form factor, the ratio of the proton's electric and magnetic form factors, and the behaviour of the form factors that characterise transitions between the nucleon and its excited states.

It is important to highlight that the computed PDAs in Fig.\,\ref{FigpionPDA} are concave functions.  Such pointwise behaviour contrasts markedly with the ``humped'' or ``bimodal'' distributions which have been favoured by some authors in phenomenological applications \cite{Chernyak:1983ej}.  In this connection it must be understood, following Ref.\,\cite{Chang:2013pqS}, that a double-humped form for the twist-two PDA lies within the class of distributions produced by a meson Bethe-Salpeter amplitude which may be characterised as vanishing at zero relative momentum, instead of peaking thereat.  No ground-state pseudoscalar or vector meson Bethe-Salpeter equation solution exhibits corresponding behaviour \cite{Maris:1997tm, Maris:1999nt, Qin:2011xq}.  Thus, a bimodal distribution cannot be a pointwise-accurate representation of the PDA for a ground-state meson.  On the other hand, if one is using such a distribution in a practical phenomenological application for which only the lowest few moments are important, where the moments are defined via
\begin{equation}
\langle x^n \rangle = \int_0^1 dx \, x^n\,\varphi(x)\,,
\end{equation}
then some carefully-constrained bimodal distributions may serve to provide an approximation to the moments of a broad, concave PDA and can therefore provide useful information nevertheless.

A question of more than thirty-years standing can be answered using Fig.\,\ref{FigpionPDA}; namely, when does $\varphi^{\rm asy}_\pi(x)$ provide a good approximation to the pion PDA?  Plainly, not at $\zeta=2\,$GeV.  The ERBL evolution equation \cite{Efremov:1979qk, Lepage:1979zb} describes the $\zeta$-evolution of $\varphi_\pi(x;\zeta)$; and applied to $\varphi_\pi(x;\zeta)$ in Fig.\,\ref{FigpionPDA}, one finds \cite{Cloet:2013tta,Segovia:2013ecaS, Shi:2014uwaS} that $\varphi^{\rm asy}_\pi(x)$ is a poor approximation to the true result even at $\zeta=200\,$GeV.  Thus at empirically accessible energy scales, the twist-two PDAs of ground-state hadrons are ``squat and fat''.  Evidence supporting this picture had long been accumulating \cite{Mikhailov:1986be, Petrov:1998kg, Braun:2006dg, Brodsky:2006uqa}; and the dilation is verified by simulations of lattice-QCD \cite{Cloet:2013tta, Segovia:2013ecaS, Shi:2014uwaS, Shi:2015esaS}.

\subsection{PDAs and Hard Exclusive Processes}
It is now possible to add flesh to the bones of Eq.\,\eqref{QCDtest}.  In the theory of strong interactions, the cross-sections for many hard exclusive hadronic reactions can be expressed in terms of the PDAs of the hadrons involved.  An example is the pion's elastic electromagnetic form factor, for which the prediction can be stated succinctly  \cite{Farrar:1979aw, Efremov:1979qk, Lepage:1979zb, Lepage:1980fj}:
\begin{equation}
\label{pionUV}
\exists Q_0>\Lambda_{\rm QCD} \; |\;  Q^2 F_\pi(Q^2) \stackrel{Q^2 > Q_0^2}{\approx} 16 \pi \alpha_s(Q^2)  f_\pi^2 \mathpzc{w}_\varphi^2,
\quad
\mathpzc{w}_\varphi = \frac{1}{3} \int_0^1 dx\, \frac{1}{x} \varphi_\pi(x)\,,
\end{equation}
where the running coupling is defined in Eq.\,\eqref{eqalphaQCD}.  As noted in Sec.\,\ref{secJLab}, the value of $Q_0$ is not predicted by perturbative QCD.

\begin{figure}[t]
\begin{minipage}[t]{\textwidth}
\begin{minipage}{0.5\textwidth}
\centerline{\includegraphics[clip,width=0.95\textwidth]{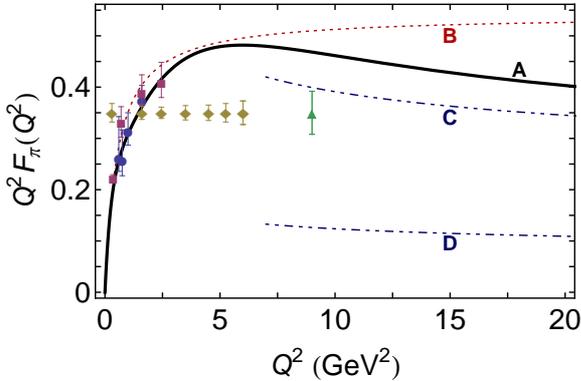}}
\end{minipage}
\begin{minipage}{0.5\textwidth}{\small
\caption{\label{figWPFpi} \small
$Q^2 F_\pi(Q^2)$.  Solid curve (A) -- Theoretical prediction \cite{Chang:2013nia}; dotted curve (B) -- monopole form fitted to data;  dot-dot-dashed curve (C) -- perturbative QCD (pQCD) prediction computed with the modern, dilated pion PDA described in Sec.\,\protect\ref{sec-1}; and dot-dot-dashed curve (D) -- pQCD prediction computed with the conformal-limit PDA, which had previously been used to guide expectations for the asymptotic behaviour of $Q^2 F_\pi(Q^2)$.  The filled-circles and -squares represent existing JLab data \cite{Huber:2008id}; and the filled diamonds and triangle, whose normalisation is arbitrary, indicate the projected $Q^2$-reach and accuracy of forthcoming experiments \cite{E1206101,E1207105}.}}
\end{minipage}
\end{minipage}
\end{figure}

It was anticipated that JLab would verify this fundamental Standard Model prediction, Eq.\,\eqref{pionUV}; and in 2001, seven years after commencing operations, the facility provided the first high-precision pion electroproduction data for $F_\pi(Q^2)$ between $Q^2$ values of 0.6 and 1.6\,GeV$^2$ \cite{Volmer:2000ek}.  In 2006 and 2007, this domain was revisited \cite{Tadevosyan:2007yd} and a new result was reported, this time at $Q^2=2.45\,$GeV$^2$ \cite{Horn:2006tm}.  However, there was disappointment and surprise, with the collaboration stating that experiment is ``still far from the transition to the $Q^2$ region where the pion looks like a simple quark-antiquark pair.''  This data is represented in Fig.\,\ref{figWPFpi} by the filled circles and squares, drawn from Ref.\,\cite{Huber:2008id}, and the collaboration were comparing with Curve-D in the figure, which is the prediction obtained when $\varphi^{\rm asy}_\pi(x)$, the asymptotic PDA appropriate to the conformal limit of QCD, is used in Eq.\,\eqref{pionUV}.

The confusion was compounded by the fact that JLab's data confirmed the behaviour predicted by a DSE-RL prediction for the pion form factor, computed in 2000 \cite{Maris:2000sk}.  That prediction could not resolve the difficulty because it relied on brute numerical methods and hence could not produce a result at $Q^2 > 4\,$GeV$^2$.\footnote{Notably, modern lQCD computations are restricted to the domain $Q^2 < 3\,$GeV$^2$ \protect\cite{Brandt:2013ffb}.}  In appearance, however, the shape of the prediction suggested to many that one might never see parton model scaling and QCD scaling violations at the momentum transfers accessible to JLab, even after its upgrade.

This conundrum was recently resolved \cite{Chang:2013nia}.  Using a refinement of known methods \cite{Nakanishi:1963zz, Nakanishi:1969ph}, also employed in the successful analysis of $\varphi_\pi(x;\zeta)$, described in Sec.\,\ref{sec-1}, a reliable prediction of $F_\pi(Q^2)$ is now available on the entire domain of spacelike $Q^2$.  This is Curve-A in Fig.\,\ref{figWPFpi}.  Moreover, the analysis enables correlation of that result with Eq.\,\eqref{pionUV}, using the modern PDA computed in precisely the same framework, which is Curve-C in the figure.  This leading-order, leading-twist QCD prediction, obtained with a pion valence-quark PDA evaluated at a scale appropriate to the experiment, underestimates the full DSE-RL computation by merely an approximately uniform 15\% on the domain depicted.  The small mismatch is explained by a combination of higher-order, higher-twist corrections to Eq.\,\eqref{pionUV} in pQCD on the one hand and, on the other hand, shortcomings in the rainbow-ladder truncation (see Sec.\,\ref{secAbInitio}), which predicts the correct power-law behaviour for the form factor but not precisely the right anomalous dimension (exponent on the logarithm) in the strong-coupling calculation.  Hence, disappointment is now transformed into optimism because the comparison of Curves A and C in Fig.\,\ref{figWPFpi} predicts that the upgraded JLab facility will reveal a maximum at $Q^2 \approx 6\,$GeV$^2$ and an experiment at $Q^2=9\,$GeV$^2$ will see a clear sign of parton model scaling for the first time in a hadron elastic form factor.

The implications of these results are far-reaching.  They indicate that hadro-particle physics is on the cusp of verifying the theory of factorisation in hard exclusive processes, with dominance of hard contributions to $F_\pi(Q^2)$ on $Q^2\gtrsim8\,$GeV$^2$.  Notwithstanding that, the normalisation of this form factor is fixed by a pion wave-function whose dilation with respect to $\varphi^{\rm cl}(x)=6 x (1-x)$ is a definitive signature of DCSB; and this affords the opportunity for an empirical measurement of the strength of DCSB, \emph{i.e}.\ the power behind the origin of visible mass.  This series of outcomes will be important pages in a book on the Standard Model, in which the first lines were written forty years ago; and they pave the way for a dramatic reassessment of pictures of proton and neutron structure, which is already well underway, as we shall see in Sec.\,\ref{secBaryons}.

\subsection{Light-front Wave Function}
A modern perspective on the pion's PDA was presented in Sec.\,\ref{sec-1}.  The more general quantity is the pion's light-front wave function:
\begin{equation}
f_\pi \, \varphi_\pi(x,k_\perp^2) =
Z_2 {\rm tr}_{\rm CD} \int^\Lambda \! \frac{dq_3dq_4}{(2\pi)^2} \,
\delta(n\cdot q_+ - x n\cdot P)\gamma_5\gamma\cdot n \,
S(q_+)\Gamma(q;P)S(q_{-})\,,
\label{varphiukp}
\end{equation}
from which one may readily recover the PDA by integrating: $\int dq_1 dq_2$, with a careful redefinition of the renormalisation constant and regularisation scale, as is plain by comparison with Eq.\,\eqref{pionPDA}.  If one proceeds from Eq.\,\eqref{varphiukp} using precisely the analytical framework in Ref.\,\cite{Chang:2013pqS}, then one will obtain a factorised result, \emph{viz}.\
\begin{equation}
\label{productphi}
\varphi_\pi(x,k_\perp^2) = \varphi_\pi^1(x) \times \varphi_\pi^2(k_\perp^2)\,,
\end{equation}
an outcome which cannot be true in any nontrivial theory.

To expose how this factorisation arises, consider the following illustrative, algebraic representations for the propagators and amplitudes:
\begin{subequations}
\label{AlgebraicSetc}
\begin{eqnarray}
\label{pointS}
S(p) &=& [-i\gamma \cdot p + M] \Delta_M(p^2)\,, \\
\label{rhoznu}
\rho_\nu(z) &=& \frac{1}{\surd \pi}\frac{\Gamma(\nu + 3/2)}{\Gamma(\nu+1)}\,(1-z^2)^\nu\,,\\
\label{rhoEpi}
\Gamma_\pi(k;q) & = &
i\gamma_5 \frac{M^{1+2\nu}}{f_\pi} \!\! \int_{-1}^{1}\!\! \!dz \,\rho(z)
\Delta_M^\nu(k_{+z}^2)\,,
\end{eqnarray}
\end{subequations}
where $\Delta_M(s) = 1/[s+M^2]$, $k_{\pm z} = k-(1 \mp z )q/2$, and I have set $P^2=0$ and used a momentum partitioning parameter $\eta=0$.  Inserting these expressions into Eq.\,\eqref{varphiukp}, one obtains an integrand denominator that is a product of $k$-quadratic forms, each raised to some power.  The denominator product can be combined into a single quadratic form, raised to a unique power, by using a Feynman parametrisation, so that Eq.\,\eqref{varphiukp} yields
\begin{align}
\nonumber
\langle x^m(k_\perp^2) \rangle &\sim  \int_{-1}^1 dz\,\rho_\nu(z)
\nu (\nu+1)\, \int_0^1 dx_1 dx_2\,x_1 (1-x_1)^{\nu-1}\\
& \times
Z_2 {\rm tr}_{\rm CD} \int^\Lambda \! \frac{d^2k_\|}{(2\pi)^2} \,
\left(\frac{n\cdot k}{n\cdot P}\right)^m
\frac{M^{2\nu}}
{[(k_\|^2- f(x_1,x_2,z)P)^2+k_\perp^2+M^2]^{2+\nu}}\,,
\end{align}
where $k_\|=(k_3,k_4)$ and $f(x_1,x_2,z)$ is some simple, algebraic function of its arguments.  At this point one would shift variables to obtain
\begin{eqnarray}
\nonumber
\langle x^m(k_\perp^2) \rangle &\sim &
Z_2 {\rm tr}_{\rm CD} \int^\Lambda \! \frac{d^2k_\|}{(2\pi)^2} \,\frac{M^{2\nu}}
{[k_\|^2+k_\perp^2+M^2]^{2+\nu}}\\
&& \times
\int_{-1}^1 dz\,\rho_\nu(z)
\nu (\nu+1)\, \int_0^1 dx_1 dx_2\,x_1 (1-x_1)^{\nu-1}
[f(x_1,x_2,z)]^m\,,
\end{eqnarray}
and the factorisation is now apparent: the rhs of the first line in this equation is independent of the second line, which contains all nontrivial information about the moments.

In order to learn how one might correct this, return to the expression for the PDA's moments:
\begin{equation}
\varphi_\pi(x) = \int^{\Lambda^\prime} \! \frac{d^2 k_\perp}{(2\pi)^2} \, \varphi_\pi(x,k_\perp^2) \,;
\end{equation}
namely,
\begin{eqnarray}
\nonumber
\langle x^m \rangle &\sim & \int_{-1}^1 dz\,\rho_\nu(z)
\nu (\nu+1)\, \int_0^1 dx_1 dx_2\,x_1 (1-x_1)^{\nu-1}\\
&& \times
Z_2 {\rm tr}_{\rm CD} \int^\Lambda \! \frac{d^4k}{(2\pi)^4} \,
\left(\frac{n\cdot k}{n\cdot P}\right)^m
\frac{M^{2\nu}}
{[(k-f(x_1,x_2,z)P)^2+M^2]^{2+\nu}} \label{momentsO}\\
\nonumber
&=& Z_2 {\rm tr}_{\rm CD} \int^\Lambda \! \frac{d^4k}{(2\pi)^4} \,
\frac{M^{2\nu}}
{[k^2+M^2]^{2+\nu}}\\
&& \times \int_{-1}^1 dz\,\rho_\nu(z)\,
\nu (\nu+1)\, \int_0^1 dx_1 dx_2\,x_1 (1-x_1)^{\nu-1} [f(x_1,x_2,z)]^m.
\end{eqnarray}
Now compare this with a slightly modified expression:
\begin{align}
\nonumber
\langle x^m \rangle_M &\sim \int_{-1}^1 dz\,\rho_\nu(z)(1-z^2)^\nu
\nu (\nu+1)\, \int_0^1 dx_1 dx_2\,x_1 (1-x_1)^{\nu-1}\\
& \times
Z_2 {\rm tr}_{\rm CD} \int^\Lambda \! \frac{d^4k}{(2\pi)^4} \,
\left(\frac{n\cdot k}{n\cdot P}\right)^m
\frac{M^{2\nu}}
{[(k-f(x_1,x_2,z)P)^2+ (1-z^2) M^2]^{2+\nu}}
\label{momentMod}\\
\nonumber
&=
Z_2 {\rm tr}_{\rm CD} \int^\Lambda \! \frac{d^4k}{(2\pi)^4} \,
\frac{M^{2\nu}} {[k^2+(1-z^2) M^2]^{2+\nu}}
\\
& \times
\int_{-1}^1 dz\,\rho_\nu(z) (1-z^2)^\nu\,
\nu (\nu+1)\, \int_0^1 dx_1 dx_2\,x_1 (1-x_1)^{\nu-1} [f(x_1,x_2,z)]^m
\!.
\end{align}
A simple change of variables: $k^2 \to \tilde k^2 = (1-z^2) k^2$, converts this expression into
\begin{eqnarray}
\nonumber
\langle x^m \rangle_M &\sim &
\int_{-1}^1 dz\,\rho_\nu(z) (1-z^2)^\nu\,
\nu (\nu+1)\, \int_0^1 dx_1 dx_2\,x_1 (1-x_1)^{\nu-1} [f(x_1,x_2,z)]^m\\
&& \times Z_2 {\rm tr}_{\rm CD} \int^\Lambda \! \frac{d^4k}{(2\pi)^4} (1-z^2)^2\,
\frac{M^{2\nu}}
{[k^2 (1-z^2) +(1-z^2) M^2]^{2+\nu}}\\
\nonumber &=& \times
Z_2 {\rm tr}_{\rm CD} \int^\Lambda \! \frac{d^4k}{(2\pi)^4}
\frac{M^{2\nu}}{[k^2  + M^2]^{2+\nu}}\\
&& \int_{-1}^1dz\, \rho_\nu(z) \,
\nu (\nu+1)\, \int_0^1 dx_1 dx_2\,x_1 (1-x_1)^{\nu-1} [f(x_1,x_2,z)]^m
\\
&=& \langle x^m \rangle\,.
\end{eqnarray}

Plainly, so far as a computation of the PDA's moments is concerned, Eq.\,\eqref{momentMod} is actually the same as Eq.\,\eqref{momentsO}.  However, a difference appears if one reverts to the light-front wave function in Eq.\,\eqref{varphiukp} and begins with Eq.\,\eqref{momentMod}, \emph{viz}.
{\allowdisplaybreaks
\begin{align}
\nonumber
\langle x^m(k_\perp^2) \rangle &\sim  \int_{-1}^1dz\, \rho_\nu(z)(1-z^2)^\nu
\nu (\nu+1)\, \int_0^1 dx_1 dx_2\,x_1 (1-x_1)^{\nu-1}\\
& \times
Z_2 {\rm tr}_{\rm CD} \int^\Lambda \!
\frac{d^2k_\|}{(2\pi)^2} \,
\left(\frac{n\cdot k}{n\cdot P}\right)^m
\frac{M^{2\nu}}
{[(k_\|-f(x_1,x_2,z)P)^2 + k_\perp^2 + (1-z^2) M^2]^{2+\nu}}
\label{momentMod2} \\
&= \nonumber
\int_{-1}^1 dz\,\rho_\nu(z)(1-z^2)^\nu
\nu (\nu+1)\, \int_0^1 dx_1 dx_2\,x_1 (1-x_1)^{\nu-1}\\
& \times
Z_2 {\rm tr}_{\rm CD} \int^\Lambda \! \frac{d^2k_\| }{(2\pi)^2} \,
\left(\frac{n\cdot (k_\|+f(x_1,x_2,z)P)}{n\cdot P}\right)^m
\frac{M^{2\nu}}
{[k_\|^2 + k_\perp^2 + (1-z^2) M^2]^{2+\nu}}\\
&= \nonumber
\int_{-1}^1dz\, \rho_\nu(z)(1-z^2)^\nu
\nu (\nu+1)\, \int_0^1 dx_1 dx_2\,x_1 (1-x_1)^{\nu-1}\\
& \times
Z_2 {\rm tr}_{\rm CD} \int^\Lambda \! \frac{d^2k_\|}{(2\pi)^2} \,
[f(x_1,x_2,z)]^m
\frac{M^{2\nu}}
{[k_\|^2 + k_\perp^2 + (1-z^2) M^2]^{2+\nu}}\\
&= \nonumber
\int_{-1}^1dz\, \rho_\nu(z)(1-z^2)^{-1}
\nu (\nu+1)\, \int_0^1 dx_1 dx_2\,x_1 (1-x_1)^{\nu-1}\\
&\times Z_2 {\rm tr}_{\rm CD} \int^\Lambda \! \frac{d^2k_\| }{(2\pi)^2} \,
[f(x_1,x_2,z)]^m
\frac{M^{2\nu}}
{[k_\|^2 + M^2+ k_\perp^2/(1-z^2)]^{2+\nu}}\,.
\end{align}}
\hspace*{-0.4\parindent}In this way one arrives at
\begin{align}
\nonumber
\langle x^m(k_\perp^2) \rangle &\sim   \int_{-1}^1 dz\,(1+1/[2\nu])\rho_{\nu-1}(z)\,
\nu (\nu+1)\, \int_0^1 dx_1 dx_2\,x_1 (1-x_1)^{\nu-1}\\
& \times Z_2 {\rm tr}_{\rm CD} \int^\Lambda \! \frac{d^2k_\| }{(2\pi)^2} \,
[f(x_1,x_2,z)]^m
\frac{M^{2\nu}}
{[k_\|^2 + M^2+ k_\perp^2/(1-z^2)]^{2+\nu}}\\
&=   \int_{-1}^1 dz\,(1+1/[2\nu])\rho_{\nu-1}(z)\,
\nu (\nu+1)\, \int_0^1 dx_1 dx_2\,x_1 (1-x_1)^{\nu-1}\\
& \times
\frac{\nu+1}{2\pi^2}
[f(x_1,x_2,z)]^m \frac{M^{2\nu}}
{[M^2+ k_\perp^2/(1-z^2)]^{1+\nu}}
\end{align}
and now, very clearly, $\langle x^m(k_\perp^2) \rangle$ has not a factorised: the $dz$ integral is modulated by the $k_\perp^2/(1-z^2)$ part of the denominator.  Consequently, $\varphi_\pi(x,k_\perp^2)$ will not be expressible as a product of the type in Eq.\,\eqref{productphi}.

In order to provide a practical illustration, let's work with a simple expression for the pion's Bethe-Salpeter wave function:
\begin{equation}
\chi_\pi(k;P) =  S(k) \Gamma(k;P) S(k-P)
=  i \gamma_5 \, {\cal E}_\pi(k,k-P) - \frac{1}{f_\pi}\gamma_5 \gamma\cdot P\, {\cal F}_\pi(k,k-P)\,.
\label{chimodel}
\end{equation}
I have omitted the ${\cal G}_\pi(k; P)$ and ${\cal H}_\pi(k;P)$ terms for the sake of simplicity.  That won't alter the gist of the following analysis.  In terms of $\chi_\pi$, the moments of the pion's PDA are:
\begin{eqnarray}
f_\pi \, (n\cdot P)^{m+1} \, \langle x^m \rangle &=&
{\rm tr}_{\rm CD}
Z_2 \! \int_{dk}^\Lambda \!\!
(n\cdot k)^m \,\gamma_5\gamma\cdot n\, \chi_\pi(k;P)\\
&=& \frac{1}{f_\pi} \, 4 N_c Z_2 \int_{dk}^\Lambda \!\!(n\cdot k)^m (n\cdot P) {\cal F}_\pi(k,k-P)\,,
\end{eqnarray}
which is a very simple expression, that further reduces to
\begin{eqnarray}
f_\pi^2 \, (n\cdot P)^{m} \, \langle x^m \rangle
&=& \, 4 N_c Z_2 \int_{dk}^\Lambda \!\!(n\cdot k)^m  {\cal F}_\pi(k,k-P)\,.
\end{eqnarray}

Focusing on the PDA's zeroth moment, one has
\begin{eqnarray}
\label{fpiF}
f_\pi^2
&=& \, 4 N_c Z_2 \int_{dk}^\Lambda \!\!  {\cal F}_\pi(k,k-P)\,.
\end{eqnarray}
One knows that in Landau-gauge $Z_2(\zeta,\Lambda)$ behaves as $[1/\ln\ln\Lambda]$ for $\Lambda_{\rm QCD}/\Lambda \simeq 0$ \cite{Maris:1997tm}.  This means that a nonzero value for $f_\pi$ is possible if, and only if,
\begin{equation}
{\cal F}_\pi(k,k-P) \stackrel{k^2 \gg \Lambda_{\rm QCD}^2}{\sim}
\left[\frac{1}{k^2}\right]^2 \frac{1}{ \ln k^2/\Lambda_{\rm QCD}^2}\,.
\end{equation}
It follows that if one wishes to make a model of ${\cal F}_\pi(k,k-P)$ for use in algebraic analysis, then a QCD-like theory will have
\begin{equation}
{\cal F}_\pi(k,k-P) = \int_{-1}^1dz\, \rho_1(z) \frac{M^2}{[(k-\varsigma P)^2 + M^2]^2}\,,
\end{equation}
where $\varsigma = (1-z)/2$.  Inserting this into Eq.\,\eqref{fpiF}, one finds
\begin{equation}
\label{fpiF2}
f_\pi^2
= \, 4 N_c Z_2 \int_{dk}^\Lambda \!\!  \frac{M^2}{[k^2+M^2]^2}
    \int_{-1}^1dz\, \rho_1(z)
= \, 4 N_c Z_2 \int_{dk}^\Lambda \!\!  \frac{M^2}{[k^2+M^2]^2}\,.
\end{equation}
With $Z_2\to 1$, this is readily recognised as the expression for $f_\pi^2$ that is obtained in models of the Nambu--Jona-Lasinio type \cite{Nambu:1961tp} when an $O(4)$-symmetric regularisation scheme is used.\footnote{As this is being written, I have learnt that Prof.\ Yoichiro Nambu died on 5\,July\,2015, at the age of 94.  Winner of the Nobel Prize in 2008 for discovering the mechanism of DCSB, he was a great physicist; but also a fine person.  I have learnt from people who knew him well that Prof.\ Nambu was always deeply interested in exotic phenomena and intent on gathering related information from a broad range of sources.  Such habits distinguish the best in this discipline.}
If one restores the logarithms, then it's a QCD-like expression.

Consider now the expression for all PDA moments:
\begin{eqnarray}
\langle x^m \rangle
&=& \frac{1}{f_\pi^2}\, \frac{1}{(n\cdot P)^m} \, 4 N_c Z_2 \int_{dk}^\Lambda \!\!  \int_{-1}^1dz\, \rho_1(z) \,
(n\cdot k)^m\frac{M^2}{[(k-\varsigma P)^2 + M^2]^2}\\
&\stackrel{k\to k+\varsigma P}{=}& \int_{-1}^1dz\, \rho_1(z) \, \varsigma^m =  \frac{6}{(m+2) (m+3)}\,.
\end{eqnarray}
It follows immediately that
\begin{equation}
\varphi(x) = \varphi^{\rm asy}(x) = 6 x (1-x)\,,
\end{equation}
\emph{i.e}.\ Eq.\,\eqref{chimodel} is sufficient to reproduce the result first explicated in Ref.\,\cite{Chang:2013pqS}.

It is now time to return to the light-front wave function, defined via its moments, as before; but corrected as described above.  Using
\begin{eqnarray}
\label{kperp4}
{\cal F}_\pi(k,k-P) = \int_{-1}^1dz\, \rho_1(z) \,
\frac{M^2}{[(k_\parallel-\varsigma P)^2 + k_\perp^2 + (1-z^2) M^2]^2}
\end{eqnarray}
then
\begin{align}
\langle x^m(k_\perp^2) \rangle &= \frac{4N_c}{f_\pi^2}\, 4 N_c  \int^\Lambda \!\! \frac{d^2k_\|}{(2\pi)^2}\,
\frac{(n\cdot k)^m}{(n\cdot P)^m}\,\int_{-1}^1dz\, \rho_1(z) \,
\frac{M^2}{[(k_\parallel-\varsigma P)^2 + k_\perp^2 + (1-z^2) M^2]^2} \label{NakanishiNew}\\
&= \frac{1}{f_\pi^2}\,4 N_c   \int^\Lambda \!\! \frac{d^2k_\|}{(2\pi)^2}\, \int_{-1}^1dz\, \rho_1(z) \,\varsigma^m\,
\frac{M^2}{[k_\|^2 + k_\perp^2 + (1-z^2) M^2]^2}\\
&= \frac{1}{f_\pi^2} \frac{N_c}{2\pi^2} \int_{-1}^1dz\, \rho_1(z) \,\varsigma^m\,\frac{M^2}{k_\perp^2 + (1-z^2) M^2}\,.
\end{align}

One may now analyse $\langle x^m(k_\perp^2) \rangle$ via a twist expansion, \emph{viz}.\ an expansion in $1/k_\perp^2$:
\begin{eqnarray}
\langle x^m(k_\perp^2) \rangle &=&
\frac{1}{f_\pi^2} \frac{N_c}{2\pi^2} \int_{-1}^1dz\, \rho_1(z) \,\varsigma^m\,
\left[\frac{M^2}{k_\perp^2} - (1-z^2) \frac{M^4}{k_\perp^4} +\ldots \right]\,.
\end{eqnarray}
Focusing first on the leading-twist coefficient -- the factor accompanying $1/k_\perp^2$, one has
\begin{eqnarray}
\langle x^m(k_\perp^2) \rangle^{(2)} &=& \frac{1}{f_\pi^2} \frac{N_c}{2\pi^2} \int_{-1}^1dz\, \rho_1(z)\,\varsigma^m\, M^2\\
&=& \frac{M^2}{f_\pi^2} \frac{N_c}{2\pi^2} \frac{6}{(m+2) (m+3)}\,,
\end{eqnarray}
from which it is plain that
\begin{equation}
\varphi^{(2)}(x) = \frac{M^2}{f_\pi^2} \frac{N_c}{2\pi^2} \varphi^{\rm asy}(x)\,.
\end{equation}

The subleading twist coefficient -- the factor accompanying $1/k_\perp^4$ -- is
\begin{eqnarray}
\langle x^m(k_\perp^2) \rangle^{(4)} &=& - \frac{M^4}{f_\pi^2} \frac{N_c}{2\pi^2} \int_{-1}^1dz\, \rho_1(z)\,\varsigma^m\, (1-z^2) \\
&=& \frac{M^4}{f_\pi^2} \frac{N_c}{2\pi^2}\,\frac{48}{(m+3) (m+4)(m+5)}\,,
\end{eqnarray}
and hence
\begin{equation}
\varphi^{(4)}(x) = \frac{M^4}{f_\pi^2} \frac{N_c}{3\pi^2} [\varphi^{\rm asy}(x)]^2\,.
\end{equation}
One may readily establish that this pattern continues.

At the other extreme, in the neighbourhood $k_\perp^2\simeq 0$, one finds:
\begin{equation}
\langle x^m(k_\perp^2=0)\rangle = \frac{1}{f_\pi^2} \frac{N_c}{2\pi^2} \int_{-1}^1dz\, \rho_1(z)\,\varsigma^m\, \frac{1}{1-z^2}\\
= \frac{1}{f_\pi^2} \frac{N_c}{2\pi^2} \frac{3}{2}\frac{1}{1+m}\,;
\end{equation}
namely,
\begin{equation}
\varphi_\pi(x,k_\perp^2=0) = \frac{1}{f_\pi^2} \frac{3 N_c}{4\pi^2} H(1-x)\,,
\end{equation}
where $H(1-x)$ is the Heaviside step-function.  This is a natural result, at least qualitatively: it means that on $k_\perp^2\simeq 0$ the valence-quarks are free to roam with (almost) no bound: each $x\in[0,1]$ has equal probability.  \emph{N.B}.\ In QCD-like theories, this translates into a very broad distribution, that is nonetheless zero at $x=0,1$.

\begin{figure}[t]
\centerline{\includegraphics[width=0.99\textwidth,clip]{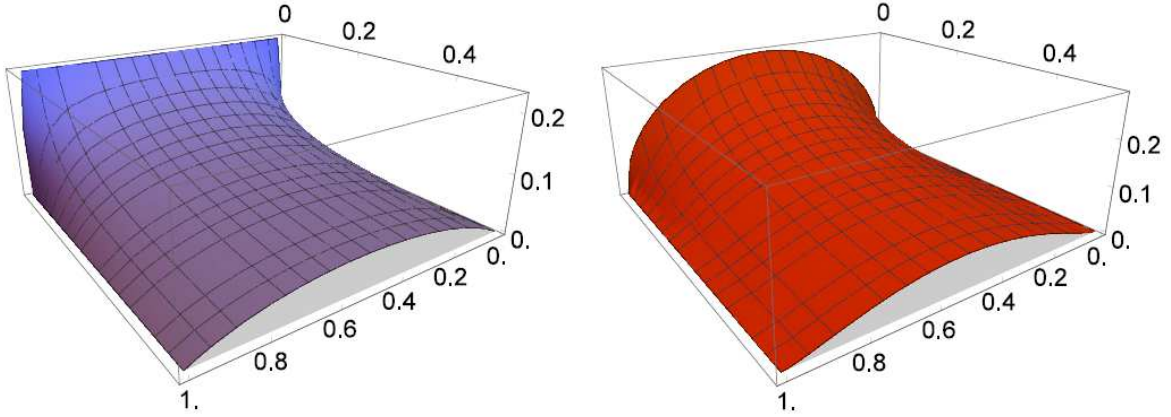}}
\caption{\label{figPsiR}
\emph{Left panel} -- Profile of light-front wave function in Eq.\,\eqref{psimodel}, $f_\pi^2 \varphi_\pi(x,k_\perp^2)$, computed with $M=0.4\,$GeV.  This expression provides an ``asymptotic''  pion wave function, in a sense similar to that in which  Eqs.\,\eqref{AlgebraicSetc} produce an asymptotic PDA.  \emph{Right panel} -- Profile of the model wave function described in Ref.\,\cite{Mezrag:2014jkaS}: the additional curvature is typical of wave functions relevant for the description of phenomena explored at a typical hadronic scale, \emph{e.g}.\ 1\,GeV. }
\end{figure}

With this information in hand, one is able to reconstruct the expression for the pion light-front wave function produced by Eqs.\,\eqref{momentMod2}, \eqref{chimodel}, \eqref{kperp4}:
\begin{equation}
\label{psimodel}
f_\pi^2 \varphi_\pi(x,k_\perp^2) = \frac{9}{4\pi^2} \frac{1}{\displaystyle 1+ \frac{k_\perp^2}{4 M^2 x(1-x)}}\,,
\end{equation}
which is depicted in Fig.\,\ref{figPsiR}.  This expression exposes the natural form for the correlated dependence of the two-body wave function on $(x,k_\perp^2)$, \emph{i.e}.\ it depends on the two-body light-front kinetic energy: $k_\perp^2/[x(1-x)]$.

\setcounter{equation}{0}
\section{Baryons -- as they really are}
\label{secBaryons}
\subsection{Grand Unification}
There are many novel challenges in the Standard Model to which one might now turn.  For example, a QCD-connected prediction of the spectrum of hybrid and exotic mesons would be very valuable given that a flagship effort at JLab\,12 is the search for these states.  Equally, or perhaps more pressing is the need to address the three valence-quark bound-state problem in QCD with the same level of sophistication that is now available for mesons, with the goal being to correlate the properties of meson and baryon ground- and excited-states within a single, symmetry-preserving framework.  Here, symmetry-preserving means that the analysis respects Poincar\'e covariance and satisfies the relevant WGTIs.  Constituent-quark models have hitherto been the most widely applied spectroscopic tools; and whilst their weaknesses are emphasised by critics and acknowledged by proponents, they are of continuing value because there is nothing better that is yet providing a bigger picture.  Nevertheless, they possess no connection with quantum field theory and therefore no connection with QCD; and they are not symmetry-preserving and therefore cannot veraciously connect meson and baryon properties.

A truly comprehensive approach to QCD will provide a unified explanation of both mesons and baryons.  I have explained that DCSB is a keystone of the Standard Model, which is evident in the momentum-dependence of the dressed-quark mass function -- Fig.\,\ref{gluoncloud}: it is just as important to baryons as it is to mesons.  The DSEs furnish the only extant framework that can simultaneously and transparently connect both meson and baryon observables with this basic feature of QCD, having provided, \emph{e.g}.\ a direct correlation of meson and baryon properties via a single interaction kernel, which preserves QCD's one-loop renormalisation group behaviour and can systematically be improved.  This is evident in Refs.\,\cite{Eichmann:2008ae, Eichmann:2008ef, Eichmann:2009qa, Nicmorus:2010sd, Eichmann:2011vu, Eichmann:2011pv, Eichmann:2011ej, Chang:2012cc}.

\subsection{Borromean Analogy}
Let us focus initially on the proton, which is the core of the hydrogen atom, lies at the heart of every nucleus, and has never been observed to decay.  It is nevertheless a composite object, whose properties and interactions are determined by its valence-quark content: $u$ + $u$ + $d$, \emph{i.e}.\ two up ($u$) quarks and one down ($d$) quark.  So far as is now known \cite{Agashe:2014kda}, bound-states seeded by two valence-quarks do not exist; and the only two-body composites are those associated with a valence-quark and -antiquark, \emph{i.e}.\ mesons.  These features are supposed to derive from colour confinement, whose complexities are discussed in Sec.\,\ref{secConfinement}.

Such observations lead one to a position from which the proton may be viewed as a Borromean bound-state \cite{Segovia:2015ufa}, \emph{viz}.\ a system constituted from three bodies, no two of which can combine to produce an independent, asymptotic two-body bound-state.  In QCD the complete picture of the proton is more complicated, owing, in large part, to the loss of particle number conservation in quantum field theory and the concomitant frame- and scale-dependence of any Fock space expansion of the proton's wave function \cite{Dirac:1949cp, Keister:1991sb, Coester:1992cg, Brodsky:1997de}, which is described briefly in Sec.\,\ref{secHP}.  Notwithstanding that, the Borromean analogy provides an instructive perspective from which to consider both quantum mechanical models and continuum treatments of the nucleon bound-state problem in QCD.  It poses a crucial question: \\[0.7ex]
\hspace*{2em}\parbox[t]{0.9\textwidth}{Whence binding between the valence quarks in the proton, \emph{i.e}.\ what holds the proton together?}\\

In numerical simulations of lQCD that use static sources to represent the proton's valence-quarks, a ``Y-junction'' flux-tube picture of nucleon structure is produced, \emph{e.g}.\ Ref.\,\cite{Bissey:2006bz, Bissey:2009gw}.  This might be viewed as originating in the three-gluon vertex, which signals the non-Abelian character of QCD and is the source of asymptotic freedom \cite{Politzer:2005kc, Gross:2005kv, Wilczek:2005az} as discussed in connection with Fig.\,\ref{F3}.  Such results and notions would suggest a key role for the three-gluon vertex in nucleon structure \emph{if} they were equally valid in real-world QCD wherein light dynamical quarks are ubiquitous.  However, as we saw in Sec.\,\ref{secConfinement}, they are not; and so a different explanation of binding within the nucleon must be found.

DCSB has numerous corollaries that are crucial in determining the observable features of the Standard Model, some of which are detailed above.  Another particularly important consequence is less well known.  Namely, any interaction capable of creating pseudo-Goldstone modes as bound-states of a light dressed-quark and -antiquark, and reproducing the measured value of their leptonic decay constants, will necessarily also generate strong colour-antitriplet correlations between any two dressed quarks contained within a nucleon.  Although a rigorous proof within QCD cannot be claimed, this assertion is based upon an accumulated body of evidence, gathered in two decades of studying two- and three-body bound-state problems in hadron physics, \emph{e.g}.\ Refs.\,\cite{Cahill:1987qr, Cahill:1988dx, Bender:1996bb, Oettel:1998bk, Bloch:1999vk, Maris:2002yu, Bender:2002as, Bhagwat:2004hn, Cloet:2008re, Eichmann:2008ef, Eichmann:2009qa, Roberts:2011cf, Eichmann:2011aa, Segovia:2014aza, Segovia:2015hraS}.  No realistic counter examples are known; and the existence of such diquark correlations is also supported by simulations of lQCD \cite{Alexandrou:2006cq, Babich:2007ahS}.

The properties of diquark correlations have been charted.  Most importantly, diquarks are confined.  However, this is not true if the RL truncation is used to define the associated scattering problem \cite{Maris:2002yu}.  Corrections to that simplest symmetry-preserving approximation are critical in quark-quark channels: they eliminate bound-state poles from the quark-quark scattering matrix but preserve the strong correlations \cite{Bender:1996bb, Bender:2002as, Bhagwat:2004hn}.

Additionally, owing to properties of charge-conjugation, a diquark with spin-parity $J^P$ may be viewed as a partner to the analogous $J^{-P}$ meson \cite{Cahill:1987qr}.  It follows that scalar, isospin-zero and pseudovector, isospin-one diquark correlations are the strongest; and whilst no pole-mass exists, the following mass-scales, which express the strength and range of the correlation and are each bounded below by the partnered meson's mass, may be associated with these diquarks \cite{Cahill:1987qr, Maris:2002yu, Alexandrou:2006cq, Babich:2007ahS}:
\begin{equation}
m_{[ud]_{0^+}} \approx 0.7-0.8\,{\rm GeV}\,,\quad
m_{\{uu\}_{1^+}}  \approx 0.9-1.1\,{\rm GeV}  \,,
\end{equation}
with $m_{\{dd\}_{1^+}}=m_{\{ud\}_{1^+}} = m_{\{uu\}_{1^+}}$ in the isospin symmetric limit.
Realistic diquark correlations are also soft.  They possess an electromagnetic size that is bounded below by that of the analogous mesonic system, \emph{viz}.\ \cite{Maris:2004bp, Roberts:2011wyS}:
\begin{equation}
\label{qqradii}
r_{[ud]_{0^+}} \gtrsim r_\pi\,, \quad
r_{\{uu\}_{1^+}} \gtrsim r_\rho\,,
\end{equation}
with $r_{\{uu\}_{1^+}} > r_{[ud]_{0^+}}$.  As with mesons, these scales are set by that associated with DCSB.

It is useful to remark that diquarks are colour-singlets in a dynamical theory based on SU$(2)$-colour.  They would thus exist as asymptotic states and form mass-degenerate multiplets with mesons composed from like-flavoured quarks.  (These properties are a manifestation of Pauli-G\"ursey symmetry \cite{Pauli:1957, Gursey:1958}.)  Consequently, the $[ud]_{0^+}$ diquark would be massless in the presence of DCSB, matching the pion, and the $\{ud\}_{1^+}$ diquark would be degenerate with the theory's $\rho$-meson.  Such identities are lost in changing the gauge group to SU$(3)$-colour; but clear and instructive similarities between mesons and diquarks nevertheless remain.

\subsection{Diquarks in the nucleon}
It should now be apparent that the bulk of QCD's particular features and nonperturbative phenomena can be traced to the evolution of the strong running coupling.  Its unique characteristics are primarily determined by the three-gluon vertex: the four-gluon vertex does not contribute dynamically at leading order in perturbative analyses of matrix elements; and nonperturbative continuum analyses of QCD's gauge sector indicate that satisfactory agreement with gluon propagator results from lQCD simulations is typically obtained without reference to dynamical contributions from the four-gluon vertex, \emph{e.g}.\ Refs.\,\cite{Aguilar:2008xm, Aguilar:2009ke, Aguilar:2009nf, Maas:2011se, Boucaud:2011ugS, Pennington:2011xs, Binosi:2012sj, Strauss:2012dg, Blum:2014gna, Meyers:2014iwa}.  The three-gluon vertex is therefore the dominant factor in producing the class of RGI running interactions that have provided both successful descriptions and predictions of many hadron observables \cite{Maris:2003vk, Chang:2011vu, Bashir:2012fs, Cloet:2013jya, Binosi:2014aea}.  It is this class of interactions that generates the strong attraction between two quarks which produces tight diquark correlations in analyses of the three valence-quark scattering problem.

\begin{figure}[t]
\begin{minipage}[t]{\textwidth}
\begin{minipage}{0.49\textwidth}
\centerline{\includegraphics[clip,width=0.9\textwidth]{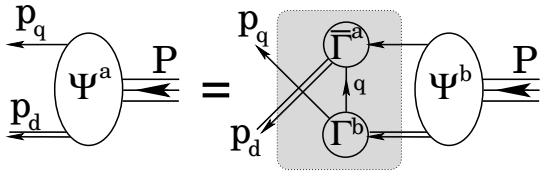}}
\end{minipage}
\begin{minipage}{0.49\textwidth}
\caption{\small
\label{fig:Faddeev} Poincar\'e covariant Faddeev equation.  $\Psi$ is the Faddeev amplitude for a baryon of total momentum $P= p_q + p_d$, where $p_{q,d}$ are, respectively, the momenta of the quark and diquark within the bound-state.  The shaded area demarcates the Faddeev equation kernel: \emph{single line},
dressed-quark propagator; $\Gamma$,  diquark correlation amplitude; and \emph{double line}, diquark propagator.}
\end{minipage}
\end{minipage}
\end{figure}

The existence of tight diquark correlations considerably simplifies analyses of the three valence-quark scattering problem and hence baryon bound states because it reduces that task to solving a Poincar\'e covariant Faddeev equation \cite{Cahill:1988dx}, depicted in Fig.\,\ref{fig:Faddeev}.  The three gluon vertex is not explicitly part of the bound-state kernel in this picture of the nucleon.  Instead, one capitalises on the fact that phase-space factors materially enhance two-body interactions over $n\geq 3$-body interactions and exploits the dominant role played by diquark correlations in the two-body subsystems.  Then, whilst an explicit three-body term might affect fine details of baryon structure, the dominant effect of non-Abelian multi-gluon vertices is expressed in the formation of diquark correlations.  Such a nucleon is then a compound system whose properties and interactions are primarily determined by the quark$+$diquark structure evident in Fig.\,\ref{fig:Faddeev}.

It is important to highlight that both scalar-isoscalar and pseudovector-isotriplet diquark correlations feature within a nucleon. Any study that neglects pseudovector diquarks is unrealistic because no self-consistent solution of the Faddeev equation in Fig.\,\ref{fig:Faddeev} can produce a nucleon constructed solely from a scalar diquark, \emph{e.g}.\ pseudovector diquarks typically provide roughly 150\,MeV of attraction \cite{Roberts:2011cf}.  The relative probability of scalar versus pseudovector diquarks in a nucleon is a dynamical statement.  Realistic computations predict a scalar diquark strength of approximately 60\% \cite{Cloet:2008re, Segovia:2014aza, Segovia:2015hraS}.  This prediction can be tested by contemporary experiments \cite{Segovia:2015ufa}.

\begin{figure}[t]
\begin{minipage}[t]{\textwidth}
\begin{minipage}{0.49\textwidth}
\centerline{\includegraphics[width=0.9\textwidth]{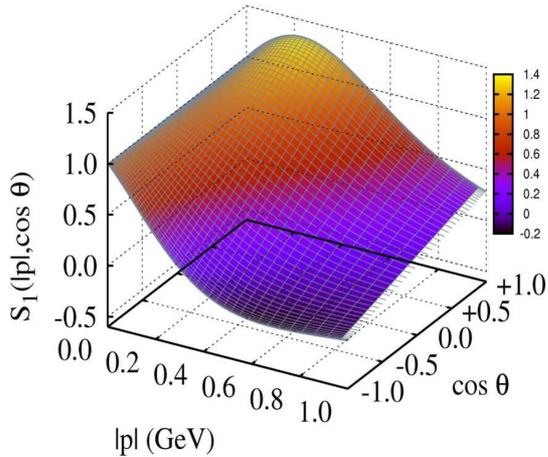}}
\end{minipage}
\begin{minipage}{0.49\textwidth}
\caption{\small
\label{figS1} Representation of the dominant piece in the nucleon's eight-component Poincar\'e-covariant Faddeev amplitude: $s_1(|p|,\cos\theta)$.  In the nucleon rest frame, this term describes that piece of the quark-diquark relative momentum correlation which possesses zero \emph{intrinsic} quark-diquark orbital angular momentum, \emph{i.e}.\ $L=0$ before the propagator lines are reattached to form the Faddeev wave function.  Referring to Fig.\,\ref{fig:Faddeev}, $p= P/3-p_q$ and $\cos\theta = p\cdot P/\sqrt{p^2 P^2}$.  (The amplitude is normalised such that its $U_0$ Chebyshev moment is unity at $|p|=0$.)}
\end{minipage}
\end{minipage}
\end{figure}

The quark$+$diquark structure of the nucleon is elucidated in Fig.\,\ref{figS1}, which provides a representation of the leading component of the nucleon's Faddeev amplitude: with the notation of Ref.\,\cite{Segovia:2014aza}, $s_1(|p|,\cos\theta)$, computed using the Faddeev kernel described therein.  This function describes a piece of the quark$+$scalar-diquark relative momentum correlation. Notably, in this solution of a realistic Faddeev equation there is strong variation with respect to both arguments.  Support is concentrated in the forward direction, $\cos\theta >0$, so that alignment of $p$ and $P$ is favoured; and the amplitude peaks at $(|p|\simeq M_N/6,\cos\theta=1)$, whereat $p_q \approx P/2 \approx p_d$ and hence the \emph{natural} relative momentum is zero.  In the antiparallel direction, $\cos\theta<0$, support is concentrated at $|p|=0$, \emph{i.e}.\ $p_q \approx P/3$, $p_d \approx 2P/3$.  A realistic nucleon amplitude is evidently a complicated function; and significant structure is lost if simple interactions and/or truncations are employed in building the Faddeev kernel, \emph{e.g}.\ extant treatments of a momentum-independent quark-quark interaction -- a contact interaction -- produce a Faddeev amplitude that is also momentum independent \cite{Wilson:2011aa, Cloet:2014rja}, a result exposed by Fig.\,\ref{figS1} as unrealistic for any probe sensitive to the nucleon interior.

A nucleon (and kindred baryons) described by Fig.\,\ref{fig:Faddeev} is a Borromean bound-state, the binding within which has two contributions.  One part is expressed in the formation of tight diquark correlations; but that is augmented by attraction generated by the quark exchange depicted in the shaded area of Fig.\,\ref{fig:Faddeev}.  This exchange ensures that diquark correlations within the nucleon are fully dynamical: no quark holds a special place because each one participates in all diquarks to the fullest extent allowed by its quantum numbers. The continual rearrangement of the quarks guarantees, \emph{inter} \emph{alia}, that the nucleon's dressed-quark wave function complies with Pauli statistics.

It is impossible to overstate the importance of appreciating that these fully dynamical diquark correlations are vastly different from the static, pointlike ``diquarks'' which featured in early attempts \cite{Lichtenberg:1967zz, Lichtenberg:1968zz} to understand the baryon spectrum and to explain the so-called missing resonance problem \cite{Ripani:2002ss, Burkert:2012ee, Kamano:2013iva}.  Modern diquarks are soft, Eq.\,\eqref{qqradii}; and enforce certain distinct interaction patterns for the singly- and doubly-represented valence-quarks within the proton.  On the other hand, the number of states in the spectrum of baryons obtained from the Faddeev equation in Fig.\,\ref{fig:Faddeev} \cite{Chen:2012qrS} is similar to that found in the three-constituent quark model, just as it is in today's lQCD calculations of this spectrum \cite{Edwards:2011jj}.

\subsection{Contact interaction}
It is only very recently that numerical algorithms have been developed that both enable one to use sophisticated DSE kernels in order to gain direct access to the large-$Q^2$ behaviour of hadron form factors \cite{Chang:2013pqS, Chang:2013epa, Chang:2013nia} and promise to pave the way for their use in treating the phenomena of deep inelastic scattering.  Absent those algorithms, a confining, symmetry-preserving DSE treatment of a vector$\,\otimes\,$vec\-tor contact interaction has proved useful in a variety of contexts \cite{GutierrezGuerrero:2010md, Roberts:2010rn, Roberts:2011wyS, Roberts:2011cf, Wilson:2011aa, Chen:2012qr, Chen:2012txaS, Pitschmann:2012byS, Wang:2013wk, Segovia:2013rca, Segovia:2013ugaS, Pitschmann:2014jxa}.  This collection of work is dwarfed, however, by the widespread application of models of the  Nambu--Jona-Lasinio (NJL) type \cite{Nambu:1961tp, Vogl:1991qt, Klevansky:1992qe, Hatsuda:1994pi, Bijnens:1995ww}. It must be remarked that whilst there is a ground-level similarity, these two approaches are different.  NJL-model practitioners allow themselves the freedom of tuning the relative strength of individual four-fermion interaction terms in a model Lagrangian in order to fit a body of data with the aim of providing a phenomenology of hadron physics that can serve both to elucidate correlations between observables and as a beacon that sheds light on novel explanations for unexpected phenomena.  The DSE approach, on the other hand, treats the contact interaction as a representation of the gluon's two-point Schwinger function.  At the outset, therefore, the DSE approach fixes the number, type and relative strength of the four-fermion interaction terms, as illustrated elsewhere \cite{Cahill:1985mh, Tandy:1997qf}, and then proceeds to identify and highlight those observables that can distinguish between different choices for the momentum-dependence of the DSE kernels.  There is true predictive and discriminative power in such analyses and hence this is the class of applications that I will illustrate.

The earliest studies in this class \cite{GutierrezGuerrero:2010md, Roberts:2010rn, Roberts:2011wyS} showed that a confining, symmetry-preserving treatment of a vector$\,\otimes\,$vec\-tor contact interaction provides results for $\pi$- and $\rho$-meson observables that are not realistically distinguishable from those obtained with the best renormalisation-group-improved one-gluon exchange interactions so long as the momentum of the probe involved, $Q$, does not exceed the zero-momentum value of the dressed-quark mass, i.e.\ on $Q^2<M^2$.  Furthermore, whilst a contact interaction typically produces form factors that are too hard \cite{Wilson:2011aa, Chen:2012txaS, Segovia:2013rca, Segovia:2013ugaS}, interpreted judiciously, even these results can be used to draw valuable insights, \emph{e.g}.\ concerning the relationships between different hadrons.  Importantly, too, studies employing a symmetry-preserving regularisation of the contact interaction serve as a useful surrogate, enabling the exploration of domains which analyses using interactions that more closely resemble those of QCD are as yet unable to enter.  They are therefore useful at present in attempts to use data as a tool for charting the nature of the quark-quark interaction at long-range, \emph{i.e}.\ for identifying clear signals in observables for the running of couplings and masses in QCD that was elucidated in Secs.\,\ref{secDCSB}, \ref{secCannibals}, \ref{secAbInitio}.

Some of the more interesting and robust studies are contact-interaction analyses which unify meson and baryon spectra \cite{Roberts:2011cf, Chen:2012qrS}.  Given that I have already described the gap and Bethe-Salpeter equations, which are crucial in computing the meson spectrum, at this point it is worth illustrating a Faddeev equation; and the simplest example is that obtained for the $\Delta$-baryon using the contact-interaction \cite{Roberts:2011cf}:
\begin{align}
1 &= 8 \frac{g_\Delta^2}{M } \frac{E_{qq_{1^+}}^2}{m_{qq_{1^+}}^2}
\int\frac{d^4\ell^\prime}{(2\pi)^4} \int_0^1 d\alpha\,
\frac{(m_{qq_{1^+}}^2 + (1-\alpha)^2 m_\Delta^2)(\alpha m_\Delta + M)}
{[\ell^{^\prime 2} + \sigma_\Delta(\alpha,M,m_{qq_{1^+}},m_\Delta)]^2}\\
&= \frac{g_\Delta^2}{M}\frac{E_{qq_{1^+}}^2}{m_{qq_{1^+}}^2}\frac{1}{2\pi^2}
\int_0^1 d\alpha\, (m_{qq_{1^+}}^2 + (1-\alpha)^2 m_\Delta^2)(\alpha m_\Delta + M)\overline{\cal C}^{\rm iu}_1(\sigma_\Delta(\alpha,M,m_{qq_{1^+}},m_\Delta))\,, \label{FEDelta}
\end{align}
where $\overline{\cal C}^{\rm iu}_1(z) = {\cal C}^{\rm iu}_1(z)/z$, ${\cal C}^{\rm iu}_1(z) = - z (d/dz){\cal C}^{\rm iu}(z)$, ${\cal C}^{\rm iu}(z)/z = \Gamma(-1,z \tau_{\rm uv}^2) - \Gamma(-1,z \tau_{\rm ir}^2)$, with $\Gamma(\alpha,y)$ being the incomplete gamma-function, and
\begin{equation}
\sigma_\Delta(\alpha,M,m_{qq_{1^+}},m_\Delta)=(1-\alpha)\, M^2 + \alpha \, m_{qq_{1^+}} - \alpha (1-\alpha)\, m_\Delta^2.
\end{equation}

Equation\,\eqref{FEDelta} is an eigenvalue problem whose solution yields the mass for the dressed-quark-core of the $\Delta$-resonance.  The primary parametric inputs to this equation are the: dressed-quark mass, $M$; axial-vector diquark mass, $m_{qq_{1^+}}$, and correlation strength, $E_{qq_{1^+}}$; and the contact-interaction regularisation parameters, $\tau_{\rm ir}$, $\tau_{\rm uv}$.  The former are computable once the latter are specified.  A final parameter is $g_\Delta$, which appears as the result of a drastic simplification, \emph{viz}.\ in the Faddeev equation for a baryon of type $B=N,\Delta$, the quark exchanged between the diquarks in the shaded region of Fig.\,\ref{fig:Faddeev} is represented as
\begin{equation}
S^{\rm T}(k) \to \frac{g_B^2}{M}\,.
\label{staticexchange}
\end{equation}
This is a variant of the so-called ``static approximation,'' which itself was introduced in Ref.\,\cite{Buck:1992wz} and has subsequently been used in studies of a range of nucleon properties \cite{Bentz:2007zs}.  In combination with diquark correlations generated by the contact-interaction, whose Bethe-Salpeter amplitudes are momentum-independent, Eq.\,(\ref{staticexchange}) generates Faddeev equation kernels which themselves are momentum-independent.  The dramatic simplifications which this produces are the merit of Eq.\,(\ref{staticexchange}).  The analogous equation for the nucleon is a five-dimensional algebraic eigenvalue problem because the nucleon also contains scalar diquark correlations.

In order to proceed it is important to highlight just what is missing in all DSE kernels that are currently available.  Namely, even the best kernels are built only from dressed-gluons and -quarks.  They omit long-range interactions.  However, QCD produces a very potent long-range interaction, \emph{viz}.\ that associated with the pion and other light mesons, without which no nuclei would be bound and we wouldn't be here.  Fortunately, since the real character of contemporary DSE kernels is understood, one knows that such resonant ``meson-cloud'' effects can be added without over-counting.  Uncovering just how that can be accomplished in a symmetry-preserving manner is where questions still lie.  Only rudimentary attempts exist (\emph{e.g}.\ Refs.\,\cite{Hecht:2002ejS, Cloet:2008fw, Sanchis-Alepuz:2014wea}).  Notwithstanding that, meson-cloud contributions must be thoughtfully considered before any comparison may be made between contemporary computations and the real world.  In this connection it is now judged that whilst resonant contributions to truncated DSE kernels may have a material impact on the masses of the nucleon and $\Delta$-baryon separately, the modification of each is approximately the same, so that the mass difference, $\delta m = m_\Delta-m_N$, is largely unaffected by such corrections.  Indeed, this perspective is consistent with an analysis \cite{Young:2002cj} that considers the effect of pion loops, which are explicitly excluded in the rainbow-ladder truncation \cite{Eichmann:2008ae}: whilst the individual masses are reduced by roughly $300\,$MeV, the mass difference, $\delta m$, increases by only $50\,$MeV.  Improving upon that analysis using insights drawn from the extensive body of work completed by the excited baryon analysis center (EBAC), which used a realistic coupled-channels model to remove meson dressing from the $\Delta$ and expose a dressed-quark-core bare-mass of $1.39\,$GeV  \cite{Suzuki:2009njS}, it is appropriate to choose
\begin{equation}
\label{gNgDelta}
g_N= 1.18 \,,\; g_\Delta = 1.56\,\; \Rightarrow m_N=1.14\,\mbox{GeV}, m_\Delta = 1.39\,\mbox{GeV}, \delta m = 0.25\,\mbox{GeV}\,.
\end{equation}

\begin{figure}[t]
\begin{minipage}[t]{\textwidth}
\begin{minipage}{0.49\textwidth}
\includegraphics[clip,width=0.95\textwidth]{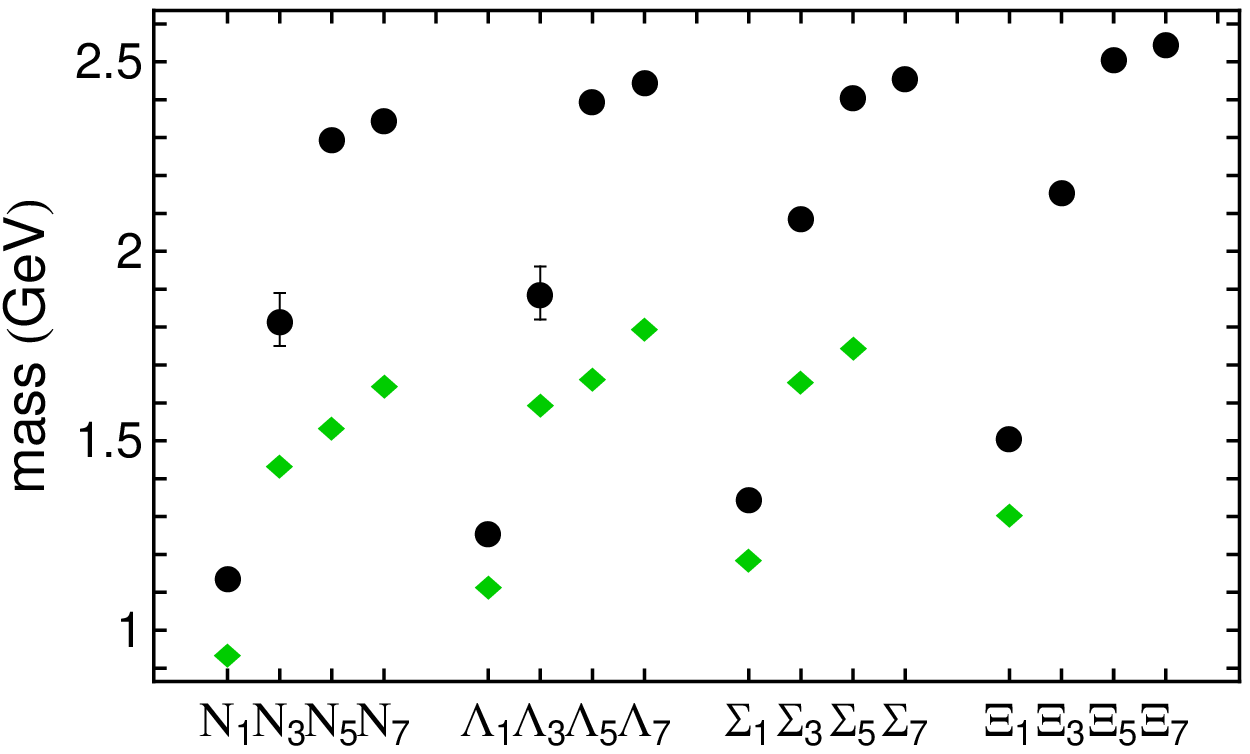}
\end{minipage}
\begin{minipage}{0.49\textwidth}
\includegraphics[clip,width=0.95\textwidth]{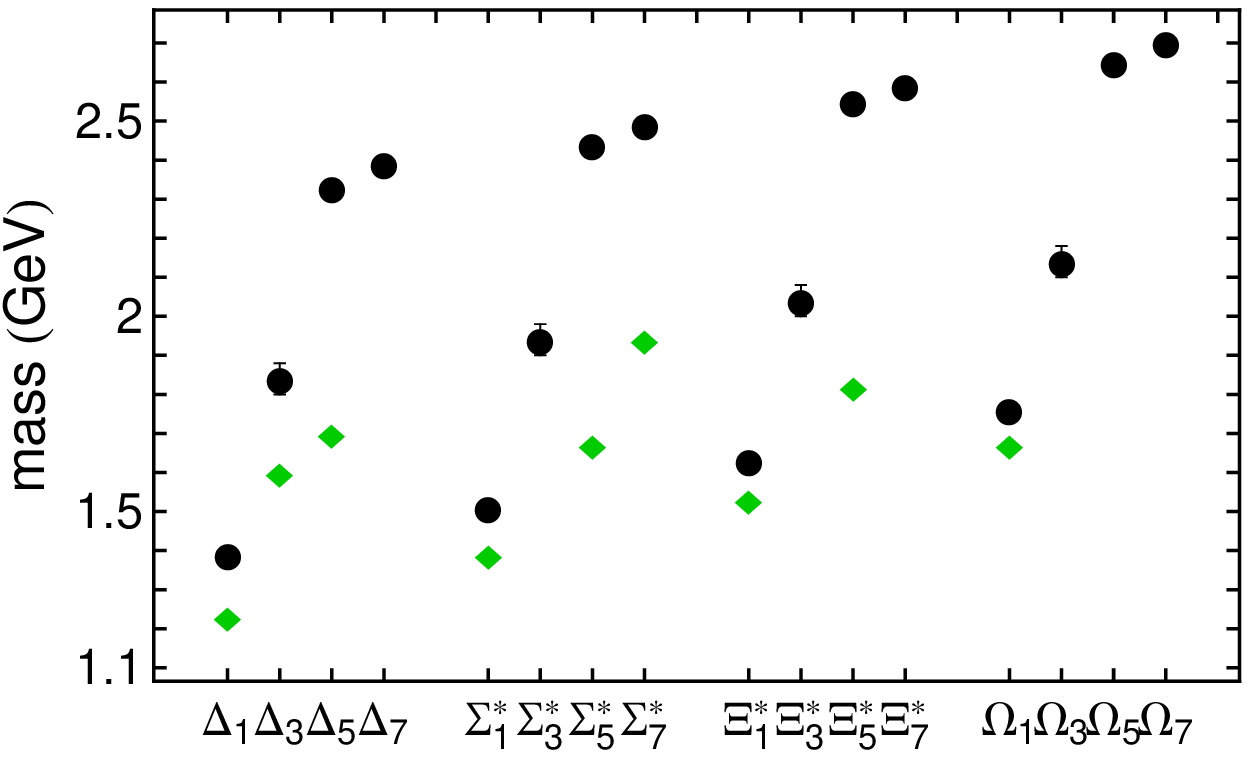}
\end{minipage}
\end{minipage}
\caption{\label{fig:OctetMasses} \small
\underline{Left panel}: Masses of members of the baryon octet: \emph{Circles} -- computed masses \cite{Chen:2012qrS}; and \emph{diamonds} -- empirical masses \cite{Agashe:2014kda}.  The horizontal axis lists a particle name with a subscript, and reading left-to-right the labels indicate ground-state, radial-excitation, ground-state's parity partner and radial excitation of that state.
\underline{Right panel}: Analogous plot for the decuplet masses.}
\end{figure}

This discussion introduces the basic elements in a contact-interaction computation of the spectrum of hadrons with strangeness \cite{Chen:2012qrS}, which is the most extensive exploration of the baryon spectrum ever accomplished using the DSEs and yields numerous insights into hadron structure.
For example, it predicts that the diquark content of baryons is largely independent of strangeness, \emph{viz}.\ that baryon structure is flavour-blind; and the computed level ordering matches that of experiment, as apparent in Fig.\,\ref{fig:OctetMasses}.  In particular, the parity-partner for each ground-state is always more massive than its first radial excitation, \emph{i.e}.\ the first $J^P=\frac{1}{2}^-$ state always lies above the second $J^P=\frac{1}{2}^+$ state.  Other approaches find this ordering difficult to achieve.  That, too, was explained in Ref.\,\cite{Chen:2012qrS}: a veracious expression of DCSB in the meson spectrum is critical to obtaining the correct level ordering in the baryon sector, so that an approach within which DCSB cannot be realised or a simulation whose parameters are such that the importance of DCSB is suppressed will both necessarily have difficulty reproducing the experimental ordering of levels.  In addition, the contact interaction study in Ref.\,\cite{Chen:2012qrS} predicts that the first radial excitation of ground-state baryons is constituted almost entirely from axial-vector diquark correlations.  A subsequent study \cite{Segovia:2015ufa}, employing a QCD-linked interaction, predicts a completely different result; namely, that radial excitations possess the same diquark content as the ground-states.  This is precisely the sort of thing one was looking for: signature empirical differences between predictions obtained with different interactions treated at the same level of approximation.

Looking closer at Fig.\,\ref{fig:OctetMasses} it is evident that the computed baryon masses lie uniformly above the empirical values.  This is a success because the computed masses are those of the baryons' dressed-quark-cores, whereas the empirical values include effects associated with meson-cloud effects, which typically produce sizable reductions \cite{Gasparyan:2003fp, Suzuki:2009njS}, as explained above.  The values in Fig.\,\ref{fig:OctetMasses} may reasonably be viewed as bare-mass inputs appropriate for dynamical coupled-channels analyses of the hadron spectrum \cite{Aznauryan:2011ub}.  This was explained and illustrated for the nucleon and $\Delta$-resonance in Sect.\,4.5 of Ref.\,\cite{Roberts:2011cf} and in particular for the Roper resonance in Ref.\,\cite{Wilson:2011aa}.  Here it is worth reiterating those instances in which a comparison can be made:
\begin{equation}
\begin{array}{l|llllll}
    & N_{940} P_{11} & N_{1440} P_{11} & N_{1535}S_{11} & N_{1650} S_{11}  & \Delta_{1232} P_{33} & \Delta_{1700} D_{33} \\\hline
{\rm Fig.}\;\ref{fig:OctetMasses}\; ({\rm DSE}) & 1.14 & 1.82 & 2.30 & 2.35 & 1.39 & 2.33
\\
M_{B}^0 \; (\mbox{Ref.\,\protect\cite{Suzuki:2009njS}})  & & 1.76 & 1.80 & 1.88 & 1.39 & 1.98\\
M_{B}^0 \; (\mbox{Ref.\,\protect\cite{Gasparyan:2003fp}}) & 1.24 &  & 2.05 & 1.92 & 1.46 & 2.25
\end{array}
\end{equation}
where $M_B^0$, when it appears, is the relevant bare mass inferred in the associated coupled-channels (DCC) analysis.  The DCC bare masses were uncertain and dependent on model details.  However, as Ref.\,\cite{Chen:2012qrS} made no attempt to fit them, their proximity to the DSE results suggests that it might now be possible to place these DCC bare masses on firmer ground, investing them with meaning within the context of hadron structure calculations that have a traceable connection with QCD.  This is the subject of an ongoing investigation.

\begin{figure}[t]
\begin{center}
\begin{tabular}{cc}
\includegraphics[clip,width=0.47\linewidth]{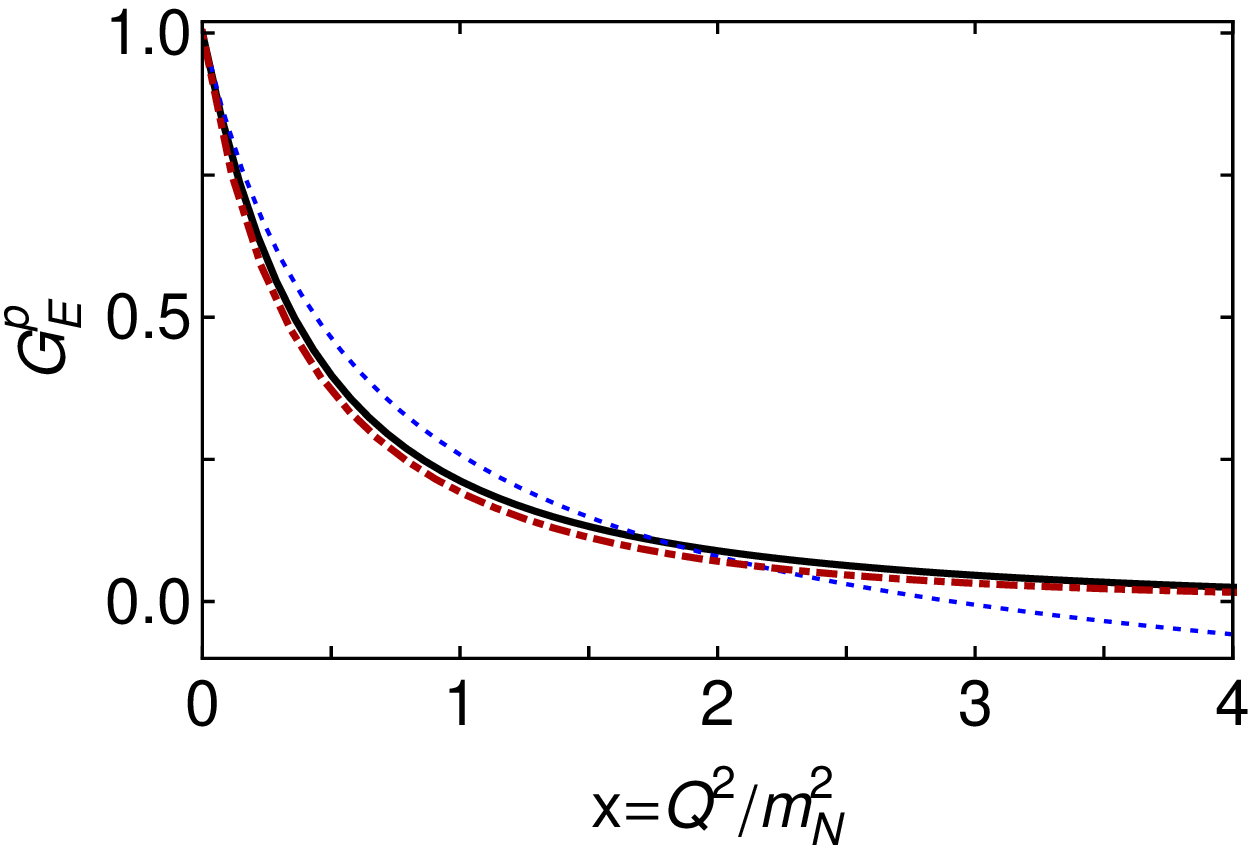}\vspace*
{-1ex } &
\includegraphics[clip,width=0.47\linewidth]{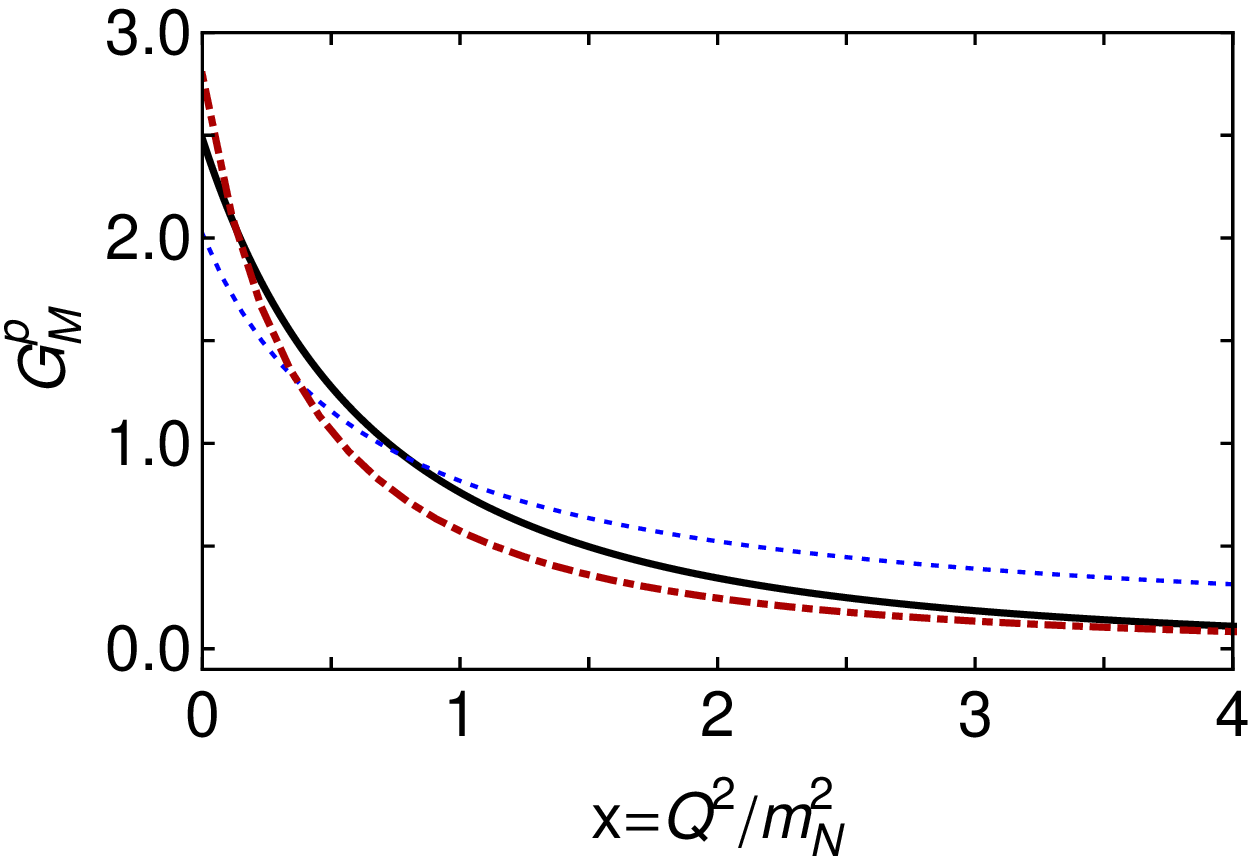}\vspace*
{-1ex}
\end{tabular}
\begin{tabular}{cc}
\includegraphics[clip,width=0.47\linewidth]{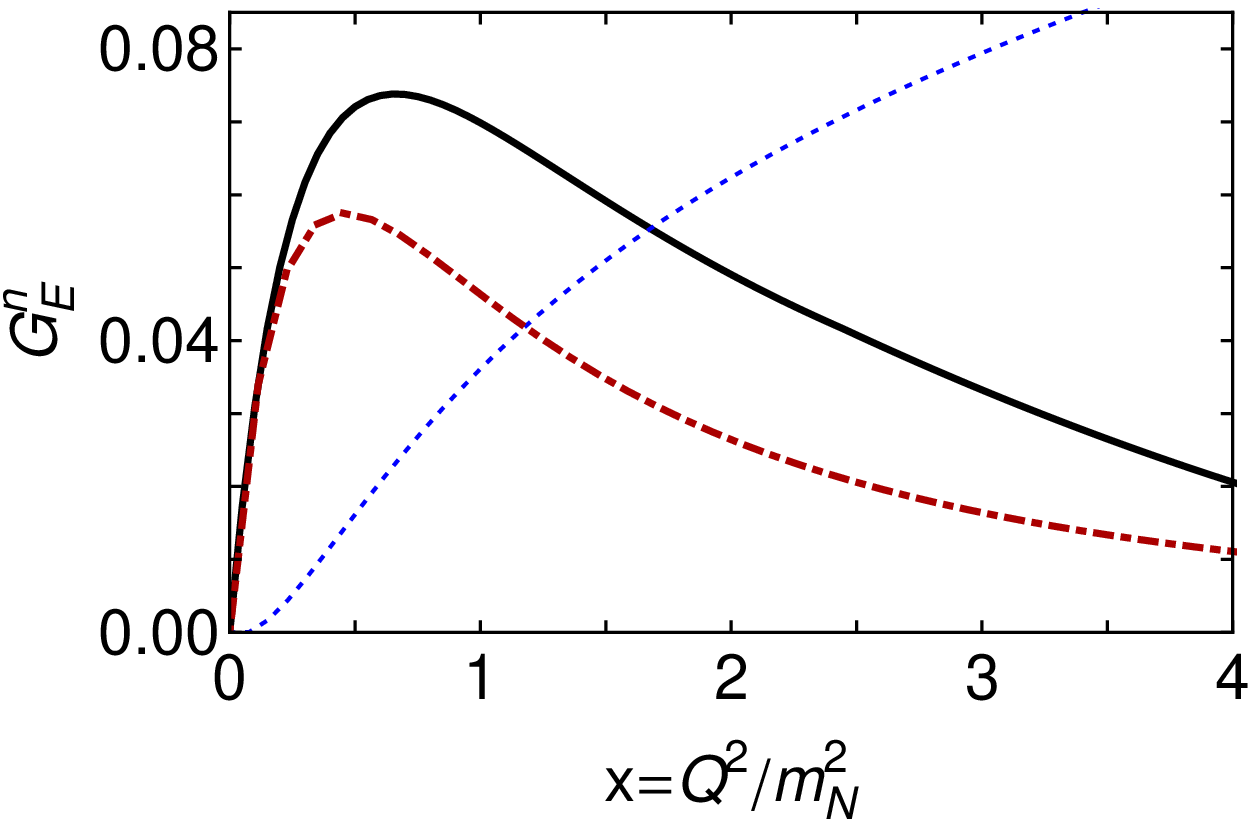}\vspace*
{-1ex } &
\includegraphics[clip,width=0.47\linewidth]{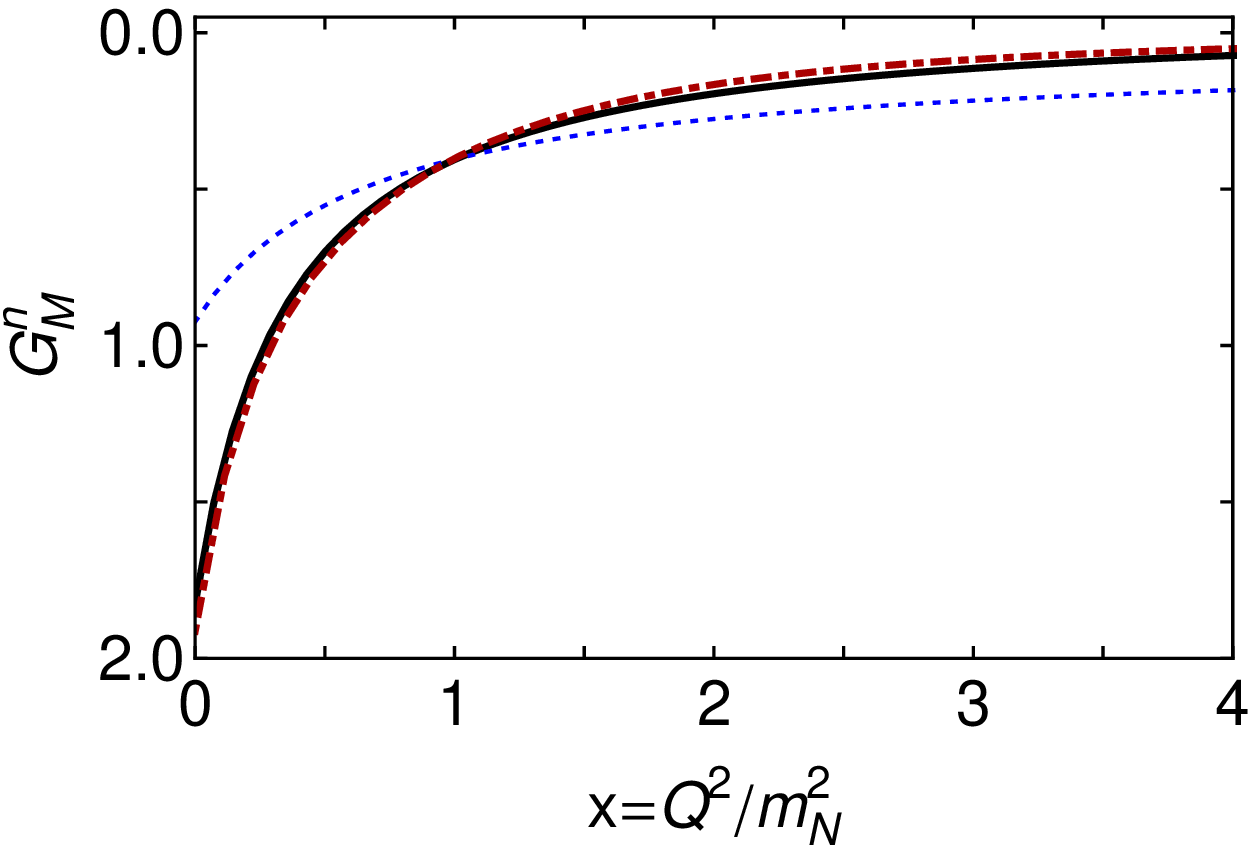}\vspace*
{-1ex}
\end{tabular}
\caption{\label{fig:FFNucleon1}
Proton (top) and neutron (bottom) electromagnetic form factors.
In both rows: {\it left panel} -- Sachs electric; {\it right panel} -- Sachs magnetic.
Curves in all panels: {\it solid, black} -- result obtained in Ref.\,\cite{Segovia:2014aza}, using a Faddeev equation kernel and interaction vertices that possess QCD-like momentum dependence, \emph{viz}. a QCD-kindred framework; {\it dotted, blue} -- result obtained with a symmetry preserving treatment of a contact interaction (CI framework) \cite{Wilson:2011aa}; {\it dot-dashed,red} -- 2004 parametrisation of experimental data \cite{Kelly:2004hm}.}
\end{center}
\end{figure}

\subsection{Predictions obtained with a Realistic Interaction}
The ultimate goal is \emph{ab initio} predictions of hadron observables within continuum QCD; and for the nucleon, $\Delta$-baryon and Roper resonance, studies of the Faddeev equation exist \cite{Eichmann:2008ef, Cloet:2008re, Eichmann:2011vu, Segovia:2014aza, Segovia:2015hraS} that are based on the one-loop renormalisation-group-improved interaction which was used efficaciously in the study of mesons.  All these analyses retain the scalar and axial-vector diquark correlations that are known to be necessary and sufficient for a reliable description of positive-parity baryons.  In order to compute baryon elastic form factors and the form factors describing transitions between them, it is necessary to know the manner through which a composite nucleon couples to a photon.  That is described in Ref.\,\cite{Oettel:1999gc}; and using this current one obtains predictions for all four nucleon elastic form factors, which are depicted in Fig.\,\ref{fig:FFNucleon1}.  The notation is explained by recalling that the Poincar\'e-covariant electromagnetic current for a spin-half nucleon is:
\begin{align}
J_\mu(K,Q) & =  ie\,\bar u(P_{f})\, \Lambda_\mu(K,Q) \,u(P_{i})
= i e \,\bar u(P_{f})\,\left( \gamma_{\mu} F_{1}(Q^{2}) +
\frac{1}{2m_{N}}\, \sigma_{\mu\nu}\,Q_\nu\,F_2(Q^2)\right) u(P_{i})\,,
\label{JnucleonB}
\end{align}
where $P_{i}$ ($P_{f}$) is the momentum of the incoming (outgoing) nucleon, $Q=
P_{f} - P_{i}$ is the photon momentum, $K=(P_{i}+P_{f})/2$ is the total momentum of the system, $K\cdot Q=0$ and $K^2 = - m_N^2 (1+\tau_N)$, $\tau_N = Q^2/(4 m_N^2)$ for elastic scattering; and $F_1$ and $F_2$ are, respectively, the Dirac and Pauli form factors, from which one obtains the nucleon's electric and magnetic (Sachs) form factors
\begin{equation}
\label{GEpeq}
G_E(Q^2)  =  F_1(Q^2) - \frac{Q^2}{4 m_{N}^2} F_2(Q^2)\,, \quad
G_M(Q^2)  =  F_1(Q^2) + F_2(Q^2)\,.
\end{equation}

It is apparent in Fig.\,\ref{fig:FFNucleon1} that the QCD-kindred results are in fair agreement with experiment, which is represented by the year-2004 parametrisation in Ref.\,\cite{Kelly:2004hm}.  (Comparisons made with a more recent parametrisation \cite{Bradford:2006yz} are not materially different.)  No parameters were tuned in order to achieve this outcome.  The most notable mismatch appears to be in the description of the neutron electric form factor at low $Q^2$.  However appearances are somewhat deceiving in this case because, on the low-$Q^2$ domain: $G_E^n$ is small and hence slight differences appear large; and $G_E^n$ is much affected by subdominant effects that were neglected in Ref.\,\cite{Segovia:2014aza}, such as so-called meson-cloud contributions.  On the other hand, as was previously observed \cite{Wilson:2011aa}, form factors obtained via a symmetry-preserving DSE treatment of a contact-interaction are typically too hard.  The defects of a contact-interaction are expressed with greatest force in the neutron electric form factor.  One thus arrives at a completely unambiguous conclusion: \\[0.7ex]
\hspace*{2em}\parbox[t]{0.9\textwidth}{\emph{Direct comparisons between experiment and sensibly formulated theory can distinguish between the momentum dependence of the interaction that underlies strong-interaction dynamics}.}\\

\hspace*{-\parindent}In this connection one should guard against models with no traceable connection to QCD, which are tuned to fit data, and practitioners inclined to obscure this empirical fact.

\begin{figure}[t]
\begin{center}
\begin{tabular}{cc}
\includegraphics[clip,width=0.47\linewidth]{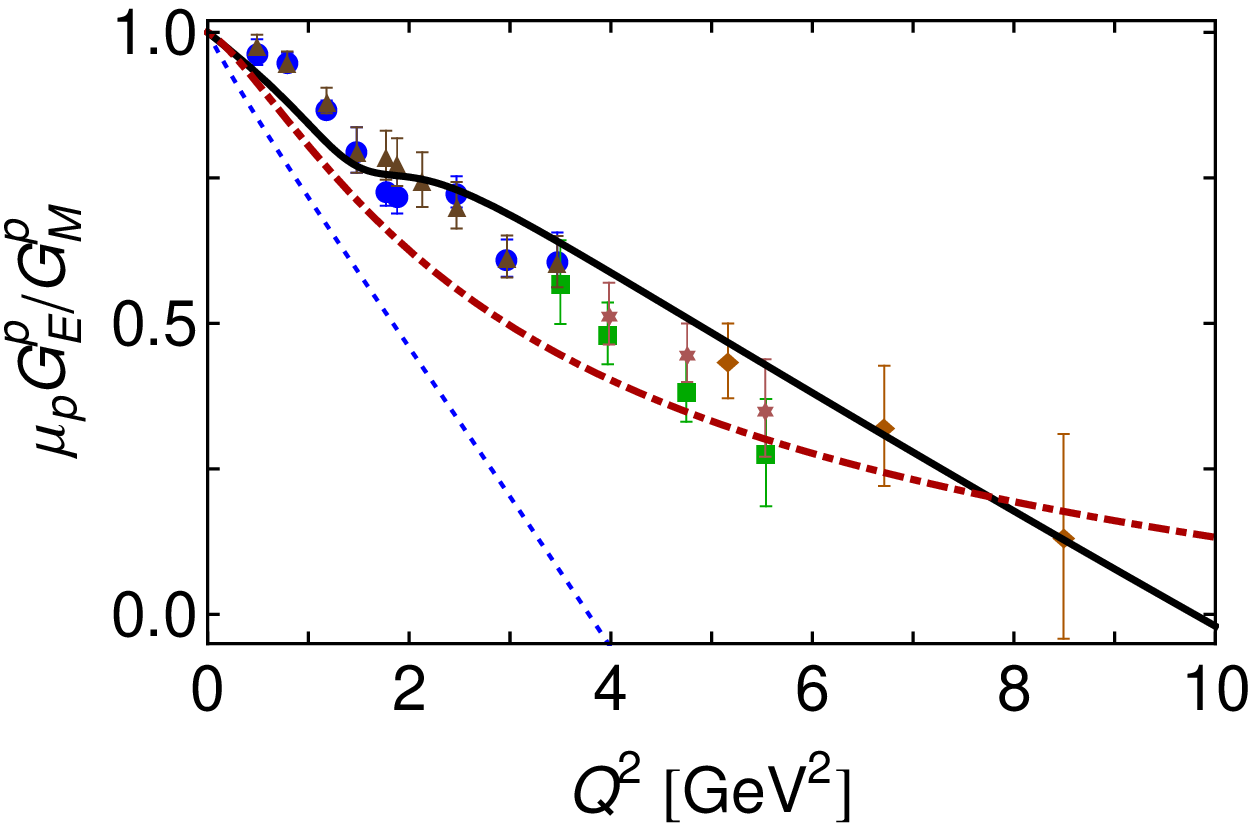}\vspace*
{-1ex } &
\includegraphics[clip,width=0.47\linewidth]{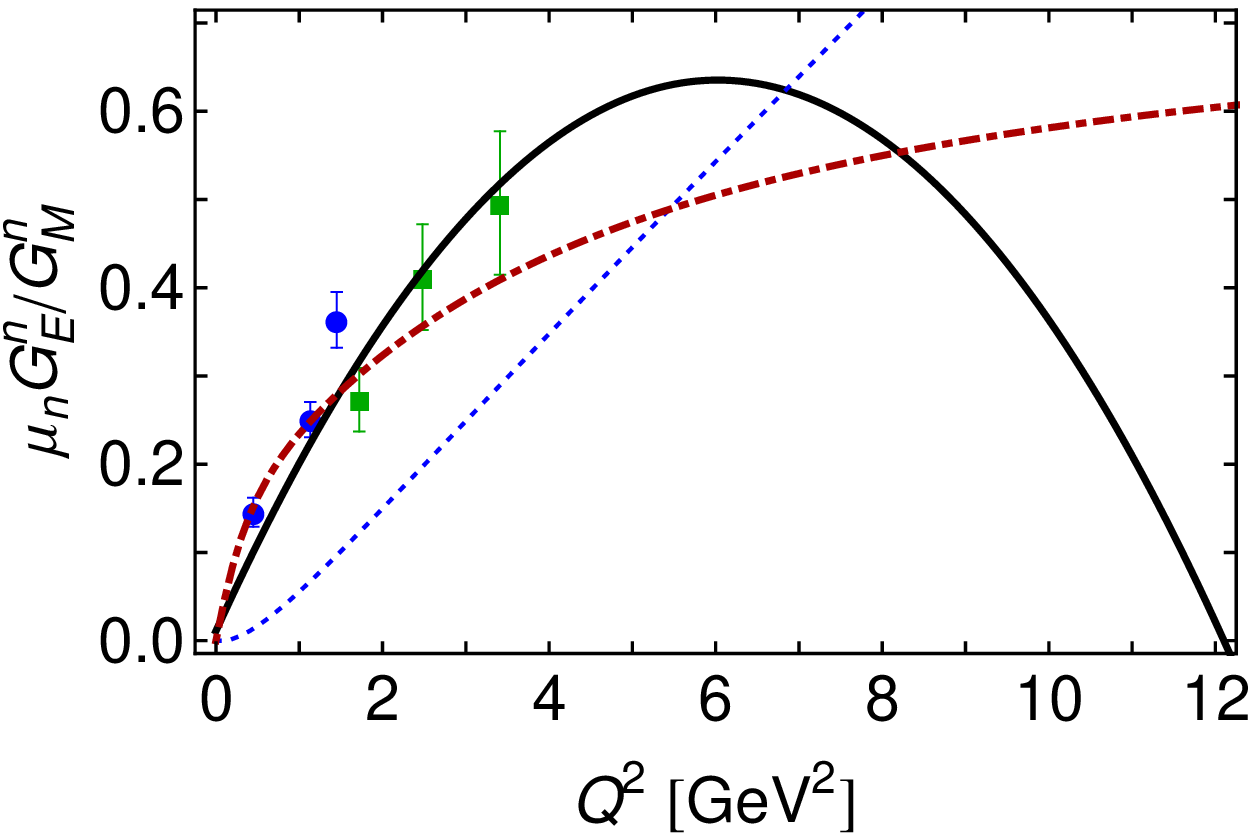}\vspace*
{-1ex}
\end{tabular}
\caption{\label{fig:FFNucleon2}
\emph{Left panel}: normalised ratio of proton electric and magnetic form factors.
Curves: {\it solid, black} -- result obtained in Ref.\,\cite{Segovia:2014aza} using the QCD-kindred framework; {\it Dashed, blue} -- CI result \cite{Wilson:2011aa}; and {\it dot-dashed, red} -- ratio inferred from 2004 parametrisation of experimental data \cite{Kelly:2004hm}.
Data:
blue circles \cite{Jones:1999rz};
green squares \cite{Gayou:2001qd};
brown triangles \cite{Punjabi:2005wqS};
purple asterisk \cite{Puckett:2010ac};
and orange diamonds \cite{Puckett:2011xgS}.
\emph{Right panel}: normalised ratio of neutron electric and magnetic form factors.  Curves: same as in left panel.
Data: blue circles \cite{Madey:2003av};
and green squares \cite{Riordan:2010idS}.}
\end{center}
\end{figure}

With predictions for the nucleon form factors in hand, one may revisit the empirical discovery described in connection with Eq.\,\eqref{GEGMRatio}, so in Fig.\,\ref{fig:FFNucleon2} I depict the unit-normalised ratio of Sachs electric and magnetic form factors for the proton and neutron.
Let us first consider the proton's ratio (left panel).  Both the contact-interaction and QCD-kindred frameworks predict a zero in $G_E^p/G_M^p$; but the comparison with extant experimental results indicates that the contact-interaction is invalid on $Q^2\gtrsim M(0)^2$, where $M(p^2)$ is the dressed-quark mass function explained in Sec.\,\ref{secDCSB}.  The result obtained with the QCD-kindred framework, on the other hand, agrees with available data and predicts a zero in this ratio at $Q^2\approx 9.5\,$GeV$^2$.   Notably, owing to the presence of strong diquark correlations, the singly-represented $d$-quark is usually sequestered inside a soft (scalar) diquark correlation.  The appearance of a zero is therefore driven primarily by the contribution to $G_E^p$ from the doubly-represented $u$-quark \cite{Cloet:2013jya}, which is four times more likely than the $d$-quark to be involved in a hard interaction.  (Additional novel insights are provided in Ref.\,\cite{Segovia:2015ufa}.)

As explained in Refs.\,\cite{Wilson:2011aa, Cloet:2013gva}, the behaviour of the dressed-quark contributions to the proton's electric form factor on $Q^2\gtrsim 5\,$GeV$^2$, and hence $G_E^p$ itself, are particularly sensitive to the rate at which the dressed-quark mass runs from the nonperturbative into the perturbative domain of QCD.  This is readily explicated using the information in the left panel of Fig.\,\ref{fig:FFNucleon2}.

The contact-interaction produces a momentum-independent dressed-quark mass; and in this counterpoint to QCD the dressed-quarks produce hard Dirac and Pauli form factors, which yield a ratio $\mu_p G_E/G_M$ that possesses a zero at $Q^2\lesssim 4\,$GeV$^2$.
Alternatively, the mass function used in Ref.\,\cite{Segovia:2014aza} is large at infrared momenta and approaches the current-quark mass as dressed-quark's momentum increases.  Such is the behaviour in QCD: dressed-quarks are massive in the infrared but become parton-like in the ultraviolet, characterised thereupon by a mass function that is modulated by the current-quark mass.  Hence, the proton's dressed-quarks possess constituent-quark-like masses at small momenta and thus have a large anomalous magnetic moment on this domain.  As the momentum transfer grows, the structure of the integrands in the computation of the elastic form factors ensures that the dressed-quark mass functions are increasingly sampled within the domain upon which the transition from nonperturbative to perturbative behaviour takes place.  This corresponds empirically to momentum transfers $Q^2 \gtrsim 5\,$GeV$^2$.  The rate at which the transition occurs determines how quickly the dressed-quarks become parton-like, \emph{i.e}.\ how rapidly they are unclothed and come to behave as light-fermion degrees of freedom.  Since light-quarks must have a small anomalous magnetic moment \cite{Chang:2010hb}, then this transition entails that the proton Pauli form factor, generated dynamically therewith, drops to zero.  This produces an interplay between the Dirac and Pauli form factors which, via Eq.\,\eqref{GEpeq}, entails that a momentum-dependent mass-function must beget a zero at larger values of $Q^2$ than is obtained with momentum-independent dressed-quark masses.

\begin{figure}[t]
\leftline{%
\includegraphics[clip,width=0.47\linewidth]{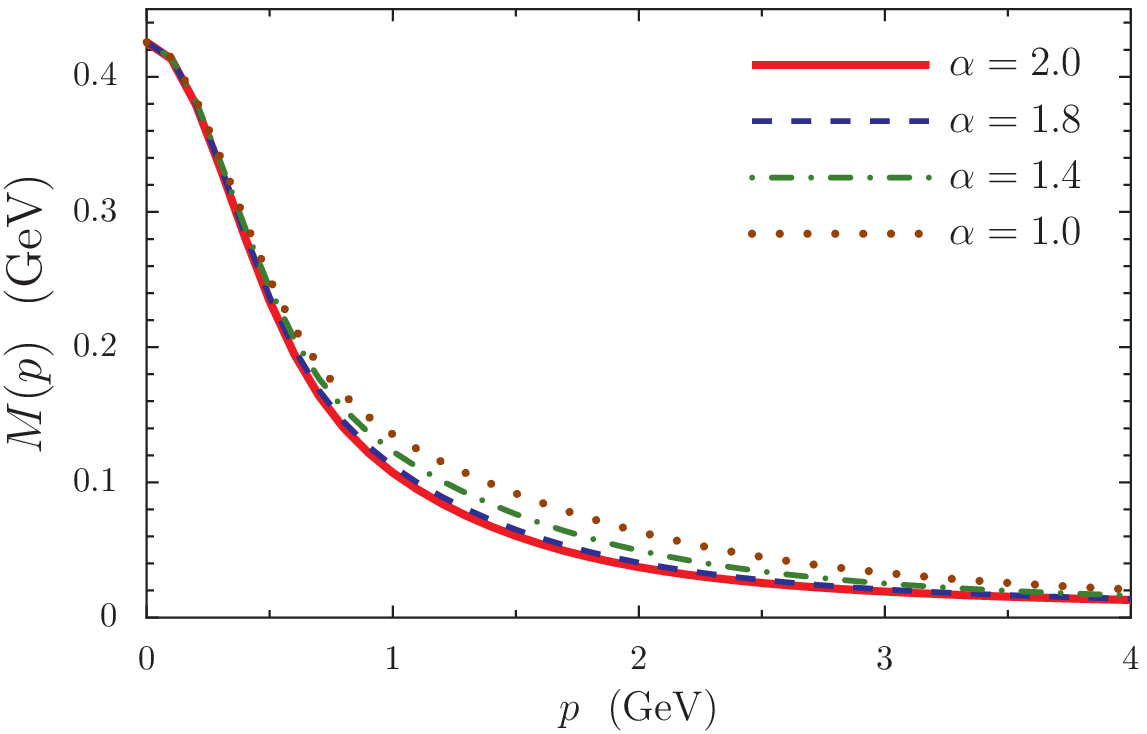}}
\vspace*{-29.5ex}

\rightline{%
\includegraphics[clip,width=0.485\linewidth]{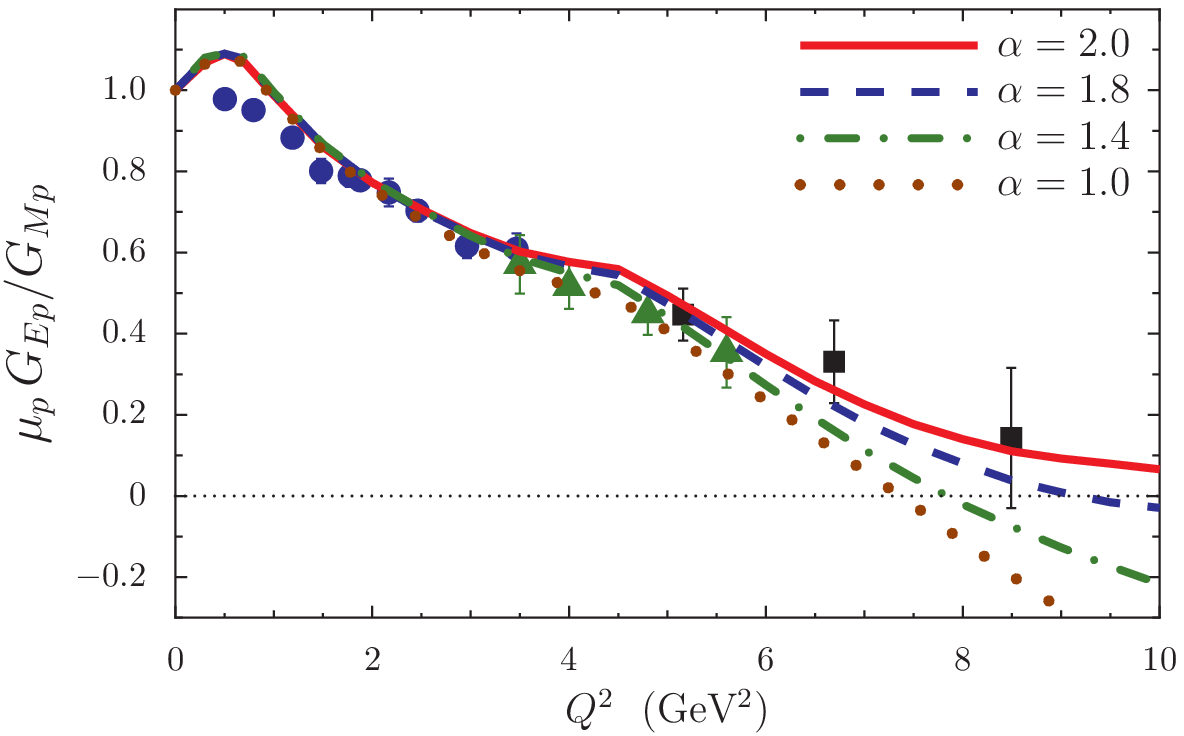}}
\caption{\label{figMpCRW} \small
\emph{Left panel}. Dressed-quark mass function employed in Ref.\,\protect\cite{Cloet:2013gva}.  $\alpha=1$ specifies the reference form, obtained in a least-squares fit to a diverse array of pion properties \protect\cite{Burden:1995ve},
and increasing $\alpha$ diminishes the domain upon which DCSB is active.
\emph{Right panel}. Response of $\mu_p G_E/G_M$ to increasing $\alpha$; i.e., to an increasingly rapid transition between constituent- and parton-like behaviour of the dressed-quarks.  Data are from Refs.\,\protect\cite{Jones:1999rz, Gayou:2001qd, Punjabi:2005wqS, Puckett:2010ac, Puckett:2011xgS}.}
\end{figure}

The dressed-quark mass function used in Ref.\,\cite{Segovia:2014aza} is characterised by a particular rate of transition between the nonperturbative and perturbative domains.  If one were to increase that rate, then the transformation to partonlike quarks would become more rapid and hence the proton's Pauli form factor would drop even more rapidly to zero.  In this case the quark angular momentum correlations, expressed by the diquark structure, remain but the individual dressed-quark magnetic moments diminish markedly.  Thus a more rapid transition pushes the zero in $\mu_p G_{Ep}/G_{Mp}$ to larger values of $Q^2$.  There is a rate of transformation beyond which the zero disappears completely \cite{Cloet:2013gva, Segovia:2015ufa}: as made plain in Fig.\,\ref{figMpCRW}, there is no zero at all in a theory in which the mass-function rapidly becomes partonic.

It follows that the possible existence and location of the zero in the ratio of proton elastic form factors [$\mu_p G_{E}^p(Q^2)/G_{M}^p(Q^2)$] are a fairly direct measure of the nature of the quark-quark interaction in the Standard Model.  Like the dilation of the meson valence-quark parton distribution amplitudes \cite{Chang:2013pqS, Cloet:2013tta, Segovia:2013ecaS, Gao:2014bca, Shi:2014uwaS}, they are a cumulative gauge of the momentum dependence of the interaction, the transition between the associated theory's nonperturbative and perturbative domains, and the width of that domain.  Hence, in extending experimental measurements of this ratio, and thereby the proton's charge form factor, to larger momentum transfers, \emph{i.e}.\ in reliably determining the proton's charge distribution, there is an extraordinary opportunity for a constructive dialogue between experiment and theory.  That feedback will assist greatly with contemporary efforts to reveal the character of the strongly interacting part of the Standard Model and its emergent phenomena.

\begin{figure}[t]
\begin{center}
\begin{tabular}{cc}
\includegraphics[clip,width=0.47\linewidth]{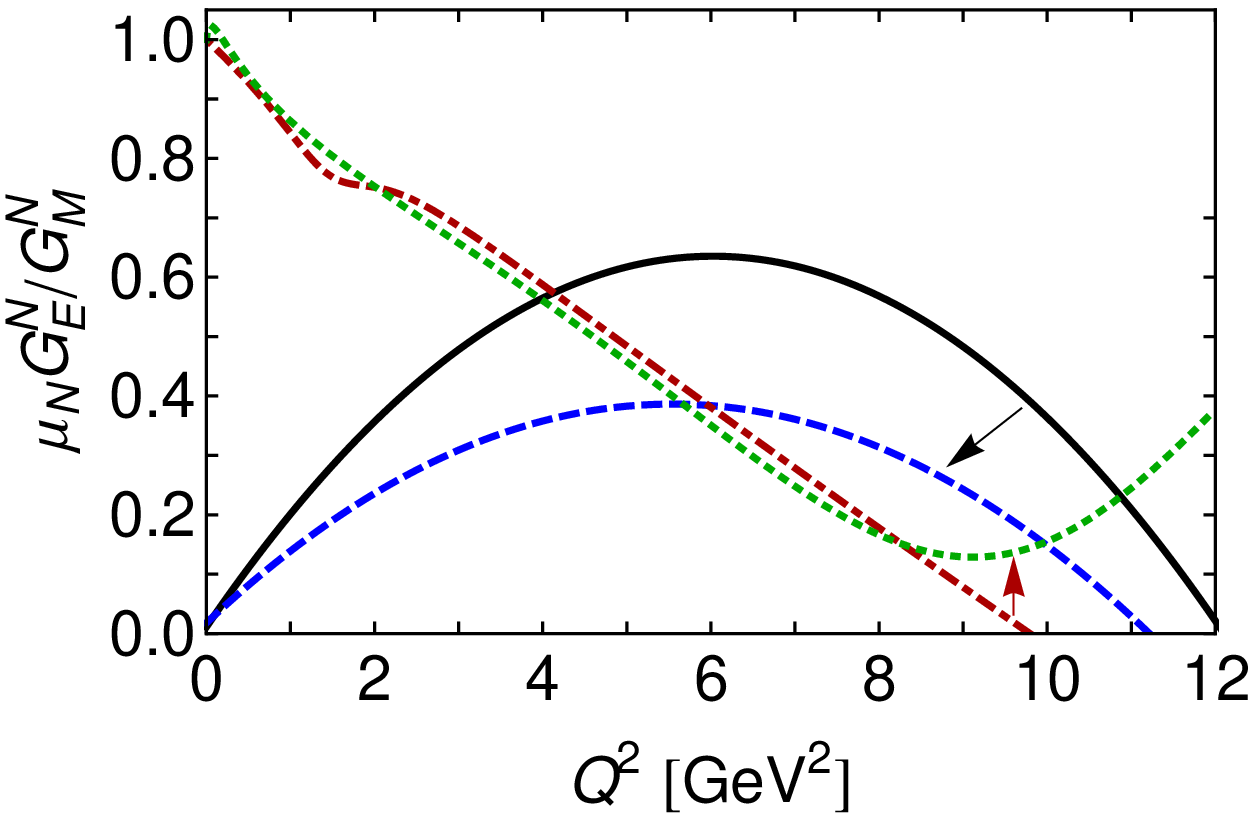}\vspace*
{-1ex } &
\includegraphics[clip,width=0.47\linewidth]{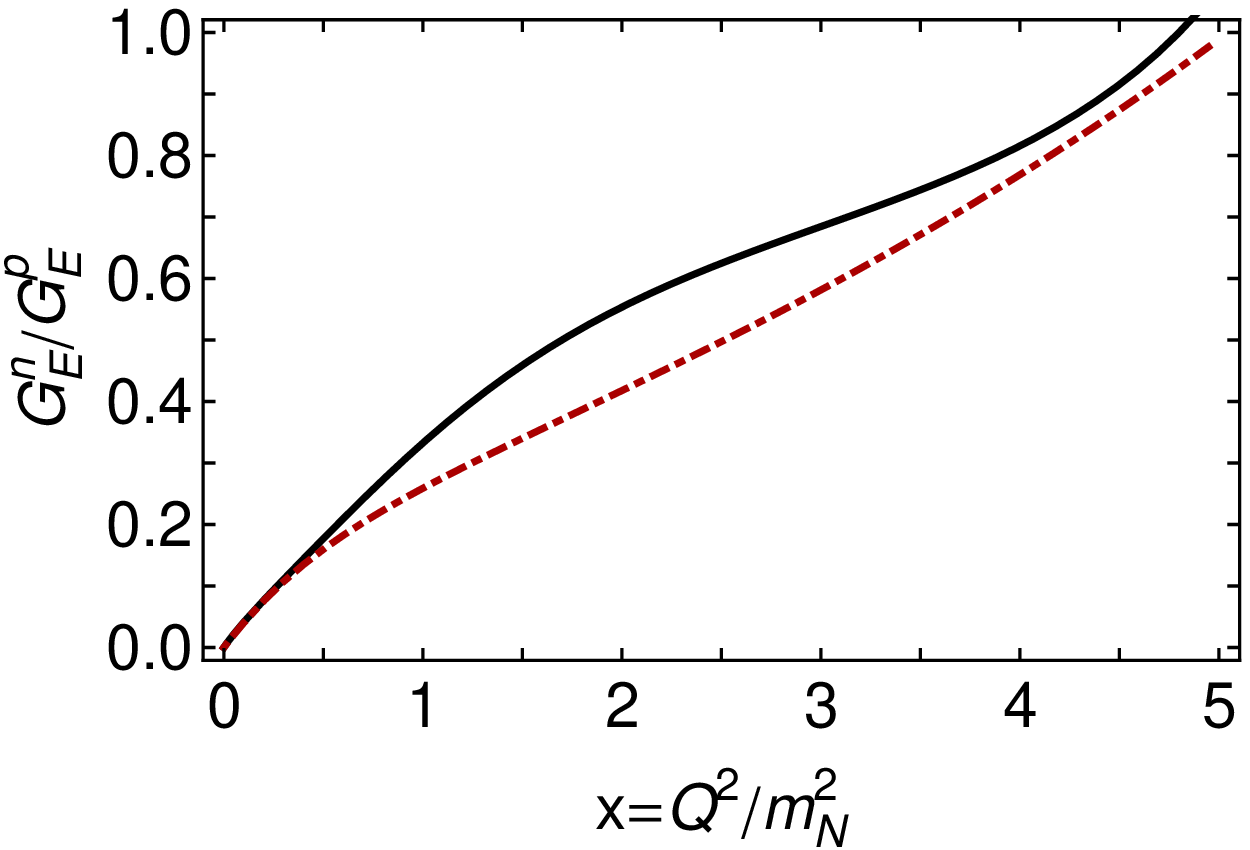} 
\end{tabular}
\caption{\label{fig:GEnGEp} \small
\emph{Left panel}.  Ratio $\mu_N G_E^N/G_M^N$ for $N=\,$neutron, proton:
{\it solid, black} -- neutron result obtained in Ref.\,\cite{Segovia:2014aza}, using QCD-like momentum-dependent quark dressing; {\it dashed, blue} -- neutron result obtained with such dressing but an accelerated rate of transition from dressed-quark to parton, \emph{viz}.\ $\alpha=2$; and \emph{dot-dashed, red} -- proton result obtained in Ref.\,\cite{Segovia:2014aza}, \emph{dotted-green} -- proton result with accelerated transition ($\alpha=2$).  (The arrows highlight the response to accelerating the dressed-quark$\,\to\,$ parton transformation.)
\emph{Right panel} -- Ratio of neutron and proton Sachs electric form factors:
{\it solid-black} -- result obtained in Ref.\,\cite{Segovia:2014aza}; and {\it dot-dashed-red} -- ratio inferred from 2004 parametrisation of experimental data \cite{Kelly:2004hm}.}
\end{center}
\end{figure}

Let us return now to the right panel of Fig.\,\ref{fig:FFNucleon2}, which displays the ratio $\mu_n G_E^n/G_M^n$.  The neutron ratio also exhibits a zero but at $Q^2\approx 12\,$GeV$^2$, \emph{i.e}. shifted to a 25\% larger value of $Q^2$ compared with the zero in the proton ratio.
The properties of the dressed-quark propagators and bound-state amplitudes which influence the appearance of a zero in $\mu_n G_E^n/G_M^n$ are qualitatively the same as those described in connection with $\mu_p G_{E}^p/G_{M}^p$.  However, owing to the different electric charge weightings of the quark contributions in the neutron, the quantitative effect is opposite to that for the proton.  Namely, when the transformation from dressed-quark to parton is accelerated, as described in Ref.\,\cite{Cloet:2013gva} and illustrated in Fig.\,\ref{figMpCRW}, the zero occurs at smaller $Q^2$.  This is depicted in the left panel of Fig.\,\ref{fig:GEnGEp}.  On the other hand, as indicated by the dotted curve in the lower-left panel of Fig.\,\ref{fig:FFNucleon1}, one typically finds that a contact interaction produces no zero in the neutron ratio \cite{Wilson:2011aa, Cloet:2014rja}.

The origin of these features is readily elucidated.  Note first that the $s$-quark contributes very little to nucleon electromagnetic form factors \cite{Aniol:2005zf, Armstrong:2005hs, Young:2006jc, Cloet:2008fw} and therefore write
\begin{equation}
\label{eqFlavourSep}
G_{E}^p = e_u G_E^{p,u} - |e_d| G_E^{p,d} \,,\quad
G_{E}^n = e_u G_E^{n,u} - |e_d| G_E^{n,d}\,,
\end{equation}
where the isolated terms denote the contribution from each quark flavour.  Consider next that charge symmetry is almost exact in QCD, so that
\begin{equation}
G_E^{n,d} = G_E^{p,u}\,, \quad G_E^{n,u} = G_E^{p,d}\,,
\end{equation}
and hence, to a very good level of approximation,
\begin{equation}
G_{E}^n = e_u G_E^{n,u} - |e_d| G_E^{n,d} =  e_u G_E^{p,d} - |e_d| G_E^{p,u}\,.
\label{GEnCS}
\end{equation}

Now, with a zero in $G_{E}^p$ at $Q^2 \approx 9.5\,$GeV$^2=:s_z$, one has $G_E^{p,d}(s_z) = 2 \,G_E^{p,u}(s_z)$ and hence $G_{E}^n(s_z) = G_E^{p,u}(s_z)>0$, where the last result is evident in Fig.\,7.3 of Ref.\,\cite{Cloet:2013jya}, which shows that although the behaviour of $G_E^{p,u}$ and $G_{E}^p$ is qualitatively similar, the zero in $G_E^{p,u}$ occurs at a larger value of $Q^2$ than that in $G_{E}^p$ itself.  Under these conditions, any zero in $G_{E}^n$ must occur at a larger value of $Q^2$ than the zero in $G_E^p$: compare the dot-dashed and solid curves in Fig.\,\ref{fig:GEnGEp}

This relative ordering of zeros can change, however, because, in contrast to $G_E^{p,u}$, $G_E^{p,d}$ evolves more slowly with changes in the rate at which the dressed-quark mass function transits from the nonperturbative to the perturbative domain, something which is also apparent in Fig.\,7.3 of Ref.\,\cite{Cloet:2013jya}.  As explained above, this inertia owes to the $d$-quark being preferentially sequestered inside a soft (scalar) diquark correlation.  Subject to these insights, consider Eq.\,\eqref{GEnCS}: with the location of a zero in $G_E^{p,d}$ shifting slowly to larger values of $Q^2$ but that in $G_E^{p,u}$ moving rapidly, one is subtracting from $G_E^{p,d}(Q^2)$ a function whose domain of positive support is becoming increasingly large.  That operation will typically shift the zero in $G_E^n$ to smaller values of $Q^2$ and eventually enable a zero in $G_E^n$ even when that in the $G_E^p$ has disappeared.

The right panel in Fig.\,\ref{fig:GEnGEp} displays a curious effect arising from the faster-than-dipole decrease of the proton's electric form factor (and possible appearance of a zero); namely, there will likely be a domain of $Q^2$ upon which the magnitude of the neutron's electric form factor exceeds that of the proton's.  This being the case, then at some value of momentum transfer the electric form factor of the neutral composite fermion becomes larger than that of its positively charged counterpart.  That occurs at $Q^2 = 4.8 M_N^2$ in the QCD-kindred analysis of Ref.\,\cite{Segovia:2014aza}.  JLab\,12 will test this prediction.

\begin{figure}[t]
\begin{minipage}[t]{\textwidth}
\begin{minipage}{0.47\textwidth}
\centerline{\includegraphics[width=0.95\textwidth]{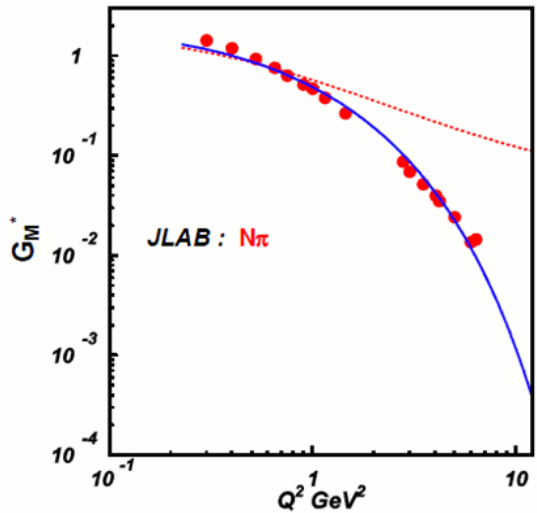}}
\end{minipage}
\begin{minipage}{0.5\textwidth}{\small
\caption{\label{figNDelta} \small
Comparison between CLAS data \cite{Aznauryan:2009mx} on the magnetic $\gamma+N\to\Delta$ transition form factor and a recent theoretical prediction \cite{Segovia:2014aza}.  The dashed curve shows the result that would be obtained if the interaction between quarks in QCD were momentum-independent \cite{Wilson:2011aa}.  The solid curve is obtained with precisely the same QCD-based formulation as was employed in a successful analysis of nucleon elastic form factors, which explains \cite{Cloet:2013gva} the behaviour of the ratio $\mu_p G_E^p(Q^2)/G_M^p(Q^2)$ described in connection with Eq.\,\eqref{GEGMRatio} and elucidated above.  The experiment-theory comparison confirms that experiments are sensitive to the momentum dependence of the running couplings and masses in QCD; and the theoretical unification of $N$ and $\Delta$ properties highlights the material progress that has been made in constraining the long-range behaviour of these fundamental quantities.  (Figure courtesy of V.\,Mokeev.)}}
\end{minipage}
\end{minipage}
\end{figure}

I would like to close this section by emphasising that given the challenges posed by non-perturbative QCD, it is insufficient to study hadron ground-states and elastic form factors alone.  Many novel perspectives and additional insights are provided by nucleon-to-resonance transition form factors, whose behaviour at large momentum transfers can reveal much about the long-range behaviour of the interactions between quarks and gluons \cite{Aznauryan:2012baS}.   Indeed, the properties of nucleon resonances are more sensitive to long-range effects in QCD than are those of hadron ground states.  The lightest baryon resonances are the $\Delta(1232)$-states; and despite possessing a width of 120\,MeV, these states are well isolated from other nucleon excitations.  Hence the $\gamma+N\to\Delta$ transition form factors have long been used to probe strong interaction dynamics.  They excite keen interest because of their use in probing, \emph{inter} \emph{alia},
the role that resonance electroproduction experiments can play in exposing non-perturbative features of QCD \cite{Aznauryan:2012baS}; the relevance of perturbative QCD to processes involving moderate momentum transfers \cite{Carlson:1985mm, Pascalutsa:2006up, Aznauryan:2011qj}; and
shape deformation of hadrons \cite{Alexandrou:2012da}.  Using the ``CLAS'' detector at JLab, precise data on the dominant $\gamma+N\to\Delta$ magnetic transition now reaches to $Q^2 = 8\,$GeV$^2$; an eventuality that poses both great opportunities and challenges for QCD theory, some of which have recently been met, as illustrated in Fig.\,\ref{figNDelta}.

\section{Epilogue}
This series of lectures has focused on quantum chromodynamics (QCD), the most interesting part of the Standard Model and Nature's only example of an essentially nonperturbative fundamental theory.  They have introduced the Dyson-Schwinger equations (DSEs), which provide a nonperturbative tool that may be used to gain insights and work toward a solution of QCD in the continuum.  In progressing through these lectures we have seen that the DSEs have enabled the proof of numerous exact results in QCD, amongst them:
\begin{itemize}
\item quarks are not simply quantum mechanical Dirac particles, they are far more complex objects;
\item gluons are nonperturbatively massive;
\item dynamical chiral symmetry breaking (DCSB) is a fact -- it's responsible for 98\% of the mass of visible matter in the Universe;
\item Goldstone's theorem is fundamentally an expression of near-equivalence between the one-body problem and the two-body problem in the pseudoscalar meson channel;
\item confinement in our real-world is a dynamical phenomenon, which can neither be expressed via any standard quantum mechanical potential nor discussed sensibly in the absence of dynamical light-quark degrees-of-freedom;
\item condensates, those quantities that have commonly been viewed historically as constant empirical mass-scales that fill all spacetime, are instead wholly contained within hadrons, {\emph i.e}.\ they are a property of hadrons themselves and expressed, for example, in their Bethe-Salpeter or light-front wave functions;
\item etc.
\end{itemize}
Moreover, we have seen the application of DSEs in explaining and predicting a wide range of experimental data.  Thus, at the conclusion of these lectures it will be clear that the DSEs are a single framework which is almost unique in providing an unambiguous path from a defined interaction $\to$ confinement and DCSB $\to$ hadron masses, radii, elastic and transition form factors, distribution functions, etc.  Namely, the DSEs are a practical, predictive, unifying tool for fundamental physics.

\medskip

\ack

The material described in these lecture notes is based on work completed by an international collaboration involving many remarkable people, to all of whom I am greatly indebted.
I would like to express my gratitude to the sponsors of \emph{Hadron Physics XIII -- XIII International Workshop on Hadron Physics}, without whom my participation would have been impossible; and to the organizers and my hosts, who ensured that the meeting was a success and my participation was enjoyable and fruitful.
Work supported by  U.S.\ Department of Energy, Office of Science, Office of Nuclear Physics, under contract no.~DE-AC02-06CH11357.


\providecommand{\newblock}{}

\end{document}